\documentclass[a4paper,11pt]{article}
\pdfoutput=1

\usepackage[a4paper,vmargin={30mm,30mm},hmargin={30mm,30mm}]{geometry}

\usepackage{lmodern}
\usepackage[T1]{fontenc}
\usepackage[utf8]{inputenc}

\usepackage[inline]{enumitem}

\usepackage{amsfonts}
\usepackage{amstext}
\usepackage{amsmath}
\usepackage{amssymb}
\usepackage{mathtools}
\usepackage{mathdots}
\usepackage{mathrsfs}
\usepackage{relsize}

\usepackage{arydshln}
\usepackage[margin=\parindent,labelfont=bf]{caption}

\usepackage{pbox}

\usepackage[usenames,dvipsnames]{xcolor}

\usepackage{tikz}
\usetikzlibrary{decorations.markings}
\usetikzlibrary{arrows}
\usetikzlibrary{backgrounds}

\numberwithin{equation}{section}

\usepackage[
colorlinks=true,
linkcolor=Red,
citecolor=Red,
urlcolor=Red,
unicode=true,
hypertexnames=false,
linktocpage=true,
bookmarksdepth=2
]{hyperref}
\hypersetup{
  pdfauthor={
    Nils Kanning, 
    Matthias Staudacher
  }, 
  pdftitle={
    Graßmannian Integrals in Minkowski Signature, Amplitudes, and Integrability
  }
}

\usepackage{cite}
\usepackage[parsep]{collref}

\usepackage{fancyhdr}
\DeclareMathOperator{\D}{d\!}
\DeclareMathOperator{\tr}{tr}
\DeclareMathOperator{\str}{str}
\DeclareMathOperator{\Real}{Re}
\DeclareMathOperator{\Imag}{Im}
\DeclareMathOperator{\sgn}{sgn}

\newcommand{\oscrep}{\mathscr{D}}

\newcommand{\smallunitary}{\mathcal{U}}

\newcommand{\indnm}{\mathcal}
\newcommand{\indssub}{\mathsf}

\usepackage[titles]{tocloft}

\begin{document} 

\baselineskip 5mm

\begin{titlepage}

  \hfill
  \pbox{5cm}{
    \texttt{HU-MATH-2018-10}\\
    \texttt{HU-EP-18/32}\\
    \texttt{LMU-ASC 72/18}
  }
  
  \vspace{2\baselineskip}

  \begin{center}
    \textbf{%
      \LARGE
      Graßmannian Integrals in Minkowski Signature,
      \\[1ex]
      Amplitudes, and Integrability
    }

    \vspace{2\baselineskip}

    Nils Kanning\textsuperscript{$\lambda$,$\overline{\lambda}$}
    and
    Matthias Staudacher\textsuperscript{$\lambda$}
 
    \vspace{2\baselineskip}

    \renewcommand{\thefootnote}{\fnsymbol{footnote}}
    
    \textit{
      \textsuperscript{$\lambda$}\!\!
      Institut für Mathematik und Institut für Physik, Humboldt-Universität zu Berlin,\\
      IRIS-Adlershof, Zum Großen Windkanal 6, 12489 Berlin, Germany\\
      \vspace{0.5\baselineskip}
      \textsuperscript{$\overline{\lambda}$}\!\!
      Arnold Sommerfeld Center for Theoretical Physics, Ludwig-Maximilians-Universität,\\
      Theresienstraße 37, 80333 München, Germany\footnotemark[1]\\
      \vspace{0.5\baselineskip}
    }
    \footnotetext[1]{Address until January 2018.}

    \vspace{2\baselineskip}

    \texttt{
      \{kanning,staudacher\}@physik.hu-berlin.de
    }

    \vspace{2\baselineskip}

    \textbf{Abstract}
    
  \end{center}

  \noindent
  We attempt to systematically derive tree-level scattering amplitudes
  in four-dimensional, planar, maximally supersymmetric Yang-Mills
  theory from integrability. We first review the connections between
  integrable spin chains, Yangian invariance, and the construction of
  such invariants in terms of Graßmannian contour integrals. Building
  upon these results, we equip a class of Graßmannian integrals for
  general symmetry algebras with unitary integration contours. These
  contours emerge naturally by paying special attention to the proper
  reality conditions of the algebras. Specializing to
  $\mathfrak{psu}(2,2|4)$ and thus to maximal superconformal symmetry
  in Minkowski space, we find in a number of examples expressions
  similar to, but subtly different from the perturbative physical
  scattering amplitudes. Our results suggest a subtle breaking of
  Yangian invariance for the latter, with curious implications for
  their construction from integrability.

\end{titlepage}

{\noindent}\hrulefill

\setcounter{tocdepth}{2}
\tableofcontents{}

\vspace{1\baselineskip}
{\noindent}\hrulefill
\vspace{1\baselineskip}

\newpage

\nocite{Kanning:2016eit}

\section{Introduction}
\label{sec:introduction}

Four-dimensional, planar, maximally supersymmetric Yang-Mills theory
is surely integrable at generic 't Hooft coupling, even though there
still is no precise formulation, let alone a completed proof, of this
statement. A presumably related problem is the absence of a systematic
derivation procedure for generic quantities from ``integrability'' in
this $\mathcal{N}=4$ SYM model. Once established, one would ideally
like to start from the underlying (algebraic?) integrable structure,
and subsequently systematically derive non-perturbatively all
quantities one fancies from a single principle: spectrum, states,
correlation functions, Wilson loops, scattering amplitudes, form
factors, etc. In practice, it is of course unlikely that one will
always succeed in completely ``solving'' for a given quantity of
interest. But at least one would like to know where to start from.

This lack of a solid starting point is particularly vexing in the case
of the scattering amplitudes of the model. Used by skillful hands, the
magic integrability machine yields, in certain kinematical limits and
employing various assumptions, highly non-trivial analytical results
for strong coupling, for an impressive number of loops at weak
coupling, and even, in some special cases, for generic coupling. What
then, precisely, is the underlying symmetry or principle for these
successes? It has been known now for nearly a decade that the
infinitesimal superconformal symmetry of tree-level amplitudes
combines with a distinct second, dual superconformal copy into a
Yangian symmetry, i.e.~into the mathematical structure underlying
(rational) integrable spin chains. But, in contradistinction to the
case of spin chains, it has not yet been possible to turn this around,
and to derive the tree-level amplitudes from this symmetry. The
present study initially set out to fill this gap. However, as we will
see below, this is far less straightforward than one might have
suspected. In fact, we shall find subtle differences between the
results obtained from Yangian invariance and the physical
amplitudes. We will discuss possible consequences of these surprising
results in our conclusion section \ref{sec:conclusions} below.

\section{Review: Amplitudes and Symmetries}
\label{sec:review}

We begin with a short review of tree-level scattering amplitudes in
planar $\mathcal{N}=4$ SYM theory. Naturally, we concentrate on those
aspects that form the foundation for our own work presented in this
paper, which is in part based on the PhD thesis \cite{Kanning:2016eit}
of one of the authors. In particular, we introduce spinor helicity
variables and highlight the reality conditions of the particle
momenta. Moreover, we discuss the integrable structure of the
amplitudes, which is closely tied to an infinite-dimensional Yangian
symmetry extending their superconformal symmetry. We recapitulate a
formulation of the amplitudes as Graßmannian integrals, which
conveniently exposes these symmetries. This formulation led to a
proposal for deformations of the amplitudes preserving Yangian
symmetry, which we also review. More extensive surveys of gauge theory
scattering amplitudes can be found in
\cite{Dixon:2013uaa,Henn:2014yza,Elvang:2015rqa} and special
properties in case of the $\mathcal{N}=4$ theory are discussed e.g.\
in \cite{Roiban:2010kk,Drummond:2010km}.

\subsection{Amplitudes in Spinor Helicity Variables}
\label{sec:spin-hel}

A scattering process in planar $\mathcal{N}=4$ SYM theory involving
the particles $i=1,\ldots, N$ is characterized by their null momenta
$p^i\in\mathbb{R}^{1,3}$ and helicities
$h^i=-1,-\frac{1}{2},0,+\frac{1}{2},+1$, whose range is determined by
the internal $\mathfrak{su}(4)$ R-symmetry. In addition, it depends on
color information originating from the gauge group $SU(N_\text{C})$.
However, this color structure can easily be stripped off from the
scattering amplitude. The total momentum is conserved in the
scattering process, $P=\sum_i p^i=0$. In general, this is not true for
the total helicity $\sum_i h^i$. The introduction of fermionic
variables $\tilde\eta^i_{\dot a}$ with $\dot a=1,2,3,4$ allows to
package all amplitudes with a common degree of helicity violation
$2K=N-\sum_i h^i$ into a single superamplitude. In the following, we
are dealing with \emph{color-stripped tree-level superamplitudes}
$\mathcal{A}_{N,K}$, which are functions of the momenta $p^i$ and the
fermions $\tilde\eta^i$. Mostly, we refer to them simply as
amplitudes.

The $\mathcal{A}_{N,K}$ can in principle be constructed employing the
textbook Feynman diagram approach. In practice, however, this is
almost infeasible because the number of diagrams grows exceedingly
fast with the number of particles $N$, see e.g.\ the discussion in
\cite{Kleiss:1988ne}. In addition, the computation of individual
diagrams completely obscures an unexpected simplicity in the
expressions for the complete amplitudes. To uncover this simplicity,
we have to choose suitable variables for the kinematics. This is
achieved by \emph{spinor helicity variables} for the particle momenta
$p^i$, which date back to the 1920s \cite{Waerden:1928,Weyl:1929}. To
introduce these variables, we use a bijection between Minkowski space
$\mathbb{R}^{1,3}$ and the space of Hermitian $2\times 2$ matrices. A
Minkowski vector $p=(p_\mu)$ is represented by the matrix
\begin{align}
  \label{eq:minkvec-hermmat}
  (p_{\alpha\dot{\beta}})=
  \begin{pmatrix}
    p_0+p_3&p_1-ip_2\\
    p_1+ip_2&p_0-p_3\\
  \end{pmatrix}\,,
\end{align}
where the indices take the values $\alpha,\dot\beta=1,2$. Using the
Minkowski inner product $p\cdot q=p_0\,q_0-\vec{p}\cdot\vec{q}$, one
verifies that $\det(p_{\alpha\dot{\beta}})=p^2$. Hence, for the null
momenta of the scattering process, the corresponding matrix is at most
of rank $1$ and can thus be factorized as
\begin{align}
  \label{eq:nullvec-spinors}
  p_{\alpha\dot{\beta}}=\lambda_\alpha\tilde{\lambda}_{\dot{\beta}}
\end{align}
with two spinors
$\lambda=(\lambda_\alpha),\tilde{\lambda}=(\tilde{\lambda}_{\dot{\beta}})\in\mathbb{C}^2$.
To turn this into a Hermitian matrix, we can without loss of
generality restrict to spinors satisfying the reality condition
\begin{align}
  \label{eq:spinors-real}
  \tilde\lambda=\pm\overline{\lambda}\,.
\end{align}
The sign here determines the sign of the energy as
$\pm 2 p_0=|\lambda_1|^2+|\lambda_2|^2$. In the field of scattering
amplitudes, one often works with complexified momenta, i.e.\
independent spinors $\lambda$ and $\tilde\lambda$ that do \emph{not}
obey the reality condition \eqref{eq:spinors-real}. This condition
will be of utmost importance for our work. Let us briefly motivate why
we prefer spinor helicity variables to twistors or momentum twistors
is this article. First, it is easy to work with real momenta by
imposing \eqref{eq:spinors-real}. What is more, the variables
$\lambda$, $\tilde\lambda$ together with the fermions $\tilde\eta$ are
associated with the superconformal algebra
$\mathfrak{psu}(2,2|4)$. They straightforwardly generalize to certain
oscillator representations of the superalgebra $\mathfrak{u}(p,q|m)$,
which will play an important role in section~\ref{sec:matrix-models}.

After setting up the formalism, we can discuss actual amplitudes
$\mathcal{A}_{N,K}$. In fact, expressions for all $\mathcal{A}_{N,K}$
are known \cite{Drummond:2008cr}. They involve the spinors in terms of
the Lorentz invariant angle and square brackets,
\begin{align}
  \label{eq:spinors-brackets}
  \langle i j\rangle=\lambda^i_1\lambda^j_2-\lambda^i_2\lambda^j_1\,,\quad
  [i j]=-\tilde\lambda^i_1\tilde\lambda^j_2+\tilde\lambda^i_2\tilde\lambda^j_1\,.
\end{align}
Making use of \eqref{eq:spinors-real}, these two brackets are related
by
\begin{align}
  \label{eq:spinors-brackets-conj}
  [ij]
  =
  -\sgn(p_0^i)\sgn(p_0^j)
  \overline{\langle ij\rangle}\,.
\end{align}
Moreover, the amplitudes contain generalized Mandelstam variables that
can be expanded in terms of the brackets,
\begin{align}
  \label{eq:spinors-mandelstam}
  s_{ij\cdots k}=(p^i+p^j+\ldots+p^k)^2
  =\!\!\!\sum_{\substack{u<v\\\in\{i,j,\ldots,k\}}}\!\!\!\langle u v\rangle[vu]\,.
\end{align}
Here we represent only the maximally helicity violating ($\text{MHV}$)
amplitudes explicitly, i.e.\ $K=2$. They are given by the unexpectedly
simple \emph{Parke-Taylor formula} \cite{Parke:1986gb}, or rather its
supersymmetric extension \cite{Nair:1988bq},
\begin{align}
  \label{eq:amp-super-mhv}
  \mathcal{A}_{N,2}
  =  
  \frac{\delta^{4}(P)\delta^{0|8}(Q)}
  {\langle 12\rangle\langle 23\rangle\cdots\langle N-1\,N\rangle\langle N1\rangle}\,,
\end{align}
which holds for $N\geq 4$. Momentum and supermomentum conservation are
implemented by
\begin{align}
  \label{eq:amp-bosferm-delta}
  \begin{aligned}
    \delta^4(P)
    &=
    \delta(P_{11})\delta(P_{22})
    \delta(\Real P_{21})\delta(\Imag P_{21})
    \quad\text{with}\quad
    P_{\alpha\dot{\beta}}=\sum_{i=1}^N\lambda_\alpha^i\tilde{\lambda}_{\dot{\beta}}^i\,,\\
    \delta^{0|8}(Q)&=
    \prod_{\alpha=1}^2\prod_{\dot{a}=1}^4Q_{\alpha\dot{a}}
    \quad\text{with}\quad
    Q_{\alpha\dot{a}}=\sum_{i=1}^N\lambda^i_\alpha\tilde\eta^i_{\dot{a}}\,.
  \end{aligned}
\end{align}
We stress that for momentum conservation to hold, both signs in
\eqref{eq:spinors-real} are needed, i.e. there have to be particles
with positive and negative energy. In a setting with complexified
momenta, \eqref{eq:amp-super-mhv} also yields a three-particle
amplitude $\mathcal{A}_{3,2}$. However, for spinors obeying
\eqref{eq:spinors-real}, this ceases to exist for purely kinematic
reasons, as explained e.g.\ in \cite{Dixon:2013uaa}.

\subsection{Yangian Symmetry}
\label{sec:yangian}

The tree-level amplitudes $\mathcal{A}_{N,K}$ are invariant under an
infinite-dimensional symmetry algebra, the \emph{Yangian} of the
superconformal algebra $\mathfrak{psu}(2,2|4)$. Let us start our
discussion a bit more general and consider the Yangian of the Lie
superalgebra $\mathfrak{gl}(n|m)$
\cite{Drinfeld:1985rx,Nazarov:1991}. An introduction to Yangians is
provided e.g.\ in the recent lectures
\cite{Loebbert:2016cdm}. Arguably, the most elegant way to define
these algebras is in the context of the \emph{quantum inverse
  scattering method} (QISM), which explores the consequences of the
Yang-Baxter equation to provide a toolbox for the study of integrable
models, see the authoritative review \cite{Faddeev:1996iy}. In this
language the generators of the Yangian are packaged into the entries
of a spin chain monodromy matrix. A Yang-Baxter equation obeyed by
this monodromy yields the commutation relations of the generators.

In what follows, we specify this monodromy matrix $M(u)$ in detail to
derive explicit formulas for the Yangian generators. It is built out
of Lax operators
\begin{align}
  \label{eq:yangian-def-lax}
  \begin{aligned}
    L_i(u-v_i)
    =
    1+(u-v_i)^{-1}\sum_{\indnm{A},\indnm{B}}E_{\indnm{AB}} J_{\indnm{BA}}^i(-1)^{|\indnm{B}|}
    =
    \enspace
  \end{aligned}
 \begin{aligned}
    \begin{tikzpicture}
      \draw[thick,densely dashed] 
      (0,0) 
      node[left] {$\square$} -- 
      (1,0);
      \draw[thick] 
      (0.5,-0.5) 
      node[below] {$i$} -- 
      (0.5,0.5);
    \end{tikzpicture}
 \end{aligned}
  \begin{aligned}
    \vphantom{\sum_{\indnm{X}}X^{|\indnm{X}|}}\enspace,
  \end{aligned}
\end{align}
where $u$ and $v_i$ are complex parameters referred to as
\emph{spectral parameter} and \emph{inhomogeneity}, respectively. The
Lax operator acts on the tensor product $\square\otimes\mathscr{V}_i$
of $\mathfrak{gl}(n|m)$ representations. The generators
$E_{\indnm{AB}}$ of the defining representation
$\square=\mathbb{C}^{n|m}$ are supermatrices satisfying in particular
$E_{\indnm{AB}}E_{\indnm{CD}}=\delta_{\indnm{BC}}E_{\indnm{AD}}$. The
generators of the representation $\mathscr{V}_i$ are denoted
$J_{\indnm{AB}}^i$. Both sets of generators obey the
$\mathfrak{gl}(n|m)$ algebra
\begin{align}
  \label{eq:gl-superalg}
  [J_{\indnm{AB}},J_{\indnm{CD}}\} =\delta_{\indnm{CB}}J_{\indnm{AD}}-
  (-1)^{(|\indnm{A}|+|\indnm{B}|)(|\indnm{C}|+|\indnm{D}|)}\delta_{\indnm{AD}}J_{\indnm{CB}}\,
\end{align}
with superindices such as $\indnm{A}=1,\ldots,n+m$ whose degree
$|\indnm{A}|=0,1$ depends on the grading. Out of these Lax operators,
we construct the monodromy matrix of an inhomogeneous spin chain with
$N$ sites,
\begin{align}
  \label{eq:yangian-mono-spinchain}
  \begin{aligned}
    \vphantom{\sum_{\indnm{X}}X^{|\indnm{X}|}}
    M(u)
    =
    L_1(u-v_1)L_2(u-v_2)
    \cdots L_N(u-v_N)
    =
    \enspace
  \end{aligned}
  \begin{aligned}
    \begin{tikzpicture}
      \draw[thick,densely dashed] 
      (0,0) 
      node[left] {$\square$} -- 
      (1.5,0);
      \draw[thick,densely dashed] 
      (2.5,0)  -- (3.5,0);
      \draw[thick] 
      (0.5,-0.5) 
      node[below] {$1$} -- 
      (0.5,0.5);
      \draw[thick] 
      (1.0,-0.5) 
      node[below] {$2$} -- 
      (1.0,0.5);
      \node at (2.0,0) {$\ldots$};
      \draw[thick] 
      (3,-0.5) 
      node[below] {$N$} -- 
      (3,0.5);
    \end{tikzpicture}
  \end{aligned}
  \begin{aligned}
    \vphantom{\sum_{\indnm{X}}X^{|\indnm{X}|}}\enspace.
  \end{aligned}
\end{align}
The product of Lax operators here is considered to be a matrix product
in the space $\square$ and a tensor product in the spaces
$\mathscr{V}_i$. The generators $M^{(l)}_{\indnm{AB}}$ with
$l=1,2,3,\ldots$ of the Yangian are obtained by expanding\footnote{The
  powers of $u$ in this expansion motivate our labeling of the
  generators by $l=1,2,3,\ldots$ despite the frequent use of the
  ``levels'' $l-1=0,1,2,\ldots$ for this purpose.} the elements of the
monodromy matrix in the spectral parameter $u$,
\begin{align}
  \label{eq:yangian-mono}
  \begin{aligned}
    M(u)&=\sum_{\indnm{A},\indnm{B}}E_{\indnm{AB}}M_{\indnm{AB}}(u)(-1)^{|\indnm{B}|}\,,\\
    M_{\indnm{AB}}(u)&=\delta_{\indnm{AB}}(-1)^{|\indnm{B}|}+u^{-1}M^{(1)}_{\indnm{AB}}+u^{-2}M^{(2)}_{\indnm{AB}}+\ldots\,.
  \end{aligned}
\end{align}
With the Lax operators in \eqref{eq:yangian-def-lax},
these generators acting in
$\mathscr{V}_1\otimes\cdots\otimes\mathscr{V}_N$ read
\begin{align}
  \label{eq:yangian-mono-coeff-glnm}
  M_{\indnm{AB}}^{(1)}=\sum_{i=1}^N J_{\indnm{BA}}^i\,,\quad
  M_{\indnm{AB}}^{(2)}=\sum_{i=1}^N v_iJ_{\indnm{BA}}^i+\sum_{\substack{i,j=1\\i<j}}^N\sum_{\indnm{C}} (-1)^{|\indnm{C}|}J_{\indnm{BC}}^jJ_{\indnm{CA}}^i\,,\quad
  \ldots\,.
\end{align}
We will encounter this form of the Yangian generators in
sections~\ref{sec:unitary-integral} and~\ref{sec:matrix-models}. In
the context of $\mathcal{N}=4$ SYM scattering amplitudes, one usually
works with slightly different looking generators
$M^{[l]}_{\indnm{AB}}$. These are obtained from another expansion of
the monodromy matrix,
\begin{align}
  \label{eq:yangian-exp-alternative}
  \begin{aligned}
    M(u)
    &=1+u^{-1}M^{(1)}+u^{-2}M^{(2)}+\ldots
    =\exp{\Big(u^{-1}M^{[1]}+u^{-2}M^{[2]}+\ldots\Big)}\,.
  \end{aligned}
\end{align}
We will provide explicit formulas for the generators
$M^{[l]}_{\indnm{AB}}$ momentarily, see \eqref{eq:sym-alg-amp-def} below.

Of central importance are \emph{Yangian invariants}. These are states
$|\Psi\rangle\in\mathscr{V}_1\otimes\cdots\otimes\mathscr{V}_N$ that
are annihilated by all Yangian generators,
\begin{align}
  \label{eq:yi-exp-1}
  M_{\indnm{AB}}^{(l)}|\Psi\rangle=0
\end{align}
for all $l=1,2,3,\ldots\,$. Due to the commutation relations of the
generators, see e.g.\ \cite{Kanning:2016eit}, it is actually
sufficient to verify this condition only for the first two sets of
generators with $l=1,2$. The Yangian invariance condition
\eqref{eq:yi-exp-1} can be expressed equivalently in terms of the
generators $M_{\indnm{AB}}^{[l]}$. In addition, it takes a natural
form when written employing the spin chain monodromy
\eqref{eq:yangian-mono-spinchain}
\cite{Frassek:2013xza,Chicherin:2013ora},
\begin{align}
  \label{eq:yi}
  M(u)|\Psi\rangle=1\,|\Psi\rangle\,,
\end{align}
where the identity operator on the right-hand side acts on
$\square=\mathbb{C}^{n|m}$.
We may represent this equation graphically as
\begin{align}
  \label{eq:yi-inv-rmm-pic}
  \begin{aligned}
    \begin{tikzpicture}
      \draw[thick,densely dashed] 
      (0,0) 
      node[left] {$\square$}
      node[left] {} -- 
      (1.5,0)
      node[right] {};
      \draw[thick,densely dashed] 
      (2.5,0) -- (3.5,0)
      node[right] {};
      \draw[thick] 
      (0.5,-0.5) 
      node[below] {$1$} -- 
      (0.5,0.5)
      node[above] {\phantom{$1$}};
      \draw[thick] 
      (1.0,-0.5) 
      node[below] {$2$} -- 
      (1.0,0.5)
      node[above] {\phantom{$2$}};
      \node at (2.0,0) {$\ldots$};
      \draw[thick] 
      (3,-0.5) 
      node[below] {$N$} -- 
      (3,0.5);
      \draw (1.75,1) 
      node[minimum height=1cm,minimum width=3.0cm,draw,
      thick,rounded corners=8pt,densely dotted] 
      {$|\Psi\rangle$};
      \path
      (0,2) 
      node[left] {\phantom{$\square$}} -- (3,2);
    \end{tikzpicture}
  \end{aligned}
  \,\,\,=
  \begin{aligned}
    \begin{tikzpicture}
      \draw[thick,densely dashed] 
      (0,2) 
      node[left] {$\square$}
      node[left] {} -- 
      (3.5,2)
      node[right] {};
      \draw[thick] 
      (0.5,-0.5) 
      node[below] {$1$} -- 
      (0.5,0.5)
      node[above] {\phantom{$1$}};
      \draw[thick] 
      (1.0,-0.5) 
      node[below] {$2$} -- 
      (1.0,0.5)
      node[above] {\phantom{$1$}};
      \node at (2.0,0) {$\ldots$};
      \draw[thick] 
      (3,-0.5) 
      node[below] {$N$} -- 
      (3,0.5);
      \draw (1.75,1) 
      node[minimum height=1cm,minimum width=3.0cm,draw,
      thick,rounded corners=8pt,densely dotted] 
      {$|\Psi\rangle$};
    \end{tikzpicture}
  \end{aligned}\,.
\end{align}
This brings Yangian invariants inside the realm of the QISM. Hence, it
potentially allows to construct them using the associated tools, such
as the algebraic Bethe ansatz \cite{Frassek:2013xza}.

After this detour, we return to the tree-level amplitudes
$\mathcal{A}_{N,K}$ of planar $\mathcal{N}=4$ SYM. Their Yangian
invariance was discovered in \cite{Drummond:2009fd} by combining the
well-known invariance under the superconformal algebra
$\mathfrak{psu}(2,2|4)$ with a rather unexpected occurrence of a
second copy of $\mathfrak{psu}(2,2|4)$ termed \emph{dual
  superconformal symmetry}. Reviews on the Yangian symmetry of
amplitudes and on its relevance for other observables of planar
$\mathcal{N}=4$ SYM are provided in
\cite{Beisert:2010jq,Loebbert:2016cdm,Ferro:2018ygf}. Instead of
following the historic route, we apply the formalism introduced in the
preceding paragraphs to arrive at Yangian generators that annihilate
the amplitudes $\mathcal{A}_{N,K}$.

A function annihilated by the generators of $\mathfrak{psu}(2,2|4)$ is
also annihilated by any complex linear combination thereof and hence
by the complexified algebra
$\mathfrak{psl}(\mathbb{C}^{4|4})\equiv\mathfrak{psl}(4|4)$. Generators
$\mathfrak{J}_{\indnm{AB}}$ of
$\mathfrak{gl}(4|4)\supset \mathfrak{psl}(4|4)$ are easily realized in
terms of spinor helicity variables. Arranged into a supermatrix they
read
\begin{align}
  \label{eq:sym-gen-gl44}
  (\mathfrak{J}_{\indnm{AB}})
  =
  \left(
    \begin{array}{c:c:c}
    \lambda_\alpha\partial_{\lambda_\beta}&
    \lambda_\alpha\tilde{\lambda}_{\dot\beta}&
    \lambda_\alpha\tilde\eta_{\dot b}\\[0.3em]
    \hdashline&&\\[-1.0em]
    -\partial_{\tilde{\lambda}_{\dot\alpha}}\partial_{\lambda_\beta}&
    -\partial_{\tilde{\lambda}_{\dot\alpha}}\tilde{\lambda}_{\dot\beta}&
    -\partial_{\tilde{\lambda}_{\dot\alpha}}\tilde\eta_{\dot b}\\[0.3em]
    \hdashline&&\\[-1.0em]
    \partial_{\tilde\eta_{\dot a}}\partial_{\lambda_\beta}&
    \partial_{\tilde\eta_{\dot a}}\tilde{\lambda}_{\dot\beta}&
    \partial_{\tilde\eta_{\dot a}}\tilde\eta_{\dot b}\\
    \end{array}
  \right)\,.
\end{align}
Here we split the superindices such as $\indnm{A}=1,\ldots,8$ into
bosonic indices $\alpha,\dot\alpha=1,2$ and a fermionic index
$\dot a=1,2,3,4$. In order to restrict to the algebra
$\mathfrak{psl}(4|4)$, we have to impose
$\mathfrak{C}=\tr(\mathfrak{J}_{\indnm{AB}})=\sum_{\indnm{A}}\mathfrak{J}_{\indnm{AA}}=0$
and
$\mathfrak{B}=\str(\mathfrak{J}_{\indnm{AB}})=\sum_{\indnm{A}}(-1)^{|\indnm{A}|}\mathfrak{J}_{\indnm{AA}}=0$.
Generators of $\mathfrak{sl}(4|4)$ are obtained from
\eqref{eq:sym-gen-gl44} by defining
\begin{align}
  \label{eq:sym-alg-gen-sl}
  \mathfrak{J}'_{\indnm{AB}}
  =
  \mathfrak{J}_{\indnm{AB}}-\frac{1}{8}(-1)^{|A|}\delta_{\indnm{AB}}\mathfrak{B}\,.
\end{align}
They satisfy $\mathfrak{C}'=\mathfrak{C}$ and $\mathfrak{B}'=0$. Let
us consider a spin chain monodromy matrix $M(u)$ as in
\eqref{eq:yangian-mono-spinchain} with generators
$J_{\indnm{AB}}^i=\mathfrak{J}^i_{\indnm{AB}}$ of the form
\eqref{eq:sym-gen-gl44} in the Lax operators. Expanding this monodromy
as in \eqref{eq:yangian-exp-alternative} yields the Yangian generators
\begin{align}
  \label{eq:sym-alg-amp-def}
  \begin{aligned}
    M^{[1]}_{\indnm{AB}}=\sum_{i=1}^N\mathfrak{J}_{\indnm{BA}}^i\,,\quad
    M_{\indnm{AB}}^{[2]}
    =\frac{1}{2}\sum_{\substack{i,j=1\\i<j}}^N
    \sum_{\indnm{C}} (-1)^{|\indnm{C}|}\Big(\mathfrak{J}_{\indnm{BC}}^j\mathfrak{J}_{\indnm{CA}}^i
    -\mathfrak{J}_{\indnm{BC}}^i\mathfrak{J}_{\indnm{CA}}^j\Big)
    +\sum_{i=1}^N\hat{v}_i\mathfrak{J}^i_{\indnm{BA}}\,.\\
  \end{aligned}
\end{align}
To obtain these expressions, we used the form of the generators in
\eqref{eq:sym-gen-gl44}, assumed $\sum_{i=1}^N\mathfrak{C}^i=0$, and
introduced $\hat{v}_i=v_i-\frac{c_i}{2}+\frac{1}{2}$ with $c_i$ being
the eigenvalue of $\mathfrak{C}^i$. For the amplitudes, these
eigenvalues are closely related to the superhelicities of the
particles, and they have to vanish, $c_i=0$. In order to act with
\eqref{eq:sym-alg-amp-def} on the amplitudes, we also have to set
$\hat{v}_i=0$. Then the amplitudes are annihilated by the Yangian
generators after replacing the $\mathfrak{gl}(4|4)$ generators
$\mathfrak{J}^i_{\indnm{AB}}$ in \eqref{eq:sym-alg-amp-def} by the
$\mathfrak{sl}(4|4)$ generators from \eqref{eq:sym-alg-gen-sl}, which
we indicate by apostrophes,
\begin{align}
  \label{eq:sym-alg-annih}
  M'^{[1]}_{\indnm{AB}}\mathcal{A}_{N,K}=0\,,\quad
  M'^{[2]}_{\indnm{AB}}\mathcal{A}_{N,K}=0\,.
\end{align}
This is the form of the Yangian invariance condition typically found
in the amplitudes literature, see e.g.\ \cite{Drummond:2009fd}. The
first equation in \eqref{eq:sym-alg-annih} is the ordinary action of a
Lie superalgebra on a tensor product and corresponds to superconformal
invariance.

We remark that a careful analysis, taking into account the reality
conditions \eqref{eq:spinors-real} of the spinor variables, reveals of
a breaking of the Yangian invariance \eqref{eq:sym-alg-annih}, and
even the superconformal invariance alone, at certain singularities of
the amplitudes
\cite{Bargheer:2009qu,Sever:2009aa,Bargheer:2011mm}. Because this
issue does not occur for generic particle momenta, it is often
neglected in the discussion of tree-level amplitudes.

\subsection{Graßmannian Integral and Deformations}
\label{sec:grassdef}

In this section, we discuss a very compact formulation of the
amplitudes $\mathcal{A}_{N,K}$ in terms of certain multi-dimensional
contour integrals called \emph{Graßmannian integrals}
\cite{ArkaniHamed:2009dn,Mason:2009qx}, see also the extensive
treatise\footnote{This book project grew out of the influential
  preprint \cite{ArkaniHamed:2012nw}.} \cite{Arkani-Hamed:2016}. This
approach particularly suites our interests because it allows for an
easy investigation of symmetries. While the superconformal symmetry is
manifest in this formulation, also the Yangian symmetry can be
verified \cite{Drummond:2010qh,Drummond:2010uq}.

Before defining the Graßmannian integral, we have to discuss some bare
essentials of Graßmannian manifolds. The complex \emph{Graßmannian}
$Gr(K,N)$ is defined as the space of all $K$-dimensional linear
subspaces of $\mathbb{C}^N$. ``Homogeneous'' coordinates on this space
are provided by the complex entries of a $K\times N$ matrix $C$.
Selecting a basis within a given subspace does not alter the point in
the Graßmannian. Consequently, a generic point in $Gr(K,N)$ can be
described by the ``gauge fixed'' matrix
\begin{align}
  \label{eq:grassint-matrix}
  C=
  \left(
  \begin{array}{c:c}
    1_{K}&\mathcal{C}\\
  \end{array}
  \right)
  \quad
  \text{with}
  \quad
  \mathcal{C}=
  \begin{pmatrix}
    C_{1 K+1}&\cdots&C_{1 N}\\
    \vdots&&\vdots\\
    C_{K K+1}&\cdots&C_{K N}\\
  \end{pmatrix},
\end{align}
where $1_K$ denotes the $K\times K$ unit matrix. In what follows, we
will also encounter the $(N-K)\times N$ matrix
$C^\perp=\big(\begin{array}{c:c}-\mathcal{C}^t&1_{N-K}\end{array}\big)$
obeying $C(C^\perp)^t=0$. It is an element of
$Gr(N-K,N)$. These ingredients are sufficient to present the
\emph{Graßmannian integral} formulation of amplitudes
\cite{ArkaniHamed:2009dn},
\begin{align}
  \label{eq:grassint-amp}
  \mathcal{A}_{N,K}
  =
  \int\!\D^{\,K(N-K)}\!\mathcal{C}
  \frac{
    \delta^{2(N-K)|0}_\ast(C^\perp\boldsymbol{\lambda})
    \delta^{2K|0}_\ast(C\boldsymbol{\tilde{\lambda}})
    \delta^{0|4K}(C\boldsymbol{\tilde\eta})
  }
  {(1,\ldots,K)\cdots(N,\ldots,K-1)}\,
\end{align}
with the holomorphic form
$\D^{\,K(N-K)}\!\mathcal{C}=\bigwedge_{k,l}\D C_{k l}$. Here
$(i,\ldots,i+K-1)$ signifies the minor of the matrix $C$ constructed
from the consecutive columns $i,\ldots,i+K-1$. These are counted
modulo $N$ such that they stay within the range $1,\ldots,N$. The
external data is encoded in the $N\times 2$ matrices
$\boldsymbol{\lambda}=(\lambda^i_\alpha)$ and
$\boldsymbol{\tilde\lambda}=(\tilde\lambda^i_{\dot{\alpha}})$ as well
as the $N\times 4$ matrix
$\boldsymbol{\tilde\eta}=(\tilde\eta_{\dot{a}}^i)$. The symbol
$\delta_\ast$ denotes a formal bosonic delta function whose argument
may be complex. It can be understood as a calculation rule to set the
argument to zero and omit an integration. To evaluate the Graßmannian
integral \eqref{eq:grassint-amp}, one first uses the formal bosonic
delta functions to reduce the number of integration variables. Then
one specifies a contour for the remaining variables, which has to be
closed to ensure Yangian invariance. The resulting integral can be
evaluated by means of a multi-dimensional generalization of Cauchy's
residue theorem, the so-called ``global residue theorem'', see the
discussion in \cite{ArkaniHamed:2009dn}. Suitable contours are known
for all amplitudes $\mathcal{A}_{N,K}$. They can be specified in a
geometric fashion \cite{Bourjaily:2010kw}, generalizing partial
results in \cite{Nandan:2009cc,ArkaniHamed:2009dg}.

Let us mention a significant open problem of the Graßmannian integral
approach in the form presented in \eqref{eq:grassint-amp}. In case of
the physical Minkowski signature $(1,3)$, the spinors obey the reality
conditions \eqref{eq:spinors-real}. Hence, the spinors contained in
$\boldsymbol{\tilde\lambda}$ depend on those in
$\boldsymbol{\lambda}$. This is neglected in the standard way of
evaluating the integral, which we just sketched. The issue is commonly
evaded by working in split signature $(2,2)$ or in a complexified
momentum space, where $\boldsymbol{\lambda}$ and
$\boldsymbol{\tilde\lambda}$ are treated as independent real or
complex variables, respectively.

In discussing the Yangian invariance of the amplitudes
$\mathcal{A}_{N,K}$ around \eqref{eq:sym-alg-annih}, we noted that we
had to set $c_i=0$ and $\hat v_i=0$ in the Yangian generators
\eqref{eq:sym-alg-amp-def}. This naturally leads to the question
whether there exist deformed amplitudes
$\mathcal{A}_{N,K}^{(\text{def\/})}$ for which these parameters do not
vanish. It was first raised in \cite{Ferro:2012xw,Ferro:2013dga}, one
motivation being that complex deformation parameters might serve as
integrability-based regulators of loop amplitudes. In addition, the
deformations should be crucial to properly understand the integrable
structure of amplitudes and to eventually put this structure to use
for their efficient construction, even at all-loop level. The study of
deformed amplitudes continued in
\cite{Beisert:2014qba,Broedel:2014pia,Broedel:2014hca}. It resulted in
a \emph{deformed Graßmannian integral}
\cite{Ferro:2014gca,Bargheer:2014mxa}
\begin{align}
  \label{eq:grassint-amp-def}
  \mathcal{A}_{N,K}^{(\text{def\/})}
  =
  \int\displaylimits_{?}\!\D^{\,K(N-K)}\!\mathcal{C}\,
  \frac{
    \delta^{2(N-K)|0}_\ast(C^\perp\boldsymbol{\lambda})
    \delta^{2K|0}_\ast(C\boldsymbol{\tilde{\lambda}})
    \delta^{0|4K}(C\boldsymbol{\tilde\eta})
  }
  {(1,\ldots,K)^{1+\hat{v}_K^--\hat{v}_1^+}
    \cdots 
    (N,\ldots,K-1)^{1+\hat{v}^-_{K-1}-\hat{v}_N^+}}\,.
\end{align}
Here the exponents are defined by \cite{Beisert:2014qba}
\begin{align}
  \label{eq:amp-wpm}
  \hat{v}_i^\pm=\hat{v}_i\pm\frac{c_i}{2}\,.
\end{align}
They have to satisfy
\begin{align}
  \label{eq:amp-wperm}
  \hat{v}^-_{i+K}=\hat{v}^+_i\,
\end{align}
for $i=1,\ldots,N$, where we count modulo $N$. This condition ensures
the Yangian invariance \eqref{eq:sym-alg-annih} of the integral
\eqref{eq:grassint-amp-def}, provided a closed integration
contour. Furthermore, starting from the Yangian invariance condition
\eqref{eq:yi} involving a spin chain monodromy,
\eqref{eq:grassint-amp-def} can even be derived using tools rooted in
the QISM \cite{Ferro:2014gca}. These tools were introduced in
\cite{Chicherin:2013sqa,Chicherin:2013ora} and studied more
systematically in \cite{Kanning:2014maa,Broedel:2014pia}. However,
this method uses somewhat formal integral operators and does not yield
a suitable contour for the resulting deformed Graßmannian integral
\eqref{eq:grassint-amp-def}. Deformations of the $\text{MHV}$
amplitudes \eqref{eq:amp-super-mhv} can be obtained from
\eqref{eq:grassint-amp-def} without specifying a contour because all
integration variables are fixed by the bosonic delta functions,
\begin{align}
  \label{eq:amp-def-mhv}
  \mathcal{A}_{N,2}^{(\text{def\/})}
  =
  \frac{\delta^{4}(P)\delta^{0|8}(Q)}{
    \langle 12\rangle^{1+\hat{v}_2^--\hat{v}_1^+}
    \cdots
    \langle N1\rangle^{1+\hat{v}_1^--\hat{v}_N^+}
  }\,.
\end{align}

This brings us to the main challenge of understanding the deformed
amplitudes $\mathcal{A}_{N,K}^{(\text{def\/})}$. It is not known how to
evaluate the integral \eqref{eq:grassint-amp-def} beyond the
$\text{MHV}$ case. Because of the complex exponents of the minors in
the denominator, Cauchy's residue theorem and its multi-dimensional
generalization do not apply any longer. The exponents lead to branch
cuts and the resulting multi-valuedness of the integrand makes it very
difficult to find a closed integration contour, which is necessary for
Yangian invariance. We will address this problem in
section~\ref{sec:unitary-integral}. It turns out that the choice of the
integration contour is tightly interrelated with using the proper
Minkowski reality conditions \eqref{eq:spinors-real} for the spinors
in $\boldsymbol{\lambda}$ and $\boldsymbol{\tilde\lambda}$.

Finally, let us mention that the Yangian invariance condition for
$\mathcal{A}^{(\text{def\/})}_{4,2}$ can be shown to be equivalent to a
Yang-Baxter equation. This suggests interpreting
$\mathcal{A}^{(\text{def\/})}_{4,2}$ as an R-matrix, where one of its
deformation parameters is the spectral parameter of this R-matrix
\cite{Ferro:2012xw,Ferro:2013dga}. In this interpretation, the
undeformed amplitude $\mathcal{A}_{4,2}$ corresponds to the R-matrix
evaluated at a special point of the spectral parameter. Interestingly,
this interpretation appears to be at odds with \cite{Zwiebel:2011bx},
where $\mathcal{A}_{4,2}$ was related to the one-loop dilatation
operator of the planar $\mathcal{N}=4$ SYM spectral problem. This
operator is not an R-matrix itself but can be constructed from one. We
believe that this conceptual difference deserves further attention and
will revisit it in section~\ref{sec:r-matrix}.

\section{Unitary Graßmannian Integral}
\label{sec:unitary-integral}

Our aim in this section is to construct a refined version of the
deformed Graßmannian integral \eqref{eq:grassint-amp-def} that
respects the reality conditions \eqref{eq:spinors-real} of the spinor
helicity variables in Minkowski signature. We will start with what
seems to be the most involved step, finding a suitable integration
contour. Then we can formulate the sought after refined Graßmannian
formula. In particular, we discuss the crucial single-valuedness of
its integrand. Finally, we define Yangian generators that annihilate
our integral formula and specify the occurring representations.

\subsection{Reality Conditions and Unitary Contour}
\label{sec:contour}

First, we have to introduce our setting. From now on, we restrict
ourselves to the case $N=2K$, which corresponds to \emph{helicity
  conserving amplitudes}. It has the technical advantage that the
integration variable $\mathcal{C}$ defined in
\eqref{eq:grassint-matrix} is a complex $K\times K$ \emph{square}
matrix. Instead of focusing on $\mathfrak{psu}(2,2|4)$ Yangian
invariants relevant for the amplitudes $\mathcal{A}_{2K,K}$, we
immediately generalize to $\mathfrak{u}(p,p|m)$. This generalization
can be done almost effortlessly, and it will help us to gain important
insights\footnote{In particular, it will be instructive to study the
  simpler algebra $\mathfrak{u}(1,1)$ in section~\ref{sec:psi42u11}
  below. This may remind the reader of the Amplituhedron
  \cite{Arkani-Hamed:2013jha}. Its definition involves a parameter
  $m$, and it yields physical amplitudes for $m=4$. At times, it is
  investigated for the mathematically more accessible case $m=2$,
  which corresponds to $\mathfrak{u}(1,1|2)$ in our language.}  into
the structure of our integrals later. It implies replacing the bosonic
variables $\lambda,\tilde{\lambda}\in\mathbb{C}^2$ defined in
\eqref{eq:nullvec-spinors} by
$\lambda,\tilde{\lambda}\in\mathbb{C}^p$. Even though in general these
variables are not associated with four-dimensional Minkowski momenta
anymore, it is often helpful to continue using this terminology. To be
able to perform concrete calculations, we have to impose the reality
conditions \eqref{eq:spinors-real} on the variables
$\lambda^i,\tilde{\lambda}^i\in\mathbb{C}^p$. We choose negative
energies for the first $K$ momenta and positive energies for the
latter $K$. This completely determines the $2K\times p$ matrix
$\boldsymbol{\tilde{\lambda}}$ containing all $\tilde{\lambda}^i$ in
terms of the matrix $\boldsymbol{\lambda}$ containing the $\lambda^i$,
\begin{align}
\label{eq:spinors-partition}
  \boldsymbol{\tilde{\lambda}}=
  \left(
  \begin{array}{c}
    \boldsymbol{\tilde{\lambda}}^{-}\\[0.3em]
    \hdashline\\[-1.0em]
    \boldsymbol{\tilde{\lambda}}^{+}
  \end{array}
  \right)
  =
  \left(
  \begin{array}{c}
    -\overline{\boldsymbol{\lambda}}^{\,-}\\[0.3em]
    \hdashline\\[-1.0em]
    \phantom{-}\overline{\boldsymbol{\lambda}}^{\,+}
  \end{array}
  \right)\,,\quad
  \boldsymbol{\lambda}=
  \left(
  \begin{array}{c}
    \boldsymbol{\lambda}^{-}\\[0.3em]
    \hdashline\\[-1.0em]
    \boldsymbol{\lambda}^{+}
  \end{array}
  \right)\,.
\end{align}
The $K\times p$ blocks of $\boldsymbol{\lambda}$ are
$\boldsymbol{\lambda}^{-}=(\lambda_\alpha^k)$ and
$\boldsymbol{\lambda}^{+}=(\lambda_\alpha^l)$, whose row indices run
over $k=1,\ldots,K$ and $l=K+1,\ldots,2K$, respectively. We will stick
to this setup throughout this article.

Now we are in a position to discuss the integration contour. A
characteristic feature of the Graßmannian integral
\eqref{eq:grassint-amp-def}, that we want to keep for our refined
version, is the linear relations among the spinors in
$\boldsymbol{\lambda}$ imposed by bosonic delta functions,
\begin{align}
  \label{eq:on-support}
  0=C^\perp\boldsymbol{\lambda}=-\mathcal{C}^t\boldsymbol{\lambda}^{-}+\boldsymbol{\lambda}^{+}\,.
\end{align}
Furthermore, we want to impose momentum conservation, recall
\eqref{eq:amp-bosferm-delta},
\begin{align}
  \label{eq:mon-consv}
  \begin{aligned}
    P_{\alpha\dot{\beta}}=\sum_{i=1}^N\lambda^i_\alpha\tilde{\lambda}^i_{\dot{\beta}}=0
    \quad\Leftrightarrow\quad
    \boldsymbol{\lambda}^t\boldsymbol{\tilde\lambda}=0\,.
  \end{aligned}
\end{align}
Here the total momentum is encoded in the Hermitian $p\times p$ matrix
$P=(P_{\alpha\dot{\beta}})$. For our purpose, it is more suitable to
work with its expression in terms of the matrices
$\boldsymbol{\lambda}$ and $\boldsymbol{\tilde{\lambda}}$. With the
help of \eqref{eq:spinors-partition} and \eqref{eq:on-support}, we
obtain
\begin{align}
  \label{eq:unitary-constr}
  0=\boldsymbol{\lambda}^t\boldsymbol{\tilde\lambda}=
  (\boldsymbol{\tilde\lambda}^{-})^\dagger
  \big(\mathcal{C}\mathcal{C}^{\dagger}-1_K\big)
  \boldsymbol{\tilde\lambda}^{-}\,.
\end{align}
This is most naturally satisfied by demanding $\mathcal{C}\in
U(K)$. It strikingly suggests that the contour of our refined version
of the Graßmannian integral \eqref{eq:grassint-amp-def} should be a
\emph{unitary group manifold}. See also appendix~\ref{sec:gluing} for
an independent argument in favor of this contour.

Clearly, in order to cover all amplitudes $\mathcal{A}_{N,K}$, it
would be necessary to extend the reasoning presented in this section
to the case $N\neq 2K$ and, in addition, to allow for arbitrary
choices of the energy signs. We comment on both of these aspects
separately. The unitary group manifold seems to generalize to a
so-called \emph{Stiefel manifold}\footnote{We will encounter this
  manifold, together with the appropriate Haar measure, for a
  different purpose in appendix~\ref{sec:gluing}.} for $N\neq
2K$. Starting with the linear constraint
$C^\perp\boldsymbol{\lambda}=0$ as in \eqref{eq:on-support}, we arrive
again at \eqref{eq:unitary-constr}, where now $\mathcal{C}$ from
\eqref{eq:grassint-matrix} is a non-square $K\times(N-K)$
matrix. Assuming $N>2K$, we can demand
$\mathcal{C}\mathcal{C}^{\dagger}=1_K$, which defines said
manifold. However, this condition has no solution in case of
$N<2K$. Here we resort to the other constraint
$C\boldsymbol{\tilde\lambda}=0$ from the Graßmannian integral
\eqref{eq:grassint-amp-def}, which leads to the Stiefel manifold
$\mathcal{C}^{\dagger}\mathcal{C}=1_{N-K}$. Note that the two
constraints become equivalent for $N=2K$ due to the unitarity of
$\mathcal{C}$. Eventually, one should also generalize the distribution
of the energy signs. As long as there are $K$ particles with negative
and $N-K$ with positive energy, we can align the gauge fixing of the
matrix $C$ in \eqref{eq:grassint-matrix} with the distribution of
signs in \eqref{eq:spinors-partition}. In this way, we obtain a
Stiefel manifold for the matrix block $\mathcal{C}$ once again.
Additional complications surface for different numbers of positive and
negative energy particles. This is a bit puzzling because the energy
signs enter the final formula \cite{Drummond:2008cr} for the amplitude
$\mathcal{A}_{N,K}$ only mildly via the reality conditions
\eqref{eq:spinors-brackets-conj} of the spinor brackets $[ij]$. We
leave the extensions discussed in this paragraph for future work as
already our specific setting will give rise to rich structures.

\subsection{Graßmannian Integral}
\label{sec:integral-formula}

Here we implement the insight on the choice of the contour. In doing
so, we generalize the superalgebra slightly further by allowing for
different gradings\footnote{This might seem unnecessarily tedious. In
  fact, we will exclusively use the grading $\mathfrak{u}(2,2|0+4)$ to
  relate our results to amplitudes in this article. However, for the
  spectral problem of planar $\mathcal{N}=4$ SYM, the grading
  $\mathfrak{u}(2,2|2+2)$ appears to be more natural at one loop
  \cite{Beisert:2003yb,Beisert:2003jj} and is the key to all-loop
  results such as \cite{Beisert:2006qh}. We will discuss the prospect
  of extending our formalism to the results of the latter reference in
  section~\ref{sec:r-formula}.} of the fermions, which we indicate by
the notation $\mathfrak{u}(p,p|m=r+s)$. We define a \emph{unitary}
Graßmannian integral computing Yangian invariants for this algebra by
\begin{align}
  \label{eq:grass-int-barg}
  \Psi_{N=2K,K}
  =
  \int\displaylimits_{U(K)}\!\![\D\mathcal{C}]\,
  \mathscr{F}(\mathcal{C})\,
  \delta_{\mathbb{C}}^{pK|0}(C^\perp\boldsymbol{\lambda})  
  \delta^{0|rK}(C^\perp\boldsymbol{\eta})
  \delta^{0|sK}(C\boldsymbol{\tilde\eta})\,,
\end{align}
where the Graßmannian matrices $C$ and $C^\perp$ are defined around
\eqref{eq:grassint-matrix}. They contain the matrix block
$\mathcal{C}$ that we impose be to unitary here. We denote the Haar
measure on the unitary group $U(K)$ by $[\D\mathcal{C}]$. The
constraint \eqref{eq:on-support} is imposed in the integrand using
\emph{complex} delta functions, which are defined by
$\delta_{\mathbb{C}}(z)=\delta(\Real z)\delta(\Imag z)$ for
$z\in\mathbb{C}$. The $2K\times p$ matrix $\boldsymbol{\lambda}$
contains the bosonic variables $\lambda^i\in\mathbb{C}^p$ associated
with the sites $i=1,\ldots,2K$, as explained around
\eqref{eq:spinors-partition}. For the representations we are working
with, each site is also associated with $r$- and $s$-dimensional
fermionic variables $\eta^i$ and $\tilde\eta^i$, respectively. They
are packaged into the $2K\times r$ matrix
$\boldsymbol{\eta}=(\eta^i_a)$ and the $2K\times s$ matrix
$\boldsymbol{\tilde\eta}=(\tilde\eta^i_{\dot{a}})$. A characteristic
part of the integrand is the function
\begin{align}
  \label{eq:grass-int-unitary-integrand}
  \begin{aligned}
    \mathscr{F}(\mathcal{C})^{-1}
    =
    \,(\det{\mathcal{C}})^{m-q-K}
    &\,\prod_{\mathclap{i=1}}^{2K}\,\,
    (i,\ldots,i+K-1)^{1+v_{i+K-1}^--v_i^+}\,,
  \end{aligned}
\end{align}
which contains the minors $(i,\ldots,i+K-1)$ of the matrix $C$. For
its use in \eqref{eq:grass-int-barg}, we have to identify $q=p$. In
section~\ref{sec:matrix-models} below, we will need
$\mathscr{F}(\mathcal{C})$ without this identification. The exponents
of the minors are given by
\begin{align}
  \label{eq:spec-redef}
  v^\pm_i=v_i'\pm\frac{c_i}{2}\,,\quad
  v_i'=v_i-\frac{c_i}{2}+
  \begin{cases}
    n-m-1&\text{for}\quad i=1,\ldots, K\,,\\
    0&\text{for}\quad i=K+1,\ldots, 2K\,,
  \end{cases}
\end{align}
where $n=2p$ and $m=r+s$. The case distinction originates from the
choice of energy signs for the sites in
\eqref{eq:spinors-partition}. The parameters $v_i\in\mathbb{C}$ are
inhomogeneities of a spin chain monodromy matrix
\eqref{eq:yangian-mono-spinchain} whose sites carry
$\mathfrak{u}(p,p|r+s)$ representations labeled by $c_i\in\mathbb{Z}$,
see the discussion in section~\ref{sec:symmetry-generators} below. To
enable the Yangian invariance of $\Psi_{2K,K}$ in
\eqref{eq:grass-int-barg}, we have to demand
\begin{align}
  \label{eq:spec-perm}
  v^-_{i+K}=v^+_i\,,
\end{align}
for $i=1,\ldots, 2K$. We denote the Yangian invariants by
$\Psi_{2K,K}$ instead of $\mathcal{A}_{2K,K}^{(\text{def\/})}$ because
we still have to establish their relation to the amplitudes
$\mathcal{A}_{2K,K}$. This will be achieved for sample invariants in
section~\ref{sec:sample-invariants}, and a further perspective on
their relation will be added in section~\ref{sec:norm-div}.

Let us compare the unitary integral \eqref{eq:grass-int-barg} with the
original deformed Graßmannian integral
\eqref{eq:grassint-amp-def}. First, there are no delta functions
explicitly constraining $\boldsymbol{\tilde{\lambda}}$ in
\eqref{eq:grass-int-barg}. This would be superfluous because the delta
functions containing $\boldsymbol{\lambda}$ together with the reality
conditions in \eqref{eq:spinors-partition} yield the constraint
$C\boldsymbol{\tilde{\lambda}}=0$. Next, the Haar measure
$[\D\mathcal{C}]$ in \eqref{eq:grass-int-barg} can easily be obtained
from the holomorphic form in \eqref{eq:grassint-amp-def},
\begin{align}
  \label{eq:measure-haar}
  [\D\mathcal{C}]\propto\frac{\D^{\,K^2}\!\mathcal{C}}{(\det\mathcal{C})^K}\,,
\end{align}
for $\mathcal{C}\in U(K)$. The proportionality constant is fixed by
demanding $\int_{U(K)}[\D\mathcal{C}]= 1$. This form of the Haar
measure is suitable for showing its left- und right-invariance, i.e.\
invariance under the transformation
$\mathcal{C}\mapsto\mathcal{V}\mathcal{C}\mathcal{W}$ with constant
matrices $\mathcal{V},\mathcal{W}\in U(K)$. What is more, the function
$\mathscr{F}(\mathcal{C})^{-1}$ from
\eqref{eq:grass-int-unitary-integrand} essentially reduces to the
product of minors in \eqref{eq:grassint-amp-def}. The additional
factors of $\det\mathcal{C}$ can be attributed to the algebra
$\mathfrak{u}(p,p|m)$ generalizing $\mathfrak{psu}(2,2|4)$, the
arguments of the bosonic delta functions, and the Haar measure. An
important change is the restriction from complex representation labels
$c_i$ in \eqref{eq:grassint-amp-def} to integer ones in
\eqref{eq:grass-int-barg}. It is required for the single-valuedness of
$\mathscr{F}(\mathcal{C})$ addressed in the next section.

\subsection{Single-Valued Integrand}
\label{sec:singlevaluedness}

The main obstruction to finding a closed contour for the original
deformed Graßmannian integral \eqref{eq:grassint-amp-def}, needed to
show its Yangian invariance, is the intricate branch cut structure of
its integrand caused by the complex exponents of the minors. Even
though the unitary group $U(K)$ is \emph{compact}, it could still fail
to yield a \emph{closed} contour for our integral
\eqref{eq:grass-int-barg}, if the integrand $\mathscr{F}(\mathcal{C})$
from \eqref{eq:grass-int-unitary-integrand} was multi-valued. Until
now, this integrand is still formal because we have not specified its
analytic structure yet. In what follows, we will do this implicitly by
manipulating it into a form that is explicitly single-valued.

We start by expressing the minors of the $K\times 2K$ matrix $C$
defined in \eqref{eq:grassint-matrix} in terms of those of the
$K\times K$ matrix $\mathcal{C}$,
\begin{align}
  \label{eq:minors-relation}
  (i,\ldots,i+K-1)=(-1)^{(K-i+1)(i-1)}
  \begin{cases}
    [1,\ldots,i-1]&\text{for}\quad i=1,\ldots,K\,,\\
    [i-K,\ldots,K]&\text{for}\quad i=K+1,\ldots,2K\,.
  \end{cases}
\end{align}
In this formula, the principal minor of $\mathcal{C}$ built from the
rows and columns $i$ to $j$ is denoted $[i,\ldots,j]$, e.g.\
$[\hphantom{1}]=1$, $[1]=C_{1\,K+1}$, and
$[1,\ldots,K]=\det{\mathcal{C}}$. Furthermore, using the unitarity of
$\mathcal{C}$ we obtain
\begin{align}
  \label{eq:prop-unitary-minors}
  [i+1,\ldots,K]=\overline{[1,\ldots,i]}\det{\mathcal{C}}\,,
\end{align}
see e.g.\ \cite{Salaff:1967}. This useful identity can be proven using
a block decomposition of $\mathcal{C}$. It is helpful to translate the
constraints on $v_i^\pm$ in \eqref{eq:spec-perm} into constraints on
the parameters $v_i,c_i$,
\begin{align}
  \label{eq:grass-int-parameters}
    v_{K+i}=v_i+n-m-1-c_i\,,\quad
    c_{K+i}=-c_i\,,
\end{align}
for $i=1,\ldots K$. Using the relations obtained here and disregarding
the analytic structure temporarily by combining products of minors
with common \emph{complex} exponents, we rewrite
\eqref{eq:grass-int-unitary-integrand}, up to a constant sign factor,
as
\begin{align}
  \label{eq:eq:grass-int-unitary-integrand-final}
  \begin{aligned}
    \mathscr{F}(\mathcal{C})^{-1}=&
    (\det\mathcal{C})^{m-q+c_K}
  \prod_{i=1}^{K-1}|[1,\ldots,i]|^{2(1+v_{i}-v_{i+1})}\overline{[1,\ldots,i]}^{\,c_{i+1}-c_{i}}\,.
  \end{aligned}
\end{align}
This function is manifestly single-valued because only non-negative
numbers are exponentiated to non-integer powers. Here we see the
paramount importance of integer representation labels
$c_i$. Henceforth, we will use the single-valued integrand
$\mathscr{F}(\mathcal{C})$ defined in
\eqref{eq:eq:grass-int-unitary-integrand-final} instead of formal
expression \eqref{eq:grass-int-unitary-integrand}. Consequently, the
unitary integration contour in \eqref{eq:grass-int-barg} is closed, as
is required to demonstrate Yangian invariance.

\subsection{Symmetry Generators and Yangian Invariance}
\label{sec:symmetry-generators}

We still have to specify the Yangian generators annihilating the
invariants $\Psi_{2K,K}$ defined by the unitary Graßmannian integral
\eqref{eq:grass-int-barg}. For this purpose, we need to introduce some
basics of two classes of $\mathfrak{u}(p,p|r+s)$ representations. In
the special case of $\mathfrak{su}(2,2)$, these are well-known
representations of the conformal algebra
\cite{Stoyanov:1968tn,Mack:1969dg} and the two classes correspond to
positive and negative energies. In general, they are equivalent to
certain oscillator representations \cite{Todorov:1966zz,Bars:1982ep},
which we will encounter in section~\ref{sec:matrix-models}.

For the first class of representations, we consider generators
$\mathfrak{J}_{\indnm{AB}}$ of the $\mathfrak{gl}(n|m)$ superalgebra
\eqref{eq:gl-superalg} built from the bosonic variables
$\lambda_\alpha\in\mathbb{C}$ and the fermions $\eta_a$ and
$\tilde\eta_{\dot{a}}$ with the index ranges $\alpha=1,\ldots,p$,
$a=1,\ldots,r$, and $\dot{a}=1,\ldots,s$. They are given by the
supermatrix
\begin{align}
  \label{eq:fermi-gen}
    (\mathfrak{J}_{\indnm{AB}})&=
    \left(
    \begin{array}{c:c:c:c}
      \lambda_\alpha\partial_{\lambda_\beta}&
      \lambda_\alpha\partial_{\eta_b}&
      \lambda_\alpha\overline{\lambda}_\beta&
      \lambda_\alpha\tilde\eta_{\dot{b}}\\[0.3em]
      \hdashline&&&\\[-1.0em]
      \eta_a\partial_{\lambda_\beta}&
      \eta_a\partial_{\eta_b}&
      \eta_a\overline{\lambda}_\beta&
      \eta_a\tilde\eta_{\dot{b}}\\[0.3em]
      \hdashline&&&\\[-1.0em]
      -\partial_{\overline{\lambda}_\alpha}\partial_{\lambda_\beta}&
      -\partial_{\overline{\lambda}_\alpha}\partial_{\eta_b}&
      -\partial_{\overline{\lambda}_\alpha}\overline{\lambda}_\beta&
      -\partial_{\overline{\lambda}_\alpha}\tilde\eta_{\dot{b}}\\[0.3em]
      \hdashline&&&\\[-1.0em]
      \partial_{\tilde\eta_{\dot{a}}}\partial_{\lambda_\beta}&
      \partial_{\tilde\eta_{\dot{a}}}\partial_{\eta_b}&
      \partial_{\tilde\eta_{\dot{a}}}\overline{\lambda}_\beta&
      \partial_{\tilde\eta_{\dot{a}}}\tilde\eta_{\dot{b}}
    \end{array}
  \right)\,
\end{align}
with superindices $\indnm{A},\indnm{B}=1,\ldots,2p+r+s=n+m$. Let
$\oscrep_c$ denote the space of those functions of the bosonic and
fermionic variables on which the central element
\begin{align}
  \label{eq:fermi-replabel-pos}
  \mathfrak{C}=
  \tr(\mathfrak{J}_{\indnm{AB}})=
  \sum_{\alpha=1}^p
  (\lambda_\alpha\partial_{\lambda_\alpha}-\overline{\lambda}_\alpha\partial_{\overline{\lambda}_\alpha})
  +\sum_{a=1}^r\eta_a\partial_{\eta_a}
  -\sum_{\dot{a}=1}^s\tilde\eta_{\dot{a}}\partial_{\tilde\eta_{\dot{a}}}
  -p+s\,
\end{align}
acts by multiplication with the eigenvalue $c$. We define conjugates
of the variables by
\begin{align}
  \label{eq:vars-dagger}
  {\lambda_\alpha}^\dagger=\overline{\lambda}_\alpha\,,\quad
  {\partial_{\lambda_\alpha}}^\dagger=-\partial_{\overline{\lambda}_\alpha}\,,\quad
  {\eta_a}^\dagger=\partial_{\eta_a}\,,\quad
  \tilde\eta_{\dot{a}}{}^\dagger=\partial_{\tilde\eta_{\dot{a}}}\,.
\end{align}
At this point, we refrain from stating the inner product leading to
this choice, which would also be required for a less formal definition
of the space $\oscrep_c$. Some details will be filled in below in
sections~\ref{sec:bargmann} and~\ref{sec:trafo-int-gen}. With
\eqref{eq:vars-dagger}, $\oscrep_c$ for $c\in\mathbb{Z}$ carries a
unitary representation of $\mathfrak{u}(p,p|r+s)$. For the algebra
$\mathfrak{u}(2,2|0+4)$, the supermatrix of generators
\eqref{eq:fermi-gen} reduces to \eqref{eq:sym-gen-gl44} with the
reality condition $\tilde{\lambda}=+\overline{\lambda}$ from
\eqref{eq:spinors-real}. Hence, in slight abuse of terminology, we
refer to the $\oscrep_c$ also for the algebra $\mathfrak{u}(p,p|r+s)$
as \emph{positive energy representation}. The second class of
representations is obtained from the $\mathfrak{gl}(n|m)$ generators
\begin{align}
  \label{eq:fermi-shift}
  \bar{\mathfrak{J}}_{\indnm{AB}}
  =
  \left.\mathfrak{J}_{\indnm{AB}}\right|_{(\lambda,\overline{\lambda})\mapsto(\lambda,-\overline{\lambda})}
  +\delta_{\indnm{AB}}(-1)^{|\indnm{A}|}\,,
\end{align}
where $\mathfrak{J}_{\indnm{AB}}$ are the generators in
\eqref{eq:fermi-gen}. We have
\begin{align}
  \label{eq:fermi-replabel-neg}
  \bar{\mathfrak{C}}=
  \tr(\bar{\mathfrak{J}}_{\indnm{AB}})=
  \sum_{\alpha=1}^p
  (\lambda_\alpha\partial_{\lambda_\alpha}-\overline{\lambda}_\alpha\partial_{\overline{\lambda}_\alpha})
  +\sum_{a=1}^r\eta_a\partial_{\eta_a}
  -\sum_{\dot{a}=1}^s\tilde\eta_{\dot{a}}\partial_{\tilde\eta_{\dot{a}}}
  +p-r\,.
\end{align}
The space of functions on which this central element acts by
multiplication with $c\in\mathbb{Z}$ is denoted $\bar{\oscrep}_c$ and
forms a unitary representation of $\mathfrak{u}(p,p|r+s)$. We refer to
$\bar{\oscrep}_c$ as \emph{negative energy representation}.

Let us remark that for $\mathfrak{u}(2,2|0+4)$ the
$\bar{\mathfrak{J}}_{\indnm{AB}}$ differ from the generators in
\eqref{eq:sym-gen-gl44} with $\tilde{\lambda}=-\overline{\lambda}$ by
the second term in \eqref{eq:fermi-shift}. This term arises naturally
in section~\ref{sec:trafo-int-gen}, where we will revisit the
representations introduced here. Practically, its inclusion has the
advantage that we do not have to change from $\mathfrak{gl}(n|m)$ to
$\mathfrak{sl}(n|m)$ generators, as it is usually the case for the
$\mathcal{N}=4$ SYM amplitudes $\mathcal{A}_{N,K}$, cf.\
\eqref{eq:sym-alg-gen-sl} and \eqref{eq:sym-alg-annih}. On a different
note, the expressions for the central elements $\mathfrak{C}$ and
$\bar{\mathfrak{C}}$ in \eqref{eq:fermi-replabel-pos} and
\eqref{eq:fermi-replabel-neg}, respectively, agree for $2p=r+s$. In
particular, this is the case for $\mathfrak{u}(2,2|0+4)$, where these
generators essentially measure the superhelicities of particles
described by the amplitudes $\mathcal{A}_{N,K}$.

Finally, given the two classes of representations, we are able to
construct a monodromy matrix $M(u)$ with $N=2K$ sites, as introduced
in \eqref{eq:yangian-mono-spinchain}, that is associated with
$\Psi_{2K,K}$ defined by the unitary Graßmannian integral
\eqref{eq:grass-int-barg}. With our choice of energy signs in
\eqref{eq:spinors-partition}, the first $K$ sites carry negative
energy representations $\bar{\oscrep}_{c_i}$ and the latter $K$ sites
carry positive energy representations $\oscrep_{c_i}$. Hence the
generators entering the Lax operators \eqref{eq:yangian-def-lax} of
the monodromy are
\begin{align}
  \label{eq:mono-spinor-gen}
  J_{\indnm{AB}}^i=
  \begin{cases}
    \bar{\mathfrak{J}}^i_{\indnm{AB}}&\text{for}\quad i=1,\ldots,K\,,\\
    \mathfrak{J}^i_{\indnm{AB}}&\text{for}\quad i=K+1,\ldots,2K\,,
  \end{cases}
\end{align}
which can be found in \eqref{eq:fermi-shift} and
\eqref{eq:fermi-gen}, respectively. The inhomogeneities $v_i$
and representations labels $c_i$ of the monodromy have to obey
\eqref{eq:spec-perm}, recall also the equivalent equation
\eqref{eq:grass-int-parameters}. Then $\Psi_{2K,K}$ defined in
\eqref{eq:grass-int-barg} satisfies the Yangian invariance condition
\eqref{eq:yi-exp-1} and thus also \eqref{eq:yi} in the QISM
language. We do not provide a direct proof of this key statement
here. Instead, we refer the reader to section~\ref{sec:matrix-models},
where we translate $\Psi_{2K,K}$ into a different basis in which the
Yangian invariance has been proven.

\section{Sample Invariants and Amplitudes}
\label{sec:sample-invariants}

Having defined Yangian invariants by the unitary Graßmannian integral
\eqref{eq:grass-int-barg}, we evaluate this integral here for several
examples. Our primary focus are sample invariants $\Psi_{2K,K}$ for
the algebra $\mathfrak{u}(2,2|4)$ whose relation to the
$\mathcal{N}=4$ SYM amplitudes $\mathcal{A}_{2K,K}$ we explore. The
first step in the evaluation of \eqref{eq:grass-int-barg} can be
performed on general grounds, even for $\mathfrak{u}(p,p|r+s)$. It is
possible to reduce\footnote{This is somewhat similar to the transition
  \cite{ArkaniHamed:2009vw} from the original $Gr(K,N)$ Graßmannian
  integral \eqref{eq:grassint-amp} for $\mathcal{A}_{N,K}$ in spinor
  helicity variables to that in terms of momentum twistors involving
  the Graßmannian $Gr(K-2,N)$.} the $U(K)$ integral to a $U(K-p)$
integral provided $K\geq p$. Technically, this reduction is based on
QR decompositions of the $K\times p$ blocks $\boldsymbol{\lambda}^\pm$
making up the kinematic data in $\boldsymbol{\lambda}$, cf.\
\eqref{eq:spinors-partition}. The unitary factors of these
decompositions can be absorbed into the integration variable
$\mathcal{C}\in U(K)$ using the left- and right-invariance of the Haar
measure. The bosonic delta functions in \eqref{eq:grass-int-barg} then
reduce the new integration variable to a $U(K-p)$ matrix. On a
different note, our evaluation of the integral
\eqref{eq:grass-int-barg} in the present section addresses only terms
with maximal kinematic support, i.e.\ the those proportional to the
momentum conserving delta function~\eqref{eq:amp-bosferm-delta}. We
neglect additional terms that can appear for special kinematic
configurations, see section~\ref{sec:norm-div} below.

\subsection{Four Particles for \texorpdfstring{$\mathfrak{u}(2,2|4)$}{u(2,2|4)}}
\label{sec:psi42u224}

As a first example, we evaluate the integral \eqref{eq:grass-int-barg}
to obtain the invariant $\Psi_{4,2}$ for the algebra
$\mathfrak{u}(2,2|0+4)$. Our naming of the fermionic variables
suggests that this grading of the algebra is required to make contact
with the amplitude $\mathcal{A}_{4,2}$ from
\eqref{eq:amp-super-mhv}. The bosonic delta functions in
\eqref{eq:grass-int-barg} fix the integration variable completely,
\begin{align}
  \label{eq:inv42-u22-e-eval}
  \mathcal{C}
  =
  \frac{1}{\langle 1 2 \rangle}
  \begin{pmatrix}
    \langle 3 2 \rangle&\langle 4 2 \rangle\\
    \langle 1 3 \rangle&\langle 1 4 \rangle\\
  \end{pmatrix}\,.
\end{align}
This is a unitary matrix because of the reality conditions of the
spinors in \eqref{eq:spinors-partition} and momentum conservation
\eqref{eq:mon-consv}. We obtain from it the Yangian invariant
\begin{align}
  \label{eq:inv42-u2204-final}
  \Psi_{4,2}=
  \frac{\delta^{4|0}(P)\delta^{0|8}(Q)}
  {\langle 12 \rangle\langle 23 \rangle\langle 34 \rangle\langle 41 \rangle}
  \left(\frac{\langle 14 \rangle}{\langle 34 \rangle}\right)^{c_1}
  \left(\frac{\langle 12 \rangle}{\langle 14 \rangle}\right)^{c_2}
  \left(\frac{\langle 34 \rangle\overline{\langle 34\rangle}}{\langle 14 \rangle\overline{\langle 14\rangle}}\right)^{v_1-v_2}\,,
\end{align}
where we dropped a numerical prefactor, as we will also do in the
following examples. The delta functions implementing momentum and
supermomentum conservation are defined in
\eqref{eq:amp-bosferm-delta}. We chose to display the result using
complex conjugates of angle brackets $\overline{\langle ij\rangle}$
instead of square brackets $[ij]$ to highlight the analytic structure,
recall the relation between these brackets from
\eqref{eq:spinors-brackets-conj}. $\Psi_{4,2}$ is a single-valued
function of the spinors $\lambda_\alpha^i$ because the factor raised
to the complex power $v_1-v_2$ is non-negative and
$c_1,c_2\in\mathbb{Z}$. Moreover, $\Psi_{4,2}$ basically agrees with
the deformed amplitude $\mathcal{A}_{4,2}^{(\text{def\/})}$ from
\eqref{eq:amp-def-mhv}, which was first obtained in
\cite{Ferro:2012xw}. This is leaving aside the crucial restriction to
integer representation labels $c_1,c_2$ and a slight, inessential
difference in the parameterization of the complex deformation
parameters. $\Psi_{4,2}$ reduces to the amplitude $\mathcal{A}_{4,2}$
from \eqref{eq:amp-super-mhv} for $c_1=c_2=0$ and $v_1=v_2$. Let us
remark that evaluating the Graßmannian integral
\eqref{eq:grass-int-barg} for $\Psi_{4,2}$ in case of the bosonic
algebra $\mathfrak{u}(2,2)$ with these deformation parameters yields
the four-particle tree-level $\text{MHV}$ gluon amplitude with the
split helicity configuration $(+1,+1,-1,-1)$.

\subsection{Six Particles for
  \texorpdfstring{$\mathfrak{u}(2,2|4)$}{u(2,2|4)}}
\label{sec:psi63u224}

Let us move on to six particles. It was argued in
\cite{Beisert:2014qba} that a deformation of the amplitude
$\mathcal{A}_{6,3}$ cannot be constructed by deforming the
\emph{individual residues} contributing to it. Does this imply that
there is no such deformation? The authors of \cite{Ferro:2014gca} put
forward the idea of deforming the \emph{entire integral} from which
the residues are extracted. This resulted in the Graßmannian integral
\eqref{eq:grassint-amp-def} for the deformed amplitude
$\mathcal{A}_{6,3}^{(\text{def\/})}$. However, as discussed in
section~\ref{sec:grassdef}, a suitable contour for this integral has
been missing so far. Here we evaluate our Graßmannian integral formula
\eqref{eq:grass-int-barg} with the unitary contour for the
$\mathfrak{u}(2,2|0+4)$ Yangian invariant $\Psi_{6,3}$. In particular,
we show how it reduces to $\mathcal{A}_{6,3}$ in the undeformed limit.

First, the bosonic delta functions in \eqref{eq:grass-int-barg}
determine the integration variable $\mathcal{C}\in U(3)$ up to a phase
$\smallunitary\in U(1)$,
\begin{align}
  \label{eq:psi63u224-cmatrix}
  \begin{aligned}
    \mathcal{C}&=
    \frac{1}{s_{123}}\left[
      \begin{pmatrix}
        \overline{\langle 56\rangle} \langle 23\rangle&
        \overline{\langle 64\rangle} \langle 23\rangle&
        \overline{\langle 45\rangle} \langle 23\rangle\\
        \overline{\langle 56\rangle} \langle 31\rangle&
        \overline{\langle 64\rangle} \langle 31\rangle&
        \overline{\langle 45\rangle} \langle 31\rangle\\
        \overline{\langle 56\rangle} \langle 12\rangle&
        \overline{\langle 64\rangle} \langle 12\rangle&
        \overline{\langle 45\rangle} \langle 12\rangle\\
      \end{pmatrix}
      \overline{\smallunitary}
      \right.\\
    &\quad\quad\quad\quad\quad\quad
    \left.-\begin{pmatrix}
        \overline{\langle 1|}2\!+\!3|4\rangle&
        \overline{\langle 1|}2\!+\!3|5\rangle&
        \overline{\langle 1|}2\!+\!3|6\rangle\\
        \overline{\langle 2|}1\!+\!3|4\rangle&
        \overline{\langle 2|}1\!+\!3|5\rangle&
        \overline{\langle 2|}1\!+\!3|6\rangle\\
        \overline{\langle 3|}1\!+\!2|4\rangle&
        \overline{\langle 3|}1\!+\!2|5\rangle&
        \overline{\langle 3|}1\!+\!2|6\rangle\\
      \end{pmatrix}\right]\,.
  \end{aligned}
\end{align}
The unitarity of this matrix can be verified using the reality
conditions in \eqref{eq:spinors-partition} and momentum conservation
\eqref{eq:mon-consv}. This reduces the $U(3)$ integral
\eqref{eq:grass-int-barg} to the $U(1)$ integral
\begin{align}
  \label{eq:psi63u224-int}
  \begin{aligned}
    \Psi_{6,3}
    =\frac{\delta^{4|0}(P)\delta^{0|8}(Q)}{s_{123}^5}
    \int\displaylimits_{U(1)}\!\![\D\smallunitary]\,
    \frac{
      \delta^{0|4}(\mathtt{a}\,\smallunitary+\mathtt{b})
    }{
      \smallunitary^{2-c_3}
      |\mathtt{A}-\smallunitary\mathtt{B}|^{2(1+v_1-v_2)}(\mathtt{A}-\smallunitary\mathtt{B})^{c_2-c_1}      
    }&\\
    \cdot\frac{
      1
    }{
      |\smallunitary\mathtt{C}-\mathtt{D}|^{2(1+v_2-v_3)}(\smallunitary\mathtt{C}-\mathtt{D})^{c_3-c_2}
    }&\,
  \end{aligned}
\end{align}
with the kinematic data encoded in the variables
\begin{align}
  \label{eq:psi63u224-para}
  \begin{gathered}
    \mathtt{A}=\frac{\langle 1|2\!+\!3\overline{|4\rangle}}{s_{123}}\,,\quad
    \mathtt{B}=\frac{\langle 56\rangle\overline{\langle 23\rangle}}{s_{123}}\,,\quad
    \mathtt{C}=\frac{\overline{\langle 3|}1\!+\!2|6\rangle}{s_{123}}\,,\quad
    \mathtt{D}=\frac{\overline{\langle 45\rangle}\langle 12\rangle}{s_{123}}\,,\\
    \mathtt{a}=\overline{\langle 23\rangle}\tilde\eta^1+\overline{\langle 31\rangle}\tilde\eta^2+\overline{\langle 12\rangle}\tilde\eta^3\,,\quad
    \mathtt{b}=\overline{\langle 56\rangle}\tilde\eta^4+\overline{\langle 64\rangle}\tilde\eta^5+\overline{\langle 45\rangle}\tilde\eta^6\,.
      \end{gathered}
\end{align}
Here $\tilde\eta^i=(\tilde\eta^i_{\dot{a}})$ are four-dimensional
fermionic variables and the momentum and supermomentum conserving
delta functions are defined in \eqref{eq:amp-bosferm-delta}. The
one-dimensional integral \eqref{eq:psi63u224-int} is our final
expression for the fully deformed Yangian invariant $\Psi_{6,3}$. It
would be desirable to express it in terms of known special functions.

\begin{figure}[!t]
  \begin{center}
    \begin{tikzpicture}
      \draw[thick,densely dotted,
      decoration={markings, mark=at position 0.0 with {\arrow{latex reversed}}},
      postaction={decorate}]
      (3,0) node[right=5] {$\Real{\smallunitary}$} -- (-3,0);
      \draw[thick,densely dotted,
      decoration={markings, mark=at position 0.0 with {\arrow{latex reversed}}},
      postaction={decorate}]
      (0,3) node[above=5] {$\Imag{\smallunitary}$} -- (0,-3);
      \draw[thick,decoration={markings, mark=at position 0.65 with {\arrow{latex}}},
      postaction={decorate}] (0,0) circle [radius=2];
      \filldraw[thick] (0,0) circle (1pt) node[below left]{$0$};
      \filldraw[thick] (-2.5,2.5) circle (1pt) node[below right]{$\infty$};
      \filldraw[thick] (55:1.1) circle (1pt) node[left]
      {$\frac{\mathtt{A}}{\mathtt{B}}$};
      \filldraw[thick] (55:3.6) circle (1pt) node[right]
      {$\frac{\overline{\mathtt{B}}}{\overline{\mathtt{A}}}$};
      \filldraw[thick] (-40:1) circle (1pt) node[left]
      {$\frac{\overline{\mathtt{C}}}{\overline{\mathtt{D}}}$};
      \filldraw[thick] (-40:4) circle (1pt) node[right]
      {$\frac{\mathtt{D}}{\mathtt{C}}$};
      \node at (2,0) [below right] {$1$};
      \draw[thick,densely dashed] (55:1.1) -- (55:3.6)
      node [midway,right=1] {$s_{234}$};
      \draw[thick,densely dashed] (-40:1) -- (-40:4)
      node [midway,right=3] {$s_{345}$};
    \end{tikzpicture}
    \caption{Sample configuration of the poles of the integrand in
      \eqref{eq:psi63u224-cint} for the undeformed
      $\mathfrak{u}(2,2|4)$ Yangian invariant $\Psi_{6,3}$. Dashed
      lines connect pairs of poles. For each pair, exactly one pole is
      inside of the contour. The signs of the Mandelstam variables
      $s_{234}$ and $s_{345}$ determine which one.}
    \label{fig:grint-inv63-u22-contour}
  \end{center}
\end{figure}
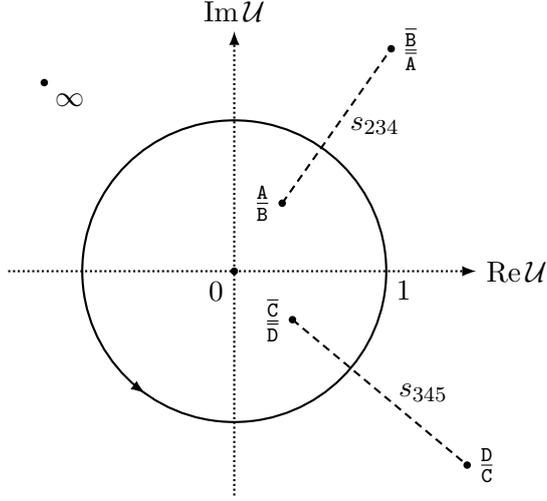

Next, to make contact with the amplitude $\mathcal{A}_{6,3}$, we study
the undeformed limit of \eqref{eq:psi63u224-int}, i.e.\ we set
$c_1=c_2=c_3=0$ and $v_1=v_2=v_3$. This yields the complex contour
integral
\begin{align}
  \label{eq:psi63u224-cint}
  \begin{aligned}
    \Psi_{6,3}
    =\frac{\delta^{4|0}(P)\delta^{0|8}(Q)}{s_{123}^5}
    \frac{1}{2\pi i\,\overline{\mathtt{A}}\mathtt{B}\mathtt{C}\overline{\mathtt{D}}}
    \oint\!\D\smallunitary\,  
    \frac{
      \delta^{0|4}(\mathtt{a}\,\smallunitary+\mathtt{b})
    }
    {
      \smallunitary
      \Big(\smallunitary-\frac{\mathtt{A}}{\mathtt{B}}\Big)
      \Big(\smallunitary-\frac{\overline{\mathtt{B}}}{\overline{\mathtt{A}}}\Big)
      \Big(\smallunitary-\frac{\mathtt{D}}{\mathtt{C}}\Big)
      \Big(\smallunitary-\frac{\overline{\mathtt{C}}}{\overline{\mathtt{D}}}\Big)
    }\,,
  \end{aligned}
\end{align}
where we wrote the Haar measure as
$[\D\smallunitary]=\frac{1}{2\pi i}\frac{\D\smallunitary}{\smallunitary}$,
cf.\ \eqref{eq:measure-haar}, and thereby parameterized $U(1)$ as the
counterclockwise unit circle in the complex $\smallunitary$-plane. This
integral can be computed by means of Cauchy's residue theorem. The
pole at $\smallunitary=0$ obviously lies inside of the contour. In
addition, there are two pairs of poles, whose positions depend on the
kinematic data. If the pole at
$\smallunitary=\frac{\mathtt{A}}{\mathtt{B}}$ is inside of the contour,
the one at
$\smallunitary=\frac{\overline{\mathtt{B}}}{\overline{\mathtt{A}}}$ is
outside, and vice versa. The same holds true for the pair at
$\smallunitary=\frac{\mathtt{D}}{\mathtt{C}},\frac{\overline{\mathtt{C}}}{\overline{\mathtt{D}}}$.
An illustration of this behavior is presented in
figure~\ref{fig:grint-inv63-u22-contour}. As a result, there are
always three residues contributing to the integral
\eqref{eq:psi63u224-cint}. Notice that here we neglect the possibility
of poles moving onto the contour because this is only possible for
special kinematics. The selection of contributing residues is
controlled by the signs of the variables
\begin{align}
  \label{eq:psi63u224-respos}
  1-\left|\frac{\mathtt{A}}{\mathtt{B}}\right|^2
  =\frac{s_{234}s_{123}}{s_{23}s_{56}}\propto s_{234}\,,
  \quad
  1-\left|\frac{\mathtt{C}}{\mathtt{D}}\right|^2
  =\frac{s_{345}s_{123}}{s_{12}s_{45}}\propto s_{345}\,.
\end{align}
We expressed them in terms of the Mandelstam variables from
\eqref{eq:spinors-mandelstam}. Our selection of energy signs in
\eqref{eq:spinors-partition} implies that $s_{123}$ and the $s_{ij}$
appearing in \eqref{eq:psi63u224-respos} are non-negative. Thus we
dropped these factors in the last step.

In order to be able to compare the undeformed invariant
\eqref{eq:psi63u224-cint} with the amplitude $\mathcal{A}_{6,3}$, we
translate the residues into the R-invariants \cite{Drummond:2008bq}
\begin{align}
  \label{eq:amp-super-r}
  \mathcal{R}^{r;st}=
  \frac{
    \langle s\,s-1\rangle\langle t\,t-1\rangle
    \delta^{0|4}\big(\langle r|x^{rs}x^{st}|\theta^{tr}\rangle+\langle r|x^{rt}x^{ts}|\theta^{sr}\rangle\big)
  }
  {
    (x^{st})^2
    \langle r|x^{rs}x^{st}|t\rangle
    \langle r|x^{rs}x^{st}|t-1\rangle
    \langle r|x^{rt}x^{ts}|s\rangle
    \langle r|x^{rt}x^{ts}|s-1\rangle
  }\,.
\end{align}
These are expressed in terms of the \emph{dual variables} $x^i$ and
$\theta^i$. They are defined by the relations
$\lambda^i_\alpha\tilde{\lambda}^i_{\dot{\beta}}=x^i_{\alpha\dot{\beta}}-x^{i+1}_{\alpha\dot{\beta}}$
and
$\lambda^i_\alpha\tilde\eta^i_{\dot{a}}=\theta^i_{\alpha\dot{a}}-\theta^{i+1}_{\alpha\dot{a}}$
with the identification $N+1\equiv 1$ for the particle indices, where
here $N=6$. Furthermore, we used the abbreviations $x^{ij}=x^i-x^j$
and $\theta^{ij}=\theta^i-\theta^j$. The dual superconformal
symmetry mentioned in section~\ref{sec:yangian} acts naturally on the
variables $x^i$ and $\theta^i$. Employing the R-invariants from
\eqref{eq:amp-super-r}, the undeformed integral
\eqref{eq:psi63u224-cint} evaluates to
\begin{align}
  \label{eq:inv63-u2204-lastint-eval}
  \Psi_{6,3}=
  \frac{\delta^{4|0}(P)\delta^{0|8}(Q)}
  {\langle 12\rangle\langle 23\rangle\langle 34\rangle
  \langle 45\rangle\langle 56\rangle\langle 61\rangle}
  \begin{cases}
    \mathcal{R}^{1;46}+\mathcal{R}^{1;35}-\mathcal{R}^{6;35}
    &\text{for}
    \quad s_{234}>0\,,\quad s_{345}<0\,,\\
    \mathcal{R}^{1;46}-\mathcal{R}^{6;25}-\mathcal{R}^{6;35}
    &\text{for}
    \quad s_{234}<0\,,\quad s_{345}<0\,,\\
    \mathcal{R}^{1;46}+\mathcal{R}^{1;35}+\mathcal{R}^{1;36}
    &\text{for}
    \quad s_{234}>0\,,\quad s_{345}>0\,,\\
    \mathcal{R}^{1;46}-\mathcal{R}^{6;25}+\mathcal{R}^{1;36}
    &\text{for}
    \quad s_{234}<0\,,\quad s_{345}>0\,.\\
  \end{cases}
\end{align}
Here the four different \emph{kinematic regions} arise from the
residue positions in \eqref{eq:psi63u224-respos}. Each
$\mathcal{R}^{r;st}$ in \eqref{eq:inv63-u2204-lastint-eval}
corresponds to one residue. For example the term with
$\mathcal{R}^{1;46}$ is associated with the residue of the integrand
in \eqref{eq:psi63u224-cint} at $\smallunitary=0$ because it appears in
all regions. The other associations can be deduced similarly. In the
kinematic region $s_{234},s_{345}>0$, the undeformed invariant
$\Psi_{6,3}$ in \eqref{eq:inv63-u2204-lastint-eval} agrees with the
amplitude $\mathcal{A}_{6,3}$
\cite{Drummond:2008vq,Drummond:2008bq}. Hence the $U(3)$ contour we
started out with in the Graßmannian integral~\eqref{eq:grass-int-barg}
automatically selects the desired residues in this region. Curiously,
it also implies the emergence of three other kinematic regions, in
which \eqref{eq:inv63-u2204-lastint-eval} does not match the
amplitude. This makes us speculate that the Yangian symmetry of the
known expression for the amplitude $\mathcal{A}_{6,3}$ could be broken
in a subtle way because it misses the region structure. We will
strengthen this point in the following section. There we compute the
unitary Graßmannian integral~\eqref{eq:grass-int-barg} for a simpler
Yangian invariant, that serves as a toy model for $\Psi_{6,3}$ of
$\mathfrak{u}(2,2|4)$ considered here. In that case, we are able to
show the necessity of the emerging kinematic regions from an
integrability perspective.

We remark that the four kinematic regions in
\eqref{eq:inv63-u2204-lastint-eval} appeared also in the study of
hexagonal light-like Wilson loops in \cite{Dorn:2012cn}. The two
fractions of Mandelstam variables in \eqref{eq:psi63u224-respos},
which define the regions, are two of the three independent \emph{dual
  conformal cross ratios} defined in equation (1) of that
reference. The authors investigate the possible values of these cross
ratios. Our choice of energy signs in \eqref{eq:spinors-partition}
corresponds to the case considered in their equation (40). In this
case, the allowed values form four distinct regions in the cross ratio
space $\mathbb{R}^3$, which match our regions in
\eqref{eq:inv63-u2204-lastint-eval} and are nicely visualized in their
figure~1.

\subsection{Toy Model and Implications: Four Particles for
  \texorpdfstring{$\mathfrak{u}(1,1)$}{u(1,1)}}
\label{sec:psi42u11}

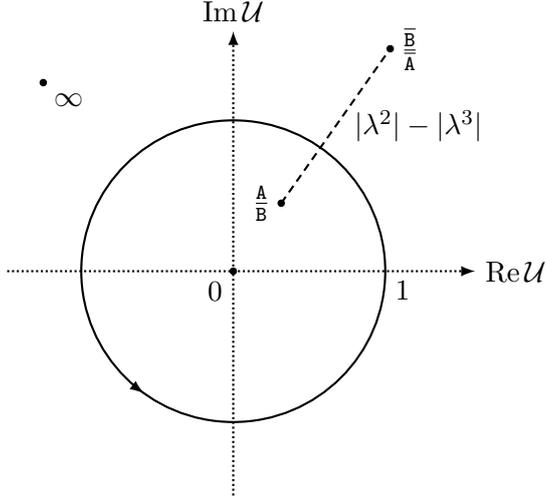
\begin{figure}[!t]
  \begin{center}
    \begin{tikzpicture}
      \draw[thick,densely dotted,
      decoration={markings, mark=at position 0.0 with {\arrow{latex reversed}}},
      postaction={decorate}]
      (3,0) node[right=5] {$\Real{\smallunitary}$} -- (-3,0);
      \draw[thick,densely dotted,
      decoration={markings, mark=at position 0.0 with {\arrow{latex reversed}}},
      postaction={decorate}]
      (0,3) node[above=5] {$\Imag{\smallunitary}$} -- (0,-3);
      \draw[thick,decoration={markings, mark=at position 0.65 with {\arrow{latex}}},
      postaction={decorate}] (0,0) circle [radius=2];
      \filldraw[thick] (0,0) circle (1pt) node[below left]{$0$};
      \filldraw[thick] (-2.5,2.5) circle (1pt) node[below right]{$\infty$};
      \filldraw[thick] (55:1.1) circle (1pt) node[left]
      {$\frac{\mathtt{A}}{\mathtt{B}}$};
      \filldraw[thick] (55:3.6) circle (1pt) node[right]
      {$\frac{\overline{\mathtt{B}}}{\overline{\mathtt{A}}}$};
      \node at (2,0) [below right] {$1$};
      \draw[thick,densely dashed] (55:1.1) -- (55:3.6)
      node [midway,right=3] {$|\lambda^2|-|\lambda^3|$};
    \end{tikzpicture}
    \caption{Sample pole configuration of the integrand in
      \eqref{eq:inv42-u11-contourint} for the undeformed
      $\mathfrak{u}(1,1)$ Yangian invariant $\Psi_{4,2}$. A pair of
      poles is connected by a dashed line. The sign of the variable
      $|\lambda^2|-|\lambda^3|$ controls which of the poles lies
      inside of the contour.}
    \label{fig:grint-inv42-u11-contour}
  \end{center}
\end{figure}

Here we evaluate the unitary Graßmannian integral
\eqref{eq:grass-int-barg} for the $\mathfrak{u}(1,1)$ Yangian
invariant $\Psi_{4,2}$. This is the simplest invariant for which
different kinematic regions emerge in the undeformed limit. In this
sense, it is a toy model for the example studied in the previous
section. We begin the evaluation by reducing the $U(2)$ integral in
\eqref{eq:grass-int-barg} to a $U(1)$ integral with the help of the
bosonic delta functions,
\begin{align}
  \label{eq:int-42-u11}
  \Psi_{4,2}
  =
  \frac{\delta(P)}
  {|\lambda^1|^2+|\lambda^2|^2}\,
  \int\displaylimits_{U(1)}\!\![\D\smallunitary]\,
    \frac{1}{\smallunitary^{1-c_2}
      |\mathtt{A}-\smallunitary\mathtt{B}|^{2(1+v_1-v_2)}
      (\mathtt{A}-\smallunitary\mathtt{B})^{c_2-c_1}}\,.
\end{align}
Here $P=-|\lambda^1|^2-|\lambda^2|^2+|\lambda^3|^2+|\lambda^4|^2$, and
we introduced
\begin{align}
  \label{eq:int-42-u11-ab}  
  \mathtt{A}=\frac{\lambda^1\overline{\lambda}^3}
  {|\lambda^1|^2+|\lambda^2|^2}\,,\quad
  \mathtt{B}=\frac{-\lambda^4\overline{\lambda}^2}
  {|\lambda^1|^2+|\lambda^2|^2}\,.
\end{align}
Notice that we write $\lambda_1^i\equiv \lambda^i$ because these
variables have only one component in case of the algebra
$\mathfrak{u}(1,1)$. The integral \eqref{eq:int-42-u11} can be
expressed in terms of a Gauß hypergeometric function and even more
specifically an associated Legendre function, see
\cite{Kanning:2016eit}. We move on to study the undeformed limit of
\eqref{eq:int-42-u11} by setting $c_1=c_2=0$ and $v_1=v_2$. This
yields
\begin{align}
  \label{eq:inv42-u11-contourint}
  \Psi_{4,2}
  =
  -\frac{\delta(P)}
  {|\lambda^1|^2+|\lambda^2|^2}\,
  \frac{1}{2\pi i\,\overline{\mathtt{A}}\mathtt{B}}\oint\!\D\smallunitary
  \frac{1}{\smallunitary\Big(\smallunitary-\frac{\mathtt{A}}{\mathtt{B}}\Big)
  \Big(\smallunitary-\frac{\overline{\mathtt{B}}}{\overline{\mathtt{A}}}\Big)}\,,
\end{align}
where, as after \eqref{eq:psi63u224-cint}, we reformulated the Haar
measure on $U(1)$ in terms of a contour integral in the complex
variable $\smallunitary$. Thus we integrate counterclockwise along the
unit circle. Just like for the example in the previous section, we
compute this integral using the residue theorem. The pole at
$\smallunitary=0$ contributes to the integral irrespective of the
kinematics. In contrast to the previous section, there is now only one
pair poles at
$\smallunitary=\frac{\mathtt{A}}{\mathtt{B}},\frac{\overline{\mathtt{B}}}{\overline{\mathtt{A}}}$
that depends on the kinematics, cf.\
figure~\ref{fig:grint-inv42-u11-contour}. For generic kinematics,
precisely one of these poles lies inside of the contour. Which one is
determined by the sign of the variable
\begin{align}
  \label{eq:psi42u11-respos}
  1-\left|\frac{\mathtt{A}}{\mathtt{B}}\right|^2
  \propto
  |\lambda^2|^2-|\lambda^3|^2\,,
\end{align}
where we used $P=0$ and dropped a non-negative factor. Summing the two
residues contributing to \eqref{eq:inv42-u11-contourint} gives the
undeformed invariant
\begin{align}
  \label{eq:psi42u11}
  \Psi_{4,2}
  =
  \frac{\delta(P)}{|\lambda^2|^2-|\lambda^3|^2}
  \begin{cases}
    \phantom{-}\frac{\lambda^4\overline{\lambda^2}}{\lambda^1\overline{\lambda^3}}
    &\text{for}\quad |\lambda^2|<|\lambda^3|\,,\\[1ex]
    -\frac{\overline{\lambda^1}\lambda^3}{\overline{\lambda^4}\lambda^2}
    &\text{for}\quad |\lambda^2|>|\lambda^3|\,
  \end{cases}
\end{align}
with two kinematic regions.

Let us now show that these two regions are required by
integrability. Our argument is based on a parity transformation
$\mathcal{P}$ of the Graßmannian integral \eqref{eq:grass-int-barg}
discussed in appendix~\ref{sec:parity}. In case of the toy model under
investigation here, this transformation boils down to exchanging the
spinors $\lambda^1\stackrel{\;\mathcal{P}}{\leftrightarrow}\lambda^2$,
$\lambda^3\stackrel{\;\mathcal{P}}{\leftrightarrow}\lambda^4$. It is a
symmetry of the $\mathfrak{u}(1,1)$ Yangian invariant $\Psi_{4,2}$ in
\eqref{eq:int-42-u11} for general deformation parameters $v_1,v_2$ and
equal representation labels $c_1=c_2$. In section~\ref{sec:r-matrix}
below, we will see that this invariant can be interpreted as an
R-matrix. Its symmetry under $\mathcal{P}$ translates into the
well-known parity invariance of this R-matrix. With this background
knowledge, we study the action of $\mathcal{P}$ on the undeformed
$\Psi_{4,2}$ in \eqref{eq:psi42u11}. We find that it \emph{exchanges}
the two kinematic regions. Let us discuss what would happen if we used
the expression of either region in \eqref{eq:psi42u11} and declared it
to be valid for all kinematics. The resulting function would still
satisfy the Yangian invariance condition \eqref{eq:yi-exp-1}, which
takes the form of a set of differential equations in $\lambda^i$, for
generic values of $\lambda^i$. However, it would clearly violate
parity symmetry.  Therefore the two kinematic regions in
\eqref{eq:psi42u11} are required for the proper Yangian invariant,
which is related to the R-matrix.

Can we exploit the parity transformation $\mathcal{P}$ from
appendix~\ref{sec:parity} also to show the necessity of the four
kinematic regions of the undeformed $\mathfrak{u}(2,2|0+4)$ Yangian
invariant $\Psi_{6,3}$ in \eqref{eq:inv63-u2204-lastint-eval} for
integrability? This is not possible because the region
$s_{234},s_{345}>0$, in which the invariant agrees with the amplitude
$\mathcal{A}_{6,3}$, is by itself invariant under $\mathcal{P}$. The
transformation $\mathcal{P}$ is part of a known dihedral symmetry of
the amplitude $\mathcal{A}_{6,3}$, cf.\ \cite{Hodges:2009hk}. In fact,
it is the only non-trivial element of that group which is compatible
with our choice of energy signs in \eqref{eq:spinors-partition}. It
would be interesting to look for a bigger discrete symmetry group of
\eqref{eq:inv63-u2204-lastint-eval} that connects all four regions and
is needed for an integrability-based reason. Because $\Psi_{6,3}$ can
be interpreted as a product of three R-matrices, see once again
section~\ref{sec:r-matrix} below, we suspect that such a bigger
symmetry group could involve parity transformations of the individual
R-matrices and the Yang-Baxter equation satisfied by the product. So
far this equation has been verified for the amplitude
$\mathcal{A}_{6,3}$ merely on the level of the original Graßmannian
integral or equivalently for on-shell diagrams
\cite{ArkaniHamed:2012nw,Ferro:2013dga}, disregarding in both cases
the integration contour.  The existence of such a discrete symmetry
group of \eqref{eq:inv63-u2204-lastint-eval} would signal a subtle
breakdown of integrability in the formula
\cite{Drummond:2008vq,Drummond:2008bq} for the
amplitude~$\mathcal{A}_{6,3}$.

It is also conceivable that the region structure of the undeformed
Yangian invariant $\Psi_{6,3}$ in \eqref{eq:inv63-u2204-lastint-eval}
is a sign of a broken conformal symmetry of the
amplitude~$\mathcal{A}_{6,3}$. This is currently under investigation
\cite{Kanning:2018}. Both quantities are certainly invariant under
\emph{infinitesimal} transformations of the conformal algebra
$\mathfrak{su}(2,2)\subset\mathfrak{u}(2,2|0+4)$ for generic values of
the spinors $\lambda^i$. However, to the best of our knowledge, the
invariance of $\mathcal{A}_{6,3}$ under \emph{finite} transformations
has not been proven. The representations of $\mathfrak{su}(2,2)$ in
terms of spinors $\lambda^i$ from
section~\ref{sec:symmetry-generators} can be exponentiated to ones of
the group $SU(2,2)$ \cite{Gross:1972,Post:1976}. In particular, some
group elements act quite non-trivially as integral transformations. We
remark that kinematic regions somewhat reminiscent of those in
\eqref{eq:inv63-u2204-lastint-eval} were introduced for $\text{NMHV}$
amplitudes in split signature $(2,2)$ to obtain amplitudes with
satisfactory superconformal properties in twistor space
\cite{Korchemsky:2009jv}.

\subsection{Eight Particles for
  \texorpdfstring{$\mathfrak{u}(2,2|4)$}{u(2,2|4)}}
\label{sec:psi84u224}

In our investigation of the four- and six-particle
$\mathfrak{u}(2,2|0+4)$ Yangian invariants in the previous sections,
we could extract the respective amplitudes, at least in one kinematic
region, by setting all deformation parameters to zero. Obviously, this
raises the question if, in general, the amplitude $\mathcal{A}_{2K,K}$
can be obtained from the Yangian invariant $\Psi_{2K,K}$ in this
simple way. Therefore we continue our studies in this section by
computing $\Psi_{8,4}$. Vexingly, we will find that the simple
procedure does not hold up for extracting the amplitude
$\mathcal{A}_{8,4}$ from it.

To begin with, we reduce the $U(4)$ Graßmannian integral formula
\eqref{eq:grass-int-barg} for the Yangian invariant $\Psi_{8,4}$ of
the algebra $\mathfrak{u}(2,2|0+4)$ to a $U(2)$ integral by making use
of the bosonic delta functions. We display the resulting integral here
only for equal integer deformation parameters $c_1=c_2=c_3=c_4$ in
order to spare the reader a slightly more cumbersome expression. In
addition, we abbreviate frequently occurring differences of complex
deformation parameters as $z_1\equiv v_1-v_2$, $z_2\equiv v_2-v_3$,
$z_3\equiv v_3-v_4$. This yields
\begin{align}
  \label{eq:psi84u224-int}
  \begin{aligned}
    \Psi_{8,4}
    =
    \frac{\delta^{4|0}(P)\delta^{0|8}(Q)}{s_{1234}^{8}}
    \int\displaylimits_{U(2)}\!\![\D\smallunitary]\,
    &\frac{
      \delta^{0|8}(\mathtt{a}\,\smallunitary+\mathtt{b})
    }{
      \det{\smallunitary}^{2-c_4}
      |\mathtt{E}_1\det(\smallunitary-\mathtt{F}_1)|^{2(1+z_1)}
    }\,\\
    \cdot
    &\frac{
      1
    }{
      |\mathtt{E}_2\det(\smallunitary-\mathtt{F}_2)|^{2(1+z_2)}
      |\mathtt{E}_3\det(\smallunitary-\mathtt{F}_3)|^{2(1+z_3)}
    }\,.
  \end{aligned}
\end{align}
Here the momentum and supermomentum conserving delta functions are
defined in \eqref{eq:amp-bosferm-delta}. In the denominator, we
rearranged the minors of the Graßmannian integrand
\eqref{eq:eq:grass-int-unitary-integrand-final} into $2\times 2$
determinants depending on the kinematic data
\begin{align}
  \label{eq:psi84u224-data-minors}
  \begin{aligned}
    \mathtt{F}_1&=
    \frac{\sqrt{s_{234}}\sqrt{s_{567}}^{\,-1}}{\overline{\langle 1|}2\!+\!3\!+\!4|5\rangle}
    \begin{pmatrix}
      0&0\\
      \sqrt{s_{1234}}\,\overline{\langle 67\rangle}&\overline{\langle 8|}6\!+\!7|5\rangle\\
    \end{pmatrix}\,,\quad
  \mathtt{E}_1=
  \frac{\overline{\langle 1}|2\!+\!3\!+\!4|5\rangle}{s_{1234}}\,,\\
  \mathtt{F}_2&=
  \frac{\sqrt{s_{234}s_{567}}^{\,-1}}{\overline{\langle 12\rangle}\langle 56\rangle}
    \begin{pmatrix}
      \langle 43\rangle \overline{\langle 1|}2\!+\!3\!+\!4|5\!+\!6\overline{|7\rangle}&
      \sqrt{s_{1234}}\langle 34\rangle\langle 56\rangle\overline{\langle 18\rangle}\\
      \sqrt{s_{1234}}\,\overline{\langle 2|}3\!+\!4|5\!+\!6\overline{|7\rangle}&
      \langle 56\rangle\overline{\langle 2|}3\!+\!4|5\!+\!6\!+\!7\overline{|8\rangle}\\
    \end{pmatrix}\,,\quad
  \mathtt{E}_2=
  \frac{\overline{\langle 12\rangle}\langle 56\rangle}{s_{1234}}\,,\\
  \mathtt{F}_3&=
  \frac{\sqrt{s_{567}}\sqrt{s_{234}}^{\,-1}}{\overline{\langle 4|}5\!+\!6\!+\!7|8\rangle}
    \begin{pmatrix}
      0&\sqrt{s_{1234}}\langle 23\rangle\\
      0&\overline{\langle 4|}2\!+\!3|1\rangle\\
    \end{pmatrix}\,,\quad
  \mathtt{E}_3=
  \frac{\overline{\langle 4|}1\!+\!2\!+\!3|8\rangle}{s_{1234}}\,.\\
  \end{aligned}
\end{align}
The numerator involves further kinematics contained in
\begin{align}
  \label{eq:psi84u224-data-fermi}
  \begin{aligned}
  \mathtt{a}&=
  \frac{1}{\sqrt{s_{234}}}
    \begin{pmatrix}
      \sqrt{s_{1234}}\big(\overline{\langle 34\rangle}\tilde\eta^2+
      \overline{\langle 42\rangle}\tilde\eta^3+
      \overline{\langle 23\rangle}\tilde\eta^4\big)
      \\
      s_{234}\tilde\eta^1+
      \langle 2|3\!+\!4\overline{|1\rangle}\tilde\eta^2+
      \langle 3|2\!+\!4\overline{|1\rangle}\tilde\eta^3+
      \langle 4|2\!+\!3\overline{|1\rangle}\tilde\eta^4
    \end{pmatrix}^t\,,\\
    \mathtt{b}&=
    \frac{1}{\sqrt{s_{567}}}
    \begin{pmatrix}
      \sqrt{s_{5678}}\big(\overline{\langle 67\rangle}\tilde\eta^5+
      \overline{\langle 75\rangle}\tilde\eta^6+
      \overline{\langle 56\rangle}\tilde\eta^7\big)
      \\
      \langle 5|6\!+\!7\overline{|8\rangle}\tilde\eta^5+
      \langle 6|5\!+\!7\overline{|8\rangle}\tilde\eta^6+
      \langle 7|5\!+\!6\overline{|8\rangle}\tilde\eta^7+
      s_{567}\tilde\eta^8\\
    \end{pmatrix}^t\,&\\
  \end{aligned}
\end{align}
with the four-dimensional fermionic variables
$\tilde\eta^i=(\tilde\eta^i_{\dot{a}})$. The $U(2)$ integral for the
deformed Yangian invariant $\Psi_{8,4}$ in \eqref{eq:psi84u224-int} is
structurally a straightforward generalization of the $U(1)$ integral
for $\Psi_{6,3}$ with $c_1=c_2=c_3$ from
\eqref{eq:psi63u224-int}. Most characteristically, the absolute values
in the denominator contain $2\times 2$ determinants in the $U(2)$
case, and there is one additional such factor compared to the $U(1)$
integral.

Despite the structural similarities, the two integrals behave
fundamentally different in the undeformed limit, where $c_i=0$ and all
$v_i$ are equal, i.e.\ $z_i=0$. The $U(1)$ integral
\eqref{eq:psi63u224-int} computing $\Psi_{6,3}$ reduces to the contour
integral \eqref{eq:psi63u224-cint}, which is finite for \emph{generic}
kinematics and can be evaluated by means of the residue
theorem. Divergencies may arise for \emph{special} kinematic
configurations, such as $s_{234}=0$ or $s_{345}=0$ where poles of the
integrand move onto the integration contour, cf.\
\eqref{eq:psi63u224-respos}. In the $U(2)$ integral
\eqref{eq:psi84u224-int} for $\Psi_{8,4}$, we can safely set
$c_i=0$. However, upon imposing $z_i=0$, we observe that the integral
diverges even for \emph{generic} kinematics.

Let us sketch this in more detail. To gain a handle on the
divergencies, we first need a criterion for which kinematic data the
integrand of $\Psi_{8,4}$ in \eqref{eq:psi84u224-int} can become
singular:
\begin{align}
  \label{eq:psi84u224-data-cond-general}
  \det(\smallunitary-\mathtt{F}_i)=0
  \quad\text{for some}\quad\smallunitary\in U(2)
  \quad\Leftrightarrow\quad
  \det(\mathtt{F}_i\mathtt{F}_i^\dagger-1_2)\leq 0\,,
\end{align}
see e.g.\ theorem~5.5.1 in \cite{Hua:1963}. For the matrices from
\eqref{eq:psi84u224-data-minors}, this criterion reads\footnote{These
  variables are combinations of dual conformal cross ratios before
  dropping the factors. Recall in this context that also the analogous
  quantities for $\Psi_{6,3}$ in \eqref{eq:psi63u224-respos} are such
  cross ratios.}
\begin{align}
  \label{eq:psi84u224-data-cond}
  \begin{aligned}
    \det{(\mathtt{F}_1\mathtt{F}_1^\dagger-1_2)}
    \propto -s_{2345}\,,\quad
    \det{(\mathtt{F}_2\mathtt{F}_2^\dagger-1_2)}
    \propto s_{3456}\,,\quad
    \det{(\mathtt{F}_3\mathtt{F}_3^\dagger-1_2)}
    \propto -s_{4567}\,,\\
  \end{aligned}
\end{align}
where we dropped non-negative factors. The signs of these
four-particle Mandelstam variables determine which factors of the
integrand in \eqref{eq:psi84u224-int} become singular while the
integration variable $\smallunitary$ traverses the $U(2)$ group
manifold. They give rise to $2^3=8$ kinematic regions. Clearly, the
integrand contains singularities in seven of these. Based on a
numerical study using random momenta, the only non-singular region
$s_{2345}<0$, $s_{3456}>0$, $s_{4567}<0$ appears to be kinematically
forbidden for our choice of energy signs in
\eqref{eq:spinors-partition}. In the undeformed limit $c_i=0$ and
$z_i\to 0$, the singularities of the integrand yield a divergent
integral \eqref{eq:psi84u224-int} for $\Psi_{8,4}$. Presently, we are
able to extract the leading divergent terms\footnote{We are grateful
  to Jacob Bourjaily for supplying us with data adapted from
  \cite{Bourjaily:2010wh,Bourjaily:2012gy} to cross-check this
  result.} from this integral by focusing on the vicinity of the
singularities \cite{Kanning:2018a}. We find that the precise
expressions of these terms depend on a rich substructure of the
kinematic regions \eqref{eq:psi84u224-data-cond}. These subregions are
defined primarily using the signs of $s_{345}$, $s_{456}$, $s_{781}$,
$s_{812}$, and those of certain determinants of Mandelstam
variables. The consecutive three-particle Mandelstam variables
appearing here and the four-particle ones in
\eqref{eq:psi84u224-data-cond} are precisely all those whose signs are
allowed to change in our setup. In most subregions the leading
divergent terms are of the form $\frac{1}{z_i}$, while in some there
exist even more singular terms with $\frac{1}{z_iz_j}$.

The occurrence of these divergent terms signals that we cannot simply
take the undeformed limit of the Yangian invariant $\Psi_{8,4}$ in
\eqref{eq:psi84u224-int} and end up with a finite expression that
agrees with the amplitude $\mathcal{A}_{8,4}$ in one kinematic region.
Finding a way to extract this amplitude from $\Psi_{8,4}$ is arguably
one of the most pressing open problems of our unitary Graßmannian
integral approach. On the one hand, it certainly requires more
sophisticated techniques for the evaluation of the $U(2)$ integral
\eqref{eq:psi84u224-int}. It could be helpful use a matrix version of
the Cayley transformation to map the $U(2)$ integration variable
$\smallunitary$ to a Hermitian $2\times 2$ matrix, see e.g.\
\cite{Shabat:1992}. This matrix can in turn be mapped as in
\eqref{eq:minkvec-hermmat} to a vector in real Minkowski space
$\mathbb{R}^{1,3}$. Our integral \eqref{eq:psi84u224-int} then
essentially takes the form of a one-loop Feynman integral with complex
exponents of the propagators and complex external momenta given in
terms of the matrices $\mathtt{F}_i$. This might make the machinery
developed for such Feynman integrals applicable to our problem. On the
other hand, also conceptually the relation between $\Psi_{8,4}$ and
$\mathcal{A}_{8,4}$ needs further attention. In this regard, we refer
the reader to section~\ref{sec:norm-div} below. There we will
demonstrate the existence of a divergent term, albeit only for special
kinematics, in case of the simple invariant $\Psi_{4,2}$ from
section~\ref{sec:psi42u224}. This will suggest a new perspective even
for the relation between $\Psi_{4,2}$ and $\mathcal{A}_{4,2}$.

Although the exact connection between the Yangian invariants
$\Psi_{2K,K}$ computed by the unitary Graßmannian integral
\eqref{eq:grass-int-barg} and the amplitudes $\mathcal{A}_{2K,K}$ is
still unclear for general $K$, we can establish a fairly complete
understanding of $\Psi_{2K,K}$ from the point of view of
integrability. We will develop it systematically in the following
sections. For this we will employ a different basis, in which the
Yangian invariants are expressed in terms of harmonic oscillators
instead of spinor helicity variables. These new variables are
instrumental in revealing fascinating relations between the unitary
Graßmannian integral and other subjects such as integrable spin chains
and matrix models. The inquisitive reader may already jump ahead and
take a first look at table~\ref{tab:interpret}, where different
interpretations of our unitary integral are summarized.

\section{Graßmannian Integral in Oscillator Basis}
\label{sec:matrix-models}

In this section, we define an analogue of the unitary Graßmannian
integral \eqref{eq:grass-int-barg} for a class of representations that
are constructed using oscillator algebras instead of spinor helicity
variables. This integral formula reveals interesting connections
between Yangian invariants and matrix models. Moreover, we show that
it is related to the original formula \eqref{eq:grass-int-barg} via a
Bargmann transformation. This integral transformation generalizes the
change of basis from Fock to position space for the harmonic
oscillator in elementary quantum mechanics.

\subsection{Oscillator Representations}
\label{sec:osc-rep}

We introduce \emph{oscillator representations} of the non-compact
superalgebra $\mathfrak{u}(p,q|r+s)$ following
\cite{Bars:1982ep}. Such representations have a long history in the
physics literature and are sometimes referred to as ``ladder
representations'', see e.g.\ \cite{Todorov:1966zz}. In case of the
algebra $\mathfrak{u}(2,2|4)$, they are discussed also in
\cite{Gunaydin:1984fk} and play a prominent role in the planar
$\mathcal{N}=4$ SYM spectral problem. For $\mathfrak{u}(p,q=p|r+s)$,
they are unitarily equivalent to the representations that we
encountered already in section~\ref{sec:symmetry-generators}, as will
be shown in section~\ref{sec:trafo-int-gen} below. This justifies our
use of the symbols $\oscrep_c$ and $\bar{\oscrep}_c$ for the two
classes of representations defined in the following.

The basic ingredient of both classes is a family of
\emph{superoscillators} obeying
\begin{align}
  \label{eq:super-osc}
  \begin{aligned}
    [\mathbf{A}_{\indnm{A}},\bar{\mathbf{A}}_{\indnm{B}}\}=\delta_{\indnm{AB}}\,,\quad
    [\mathbf{A}_{\indnm{A}},\mathbf{A}_{\indnm{B}}\}=0\,,\quad
    [\bar{\mathbf{A}}_{\indnm{A}},\bar{\mathbf{A}}_{\indnm{B}}\}=0\,,\quad
    \mathbf{A}_{\indnm{A}}^\dagger=\bar{\mathbf{A}}_{\indnm{A}}\,,\quad
    \mathbf{A}_{\indnm{A}}|0\rangle=0\,,
  \end{aligned}  
\end{align}
where the indices of the annihilation operators
$\mathbf{A}_{\indnm{A}}$ and creation operators
$\bar{\mathbf{A}}_{\indnm{A}}$ take the values
$\indnm{A}=1,\ldots,n+m$ with $n=p+q$ and $m=r+s$. It is equipped with
a conjugation $\dagger$ and acts on a Fock space $\mathcal{F}$ that is
spanned by monomials in $\bar{\mathbf{A}}_{\indnm{A}}$ acting on a
vacuum state $|0\rangle$. We illustrate the grading of the superindex
${\indnm{A}}$ by means of the creation operators,
\begin{align}
  \label{eq:osc-split}
  \left(\bar{\mathbf{A}}_{\indnm{A}}\right)=
  \left(
    \begin{array}{c}    
      \vcenter{\vspace{1.2cm}}\bar{\mathbf{A}}_{\indssub{A}}\\[0.3em]
      \hdashline\\[-1.0em]
      \vcenter{\vspace{1.2cm}}\bar{\mathbf{A}}_{\dot{\indssub{A}}}\\
    \end{array}
  \right)=
    \left(
    \begin{array}{c}    
      \bar{\mathbf{a}}_{\alpha}\\[0.3em]
      \hdashline\\[-1.0em]
      \bar{\mathbf{c}}_{a}\\[0.3em]
      \hdashline\\[-1.0em]
      \bar{\mathbf{b}}_{\dot{\alpha}}\\[0.3em]
      \hdashline\\[-1.0em]
      \bar{\mathbf{d}}_{\dot{a}}\\
    \end{array}
  \right)\,.
\end{align}
Here we first split the family of superoscillators with
$\mathfrak{gl}(n|m)$ index $\indnm{A}$ into two parts. One carries a
$\mathfrak{gl}(p|r)$ index $\indssub{A}=1,\ldots,p+r$ and the other
one a $\mathfrak{gl}(q|s)$ index $\dot{\indssub{A}}=p+r+1,\ldots,n+m$.
Then we spelled out the superoscillators
$\bar{\mathbf{A}}_{\indssub{A}},\bar{\mathbf{A}}_{\dot{\indssub{A}}}$
in terms of bosonic oscillators
$\bar{\mathbf{a}}_\alpha,\bar{\mathbf{b}}_{\dot{\alpha}}$ and
fermionic oscillators $\bar{\mathbf{c}}_a,\bar{\mathbf{d}}_{\dot{a}}$
with the index ranges $\alpha=1,\ldots,p$, $\dot{\alpha}=1,\ldots,q$,
$a=1,\ldots,r$, and $\dot{a}=1,\ldots,s$. This notation fixes the
grading and is inspired by \cite{Ferro:2013dga} and
\cite{Beisert:2003jj}.

A set of generators $\mathbf{J}_{\indnm{AB}}$ of the
$\mathfrak{gl}(n|m)$ superalgebra is provided by
\begin{align}
  \label{eq:gen-ordinary}
  (\mathbf{J}_{\indnm{AB}})=
  \left(
  \begin{array}{c:c}
    \bar{\mathbf{A}}_{\indssub{A}} \mathbf{A}_{\indssub{B}}&
    \bar{\mathbf{A}}_{\indssub{A}} \bar{\mathbf{A}}_{\dot{\indssub{B}}}\\[0.3em]
    \hdashline\\[-1.0em]
    -(-1)^{|\dot{\indssub{A}}|}\mathbf{A}_{\dot{\indssub{A}}}\mathbf{A}_{\indssub{B}}&
    -(-1)^{|\dot{\indssub{A}}|}\mathbf{A}_{\dot{\indssub{A}}}\bar{\mathbf{A}}_{\dot{\indssub{B}}}\\
  \end{array}
  \right)\,.
\end{align}
The two diagonal blocks of this supermatrix realize the subalgebras
$\mathfrak{gl}(p|r)$ and $\mathfrak{gl}(q|s)$, respectively. Let
$\oscrep_c\subset\mathcal{F}$ be the eigenspace of the central element
\begin{align}
  \label{eq:central-ord}
  \mathbf{C}=\tr(\mathbf{J}_{\indnm{AB}})
\end{align}
with eigenvalue $c$. This infinite-dimensional space forms a unitary
representation of the superalgebra $\mathfrak{u}(p,q|r+s)$ for each
$c\in\mathbb{Z}$, see \cite{Bars:1982ep}. It contains a lowest weight
state annihilated, by definition, by all generators
$\mathbf{J}_{\indnm{AB}}$ with $\indnm{A}>\indnm{B}$. Note that the
space $\oscrep_c$ is finite-dimensional in the special cases $q=0$ or
$p=0$. We define a class of \emph{dual} representations by applying
the automorphism
$\bar{\mathbf{J}}_{\indnm{AB}}=-(-1)^{|\indnm{A}|+|\indnm{A}||\indnm{B}|}\mathbf{J}_{\indnm{AB}}^\dagger$
of the $\mathfrak{gl}(n|m)$ algebra to the generators in
\eqref{eq:gen-ordinary},
\begin{align}
  \label{eq:gen-dual}
  (\bar{\mathbf{J}}_{\indnm{AB}})=
  \left(
  \begin{array}{c:c}
    -(-1)^{|\indssub{A}|+|\indssub{A}||\indssub{B}|}\bar{\mathbf{A}}_{\indssub{B}} \mathbf{A}_{\indssub{A}}&
    -(-1)^{|\indssub{A}|+|\dot{\indssub{B}}|+|\indssub{A}||\dot{\indssub{B}}|}\mathbf{A}_{\dot{\indssub{B}}}\mathbf{A}_{\indssub{A}}\\[0.3em]
    \hdashline\\[-1.0em]
    (-1)^{|\dot{\indssub{A}}|+|\dot{\indssub{A}}||\indssub{B}|}\bar{\mathbf{A}}_{\indssub{B}} \bar{\mathbf{A}}_{\dot{\indssub{A}}}&
    (-1)^{|\dot{\indssub{A}}|+|\dot{\indssub{B}}|+|\dot{\indssub{A}}||\dot{\indssub{B}}|}\mathbf{A}_{\dot{\indssub{B}}}\bar{\mathbf{A}}_{\dot{\indssub{A}}}\\
  \end{array}
  \right)\,.
\end{align}
We denote by $\bar{\oscrep}_c\subset\mathcal{F}$ the eigenspace of
the central element
\begin{align}
  \label{eq:central-dual}
  \bar{\mathbf{C}}=
  \tr{(\bar{\mathbf{J}}_{\indnm{AB}})}
\end{align}
with eigenvalue $c$. This space carries a unitary representation of
$\mathfrak{u}(p,q|r+s)$ for each $c\in\mathbb{Z}$. It contains a
highest weight state annihilated by all
$\bar{\mathbf{J}}_{\indnm{AB}}$ with $\indnm{A}<\indnm{B}$.

Let us add that, recently, all unitary representations of the
superalgebra $\mathfrak{su}(p,q|m)$ were constructed by means of a
generalized oscillator formalism in \cite{Gunaydin:2017lhg}. This
paper also contains an informative overview of the evolution of the
oscillator method with numerous further references.

\subsection{Graßmannian Integral}
\label{sec:int-osc}

The oscillator representations of $\mathfrak{u}(p,q|r+s)$ from the
previous section at hand, we construct an analogue of the unitary
Graßmannian integral \eqref{eq:grass-int-barg}. Yangian invariants for
these representations are given by
\begin{align}
  \label{eq:grass-int-unitary}
  |\Psi_{2K,K}\rangle
  =\int\displaylimits_{U(K)}\!\![\D\mathcal{C}]\,
  \mathscr{F}(\mathcal{C})(\det\mathcal{C})^r
  e^{\tr(\mathbf{I}_\bullet\mathcal{C}^\dagger+\mathcal{C}\mathbf{I}_\circ^t)}
    |0\rangle\,.
\end{align}
Here the delta functions of \eqref{eq:grass-int-barg} have been
replaced by an exponential function acting on a Fock vacuum. Its
argument contains the $K\times K$ matrices
\begin{align}
  \label{eq:osc-matrix}
  \mathbf{I}_{\mathrel{\ooalign{\raisebox{0.4ex}{$\scriptstyle\bullet$}\cr\raisebox{-0.4ex}{$\scriptstyle\circ$}}}}=
  \begin{pmatrix}
    (1\mathrel{\ooalign{\raisebox{0.7ex}{$\bullet$}\cr\raisebox{-0.3ex}{$\circ$}}} K+1)&
    \cdots&
    (1\mathrel{\ooalign{\raisebox{0.7ex}{$\bullet$}\cr\raisebox{-0.3ex}{$\circ$}}} 2K)\\
    \vdots&&\vdots\\
    (K\mathrel{\ooalign{\raisebox{0.7ex}{$\bullet$}\cr\raisebox{-0.3ex}{$\circ$}}} K+1)&
    \cdots&
    (K\mathrel{\ooalign{\raisebox{0.7ex}{$\bullet$}\cr\raisebox{-0.3ex}{$\circ$}}} 2K)\\
  \end{pmatrix}.
\end{align}
Their entries are contractions of creation operators,
\begin{align}
  \label{eq:bullets-nc-recall}
    (k\bullet l)=
    \sum_{\indssub{A}}\bar{\mathbf{A}}^l_{\indssub{A}}\bar{\mathbf{A}}^k_{\indssub{A}}\,,\quad
    (k\circ l)=
    \sum_{\dot{\indssub{A}}}\bar{\mathbf{A}}^l_{\dot{\indssub{A}}}\bar{\mathbf{A}}^k_{\dot{\indssub{A}}}\,.
\end{align}
They are $\mathfrak{gl}(p|r)$ and $\mathfrak{gl}(q|s)$ invariant,
respectively. Moreover, they are bosonic because fermionic oscillators
appear only in quadratic terms. The rest of \eqref{eq:grass-int-barg}
can be found essentially unchanged in
\eqref{eq:grass-int-unitary}. The function $\mathscr{F}(\mathcal{C})$
is the manifestly single-valued expression from
\eqref{eq:eq:grass-int-unitary-integrand-final}, which we obtained
from the formal integrand in \eqref{eq:grass-int-unitary-integrand},
now in general with $q\neq p$. It contains the inhomogeneities $v_i$
and the representation labels $c_i$. These are related via
\eqref{eq:spec-redef} to the deformation parameters $v_i^\pm$, which
have to obey \eqref{eq:spec-perm}. The $v_i$ and $c_i$ also enter the
monodromy matrix $M(u)$ defined in \eqref{eq:yangian-mono-spinchain}
that is associated with $|\Psi_{2K,K}\rangle$ from
\eqref{eq:grass-int-unitary}. The first $K$ of its $N=2K$ sites carry
the oscillator representations $\bar\oscrep_{c_i}$, and the remaining
$K$ sites have the representations $\oscrep_{c_i}$. Therefore the
$\mathfrak{gl}(n|m)$ generators in the Lax operators
\eqref{eq:yangian-def-lax} of the monodromy are
\begin{align}
  \label{eq:mono-osc-gen}
  J_{\indnm{AB}}^i=
  \begin{cases}
    \bar{\mathbf{J}}^i_{\indnm{AB}}&\text{for}\quad i=1,\ldots,K\,,\\
    \mathbf{J}^i_{\indnm{AB}}&\text{for}\quad i=K+1,\ldots,2K\,,
  \end{cases}
\end{align}
which are given in \eqref{eq:gen-dual} and \eqref{eq:gen-ordinary},
respectively. With this monodromy $M(u)$, the state
$|\Psi_{2K,K}\rangle$ satisfies the Yangian invariance condition
\eqref{eq:yi-exp-1} and therefore also \eqref{eq:yi}. A proof of this
central statement is presented in \cite{Kanning:2016eit}. It is a
straightforward extension of the proof for the bosonic algebra
$\mathfrak{u}(p,q)$ in \cite{Kanning:2014cca}. In the latter
reference, the existence of a closed contour is assumed to perform
partial integrations, and a measure of the form
\eqref{eq:measure-haar} is used. However, at the time, no explicit
example of such a contour was known. This gap is filled here by the
unitary contour and the single-valued integrand, recall
section~\ref{sec:singlevaluedness}.

\begin{table}
  \begin{center}
    \begin{tabular}{|l|l|}
      \hline
      \multicolumn{1}{|c|}{\vphantom{\pbox{1cm}{a\\a}}\emph{Parameters}}
      &\multicolumn{1}{|c|}{\emph{Interpretation}}\\
      \hline
      $v_i^\pm\to 0$ and $\mathfrak{u}(2,2|4)$
      &\vphantom{\pbox{1cm}{A\\A\\A}}\pbox{8cm}{\strut
        $\Psi_{2K,K}$ related to amplitude $\mathcal{A}_{2K,K}$?\\
      (section~\ref{sec:sample-invariants})
      \strut}\\
      \hline
      \vphantom{\pbox{1cm}{A\\A\\A\\[2ex]A}}\pbox{8cm}{\strut
      $v_i^\pm$ s.t.\ $\mathscr{F}(\mathcal{C})(\det\mathcal{C})^{r}=
      \begin{cases}
        (\det\mathcal{C})^{q-s-c_K}\\
        1
      \end{cases}\!\!\!\!\!\!\!
      $\\
      \strut}
      &\vphantom{\pbox{1cm}{A\\A\\A\\[2ex]A}}\pbox{8cm}{\strut
      $|\Psi_{2K,K}\rangle$ is $
      \begin{cases}
        \text{Leutwyler-Smilga model \cite{Leutwyler:1992yt}}\\
        \text{Brezin-Gross-Witten model \cite{Gross:1980he,Brezin:1980rk,Bars:1979xb}}
      \end{cases}\!\!\!\!\!\!
      $\\
      Link to cusp equation for $|\Psi_{4,2}\rangle$ (section~\ref{sec:cusp-equation})
      \strut}
      \\
      \hline
      $v_i^\pm$ general
      &\vphantom{\pbox{1cm}{A\\A\\A}}\pbox{8cm}{\strut
        $|\Psi_{2K,K}\rangle$ equals product of R-matrices\\
      (section~\ref{sec:r-matrix} and appendix~\ref{sec:gluing})
      \strut}
      \\
      \hline
    \end{tabular}
  \end{center}
  \caption{Interpretations of the unitary Graßmannian integrals
    \eqref{eq:grass-int-barg} and \eqref{eq:grass-int-unitary} for
    different deformation parameters $v_i^\pm$ in the integrand
    $\mathscr{F}(\mathcal{C})$.}
  \label{tab:interpret}
\end{table}

We will show in section~\ref{sec:trafo-int-gen} that the Yangian
invariants $|\Psi_{2K,K}\rangle$ in \eqref{eq:grass-int-unitary} are
related to the $\Psi_{2K,K}$ in \eqref{eq:grass-int-barg} by a change
of basis for representations of the algebra $\mathfrak{u}(p,q=p|r+s)$.
In case of $\mathfrak{u}(2,2|4)$ and vanishing deformation parameters
$v_i^\pm\to 0$, these invariants are of relevance for the
$\mathcal{N}=4$ SYM amplitudes $\mathcal{A}_{2K,K}$, as we argued in
section~\ref{sec:sample-invariants}. Is there also a significance of
the $|\Psi_{2K,K}\rangle$ directly in the oscillator basis? As is
turns out, their integral representation \eqref{eq:grass-int-unitary}
reduces to known matrix models for specific values of
$v_i^\pm$. Equation \eqref{eq:spec-perm} constraining the $v_i^\pm$
has a solution where all minors appearing in
$\mathscr{F}(\mathcal{C})$ from
\eqref{eq:eq:grass-int-unitary-integrand-final}, except for
$\det\mathcal{C}$, have a vanishing exponent,
\begin{align}
  \label{eq:ls-model}
  \mathscr{F}(\mathcal{C})=(\det\mathcal{C})^{-r+q-s-c_K}\,.
\end{align}
With this integrand, \eqref{eq:grass-int-unitary} is equivalent to the
\emph{Leutwyler-Smilga model} \cite{Leutwyler:1992yt}, which describes
aspects of quantum chromodynamics in a certain low energy regime. For
$q=s+c_K$, all factors of $\det\mathcal{C}$ disappear from
\eqref{eq:grass-int-unitary}, and it becomes the
\emph{Brezin-Gross-Witten model}
\cite{Gross:1980he,Brezin:1980rk,Bars:1979xb}, which appears in the
context of two-dimensional lattice gauge theory. Remarkably, these
integrals can be computed exactly as determinants of matrices whose
entries are Bessel functions. For two independent matrices
$\mathbf{I}_\bullet^t$ and $\mathbf{I}_\circ$, this was achieved in
\cite{Schlittgen:2002tj} using the character expansion methods of
\cite{Balantekin:2000vn}. This determinant formula will allow us in
section~\ref{sec:cusp-equation} below to establish a heuristic yet
intriguing link between $|\Psi_{4,2}\rangle$ and the cusp equation
\cite{Eden:2006rx,Beisert:2006ez}, which governs certain all-loop
results in planar $\mathcal{N}=4$ SYM.  In addition, the two
aforementioned matrix integrals provide solutions, so-called
$\tau$-functions, of the Kadomtsev-Petviashvili (KP) hierarchy, cf.\
\cite{Mironov:1994mv,Orlov:2002ka}. It would be interesting to expose
a connection between the well-studied integrable structure of this
hierarchy, see the substantial review in \cite{Miwa:2000}, and the
Yangian invariance of the integrals. Moreover, we are tempted to
speculate that \eqref{eq:grass-int-unitary} could be a KP
$\tau$-function for a wider range of the deformation parameters
$v_i^\pm$. Let us move on to another interpretation of the
$|\Psi_{2K,K}\rangle$ that clarifies their role within the QISM and is
valid even for general $v_i^\pm$ obeying \eqref{eq:spec-perm}. We will
discuss in section~\ref{sec:r-matrix} below that the Yangian invariant
$|\Psi_{4,2}\rangle$ can be understood as an R-matrix. Furthermore, in
appendix~\ref{sec:gluing} we show that $|\Psi_{2K,K}\rangle$ from
\eqref{eq:grass-int-unitary} with its $U(K)$ contour can be
constructed by ``gluing'' together multiple copies of
$|\Psi_{4,2}\rangle$ with $U(2)$ contours. Thus $|\Psi_{2K,K}\rangle$
corresponds to a product of R-matrices. The interpretations outlined
in this paragraph are summarized in table~\ref{tab:interpret}.

\subsection{Bargmann Realization and Transformation}
\label{sec:bargmann}

The equivalence of the two Graßmannian integral formulas
\eqref{eq:grass-int-barg} and \eqref{eq:grass-int-unitary} will be
established in the next section using a Bargmann transformation. For
the one-dimensional harmonic oscillator in quantum mechanics, this
integral transformation implements the change of basis between Fock
and position space. We introduce the Bargmann transformation along the
lines of the original publication \cite{Bargmann:1961gm}. From the
outset, we work in a multi-dimensional setting because it is needed
for our application.

We start out with a family of bosonic oscillators on a Fock space
obeying
\begin{align}
  \label{eq:barg-barg}
  [\mathbf{a}_{\alpha},\bar{\mathbf{a}}_{\beta}]=\delta_{\alpha\beta}\,,\quad 
  {\mathbf{a}_{\alpha}}^\dagger=\bar{\mathbf{a}}_{\alpha}\,,\quad
  \mathbf{a}_{\alpha}|0\rangle=0
\end{align}
with $\alpha,\beta=1,\ldots,p$. Let $\mathbf{a}=(\mathbf{a}_{\alpha})$
etc.\ denote $r$-component column vectors. The relations in
\eqref{eq:barg-barg} are implemented by the \emph{Bargmann
  realization}
\begin{align}
  \label{eq:barg-map-b}
  \bar{\mathbf{a}}\mapsto z\,,\quad
  \mathbf{a}\mapsto\partial_z\,,\quad
  |0\rangle\mapsto \Psi_0(z)=1
\end{align}
on the Bargmann space $\mathscr{H}_\text{B}$. This is the Hilbert
space of holomorphic functions of $z\in\mathbb{C}^p$ with the inner
product
\begin{align}
  \label{eq:barg-inner-barg}
  \langle\Psi(z),\Phi(z)\rangle_{\text{B}}=
  \int\displaylimits_{\mathbb{C}^p}
  \frac{\D^{\,p}\!\overline{z}\D^{\,p}\!z}{(2\pi i)^p}
  e^{-\overline{z}^tz}\overline{\Psi(z)}\Phi(z)\,,
\end{align}
where
$(2i)^{-p}\D^{\,p}\!\overline{z}\D^{\,p}\!z=\D^{\,p}\!\Real{z}\D^{\,p}\!\Imag{z}$
is understood as the measure on $\mathbb{R}^{2p}$. In particular, this
inner product implements the reality condition in
\eqref{eq:barg-barg}, i.e.\
${\partial_{z_{\alpha}}}^\dagger=z_{\alpha}$. The Bargmann realization
can be thought of as a concrete implementation of the formal Fock
space operators. For recent expositions of this realization see e.g.\
\cite{Takhtajan:2008,ZinnJustin:2005}, where it is called
``holomorphic representation''.

In addition, we introduce another family of canonical variables
obeying different reality conditions,
\begin{align}
  \label{eq:barg-sch}
  [\partial_{x_{\alpha}},x_{\beta}]=\delta_{\alpha\beta}\,,\quad 
  {\partial_{x_{\alpha}}}^\dagger=-\partial_{x_{\alpha}}\,,\quad
  {x_{\alpha}}^\dagger=x_{\alpha}\,.
\end{align}
These are considered as operators on the Hilbert space
$\mathscr{H}_{\text{Sch}}$ of square integrable functions of the
variable $x\in\mathbb{R}^p$ with the inner product
\begin{align}
  \label{eq:barg-inner-sch}
  \langle\Psi(x),\Phi(x)\rangle_{\text{Sch}}=
  \int\displaylimits_{\mathbb{R}^p}\!\D^{\,p}\!x\, \overline{\Psi(x)}\Phi(x)\,.
\end{align}
This implementation of \eqref{eq:barg-sch} is referred to as
\emph{Schrödinger realization}. For the example of the one-dimensional
harmonic oscillator, it may be interpreted as the realization in
position space.

We observe that by a naive counting the degrees of freedom in
$\mathscr{H}_{\text{B}}$ and $\mathscr{H}_{\text{Sch}}$ do match. A
function $\Psi(z)$ in $\mathscr{H}_{\text{B}}$ depends on $p$ complex
coordinates $z_{\alpha}$ but not on their complex conjugates
$\overline{z}_{\alpha}$. Similarly, $\Psi(x)$ in
$\mathscr{H}_{\text{Sch}}$ is a function of $p$ real coordinates
$x_{\alpha}$. Thus we want to identify the canonical variables in
$\mathscr{H}_{\text{B}}$ and $\mathscr{H}_{\text{Sch}}$. For this
purpose we make the ansatz
\begin{align}
  \label{eq:barg-rel}
  \partial_z\leftrightarrow A(x+\gamma\partial_x)\,,\quad
  z\leftrightarrow \overline{A}(x-\gamma\partial_x)\,,
\end{align}
where we allow for a constant $\gamma>0$ and a $p\times p$ matrix $A$,
whose complex conjugate we denote $\overline{A}$. The second relation
is obtained from the first one by taking the Hilbert space
adjoint. For \eqref{eq:barg-rel} to be compatible with the commutation
relations and reality conditions in \eqref{eq:barg-barg} and
\eqref{eq:barg-sch}, we have to impose
\begin{align}
  \label{eq:barg-rel-cond-simp}
  2\gamma\,AA^\dagger=1_p\,,
\end{align}
where ${}^\dagger$ stands for Hermitian conjugation of matrices. Note
that this condition can be solved trivially by taking $A\propto 1_p$,
in which case the components of the relations in \eqref{eq:barg-rel}
decouple. The identification \eqref{eq:barg-rel} of the Hilbert spaces
$\mathscr{H}_{\text{B}}$ and $\mathscr{H}_{\text{Sch}}$ is implemented
by the \emph{Bargmann transformation}
\begin{align}
  \label{eq:barg-trafo}
  \Psi(z)=\langle\overline{\mathcal{K}(z,x)},\Psi(x)\rangle_{\text{Sch}}\,,\quad
  \Psi(x)=\langle\mathcal{K}(z,x),\Psi(z)\rangle_{\text{B}}\,
\end{align}
with the kernel
\begin{align}
  \label{eq:barg-kernel}
  \mathcal{K}(z,x)=(\pi\gamma)^{-\frac{p}{4}}e^{-\gamma z^tAA^tz-\frac{1}{2\gamma}x^tx+2z^tAx}\,.
\end{align}
This kernel solves the differential equations obtained by imposing
\eqref{eq:barg-rel} on \eqref{eq:barg-trafo},
\begin{align}
  \partial_z\mathcal{K}(z,x)=A(x-\gamma\partial_x)\mathcal{K}(z,x)\,,\quad
  z \mathcal{K}(z,x)=\overline{A}(x+\gamma\partial_x)\mathcal{K}(z,x)\,.
\end{align}
The prefactor in \eqref{eq:barg-kernel} is fixed by demanding that the
transformation \eqref{eq:barg-trafo} preserves the unit norm of the
vacuum state
$\Psi_0(x)=(\pi \gamma)^{-\frac{p}{4}}e^{-\frac{1}{2\gamma}x^tx}$.
Equation \eqref{eq:barg-trafo} together with the definitions of the
inner products in \eqref{eq:barg-inner-barg} and
\eqref{eq:barg-inner-sch} yields the concrete form of the Bargmann
transformation as an integral transformation that we will use in the
following section.

We also briefly discuss a realization of the fermionic oscillator algebra
\begin{align}
  \label{eq:fermi-osc}
  \{\mathbf{c}_{a},\bar{\mathbf{c}}_b\}=\delta_{ab}\,,\quad
  {\mathbf{c}_a}^\dagger=\bar{\mathbf{c}}_a\,,\quad
  \mathbf{c}_a|0\rangle=0\,,
\end{align}
where $a,b=1,\ldots,r$. It is realized on a Graßmann algebra with $r$
variables $\eta_a$,
\begin{align}
  \label{eq:fermi-map-gr}
  \bar{\mathbf{c}}\mapsto \eta\,,\quad
  \mathbf{c}\mapsto\partial_{\eta}\,,\quad
  |0\rangle\mapsto 1\,.
\end{align}
On order to implement the adjoint in \eqref{eq:fermi-osc} as
${\eta_a}^\dagger=\partial_{\eta_a}$, one has to define an inner
product on functions $\Psi(\eta)$. This can be done in formal analogy
to \eqref{eq:barg-inner-barg} of the Bargmann realization.  The
integral is replaced by a Berezin integral, and instead of complex
conjugation one uses an antiinvolution of the Graßmann
algebra. Consequently, \eqref{eq:fermi-map-gr} can be understood as a
\emph{fermionic Bargmann realization}, see e.g.\
\cite{Takhtajan:2008,ZinnJustin:2005} for more details.

\subsection{Transformation of Generators and Integral}
\label{sec:trafo-int-gen}

Utilizing the tools from the previous section, in particular the
Bargmann transformation, we now identify the $\mathfrak{u}(p,q=p|r+s)$
representations in terms of spinor helicity variables from
section~\ref{sec:symmetry-generators} with those defined using
oscillators in section~\ref{sec:osc-rep}. This then allows us to
establish the equivalence of the two unitary Graßmannian integral
formulas \eqref{eq:grass-int-barg} and \eqref{eq:grass-int-unitary}
for these algebras. The identification of the representations goes
back to a calculation for the special case of the conformal algebra
$\mathfrak{su}(2,2)$ \cite{Stoyanov:1968tn}. In this case, it can be
associated with a transformation from a \emph{Lorentz} to a
\emph{maximally compact basis} of the algebra. The reason being that
the diagonal blocks of the matrices of spinor helicity
\eqref{eq:fermi-gen} and oscillator \eqref{eq:gen-ordinary} generators
are associated with the corresponding subalgebras
$\mathfrak{sl}(\mathbb{C}^2)$ and
$\mathfrak{su}(2)\times\mathfrak{su}(2)\times\mathfrak{u}(1)$,
respectively. In addition, the identification of representations is of
particular importance in the $\mathfrak{u}(2,2|4)$ case because it
provides a connection between the variables frequently used for the
$\mathcal{N}=4$ SYM amplitudes $\mathcal{A}_{N,K}$ and those featuring
in the spectral problem of this theory.

To begin with, we concentrate on the positive energy representations
$\oscrep_c$ of $\mathfrak{u}(p,p|r+s)$. Our aim is to relate the
generators $\mathbf{J}_{\indnm{AB}}$ from \eqref{eq:gen-ordinary} to
the $\mathfrak{J}_{\indnm{AB}}$ from \eqref{eq:fermi-gen}. In a first
step, we employ the bosonic \eqref{eq:barg-map-b} and the fermionic
\eqref{eq:fermi-map-gr} Bargmann realizations for the oscillators
appearing in the former set of generators. In a second step, we apply
the replacement \eqref{eq:barg-rel} to the bosonic variables, while
the fermionic ones remain unchanged. For the creation operators
\eqref{eq:osc-split}, this reads
\begin{align}
  \label{eq:trafo-osc}
  \left(\bar{\mathbf{A}}_{\indnm{A}}\right)=
  \left(
  \begin{array}{c}    
    \bar{\mathbf{a}}_{\alpha}\\[0.3em]
    \hdashline\\[-1.0em]
    \bar{\mathbf{c}}_{a}\\[0.3em]
    \hdashline\\[-1.0em]
    \bar{\mathbf{b}}_{\dot{\alpha}}\\[0.3em]
    \hdashline\\[-1.0em]
    \bar{\mathbf{d}}_{\dot{a}}\\
  \end{array}
  \right)
  \mapsto
  \left(
  \begin{array}{c}    
    z_{\alpha}\\[0.3em]
    \hdashline\\[-1.0em]
    \eta_{a}\\[0.3em]
    \hdashline\\[-1.0em]
    w_{\dot{\alpha}}\\[0.3em]
    \hdashline\\[-1.0em]
    \tilde\eta_{\dot{a}}\\
  \end{array}
  \right)
  \leftrightarrow
  \left(
  \begin{array}{c}    
    \frac{1}{\sqrt{2}}(\lambda_\alpha-\partial_{\overline{\lambda}_\alpha})\\[0.3em]
    \hdashline\\[-1.0em]
    \eta_{a}\\[0.3em]
    \hdashline\\[-1.0em]
    \frac{1}{\sqrt{2}}(\overline{\lambda}_\alpha-\partial_{\lambda_\alpha})\\[0.3em]
    \hdashline\\[-1.0em]
    \tilde\eta_{\dot{a}}\\
  \end{array}
  \right)\,.
\end{align}
Recall the \emph{differing} reality conditions of the superoscillators
in \eqref{eq:super-osc} compared to those of the spinor helicity
variables and fermions in \eqref{eq:vars-dagger}. They allow to derive
the transformation of the annihilation operators from
\eqref{eq:trafo-osc}. The replacement \eqref{eq:barg-rel} is
implemented by the Bargmann transformation \eqref{eq:barg-trafo}. The
latter is specified by
\begin{align}
  \label{eq:trafo-vars}
  \left(
  \begin{array}{c}
    z_{\alpha}\\[0.3em]
    \hdashline\\[-1.0em]
    w_{\dot{\alpha}}
  \end{array}
  \right)
  \in\mathbb{C}^{2p}\,,\quad
  \left(
  \begin{array}{c}
    x_{\alpha}\\[0.3em]
    \hdashline\\[-1.0em]
    y_{\dot{\alpha}}
  \end{array}
  \right)
  \in\mathbb{R}^{2p}\,,\quad
  A=
  \frac{1}{\sqrt{2}}\left(
  \begin{array}{c:c}
    1_p&-i 1_p\\[0.3em]
    \hdashline\\[-1.0em]
    1_p&i 1_p
  \end{array}
         \right)
         \,,\quad
         \gamma=\frac{1}{2}\,,
\end{align}
which are the arguments of holomorphic functions $\Psi(z,w)$ in the
Bargmann space $\mathscr{H}_{\text{B}}$, those of square integrable
functions $\Psi(x,y)$ in $\mathscr{H}_{\text{Sch}}$, and the
parameters of the transformation from \eqref{eq:barg-rel-cond-simp},
respectively. Importantly, to obtain spinor helicity variables, the
$x,y\in\mathbb{R}^p$ are packaged into
$\lambda=x+iy\in\mathbb{C}^p$. With these definitions, the Bargmann
transformation \eqref{eq:barg-trafo} from $\mathscr{H}_{\text{B}}$ to
$\mathscr{H}_{\text{Sch}}$ takes the form
\begin{align}
  \label{eq:barg-trafo-ordinary}
  \Psi(\lambda,\overline{\lambda})=
  \sqrt{\frac{2}{\pi}}^{\,p}
  e^{-\overline{\lambda}^t\lambda}
  \int\displaylimits_{\mathbb{C}^{2p}}
  \frac{\D^{\,p}\!\overline{z}\D^{\,p}\!z\D^{\,p}\!\overline{w}\D^{\,p}\!w}{(2\pi i)^{2p}}
  e^{-\overline{z}^t z
    -\overline{w}^t w
    -\overline{w}^t\overline{z}
    +\sqrt{2}(\overline{z}^t\lambda+\overline{w}^t\overline{\lambda})}
  \Psi(z,w)\,.
\end{align}
Notice that in this construction, the spinor helicity variables are a
generalization of the position operator in the Schrödinger realization
of the harmonic oscillator, cf.\ section~\ref{sec:bargmann}. Next, we
move on to the negative energy representations $\bar{\oscrep}_c$ of
$\mathfrak{u}(p,p|r+s)$, where we have to relate the generators
$\bar{\mathbf{J}}_{\indnm{AB}}$ from \eqref{eq:gen-dual} to the
$\bar{\mathfrak{J}}_{\indnm{AB}}$ from \eqref{eq:fermi-shift}. We
proceed mostly analogous to above. The required changes can be
summarized by replacements in the final transformation formulas
\eqref{eq:trafo-osc}, \eqref{eq:barg-trafo-ordinary}, and of the
fermionic vacuum state,
\begin{align}
  \label{eq:fermi-trafo-dual}
  \begin{gathered}
    \lambda\mapsto\overline{\lambda}\,,\quad
    \eta\mapsto-\partial_\eta\,,\quad
    \partial_\eta\mapsto-\eta\,,\quad
    \tilde\eta\mapsto\partial_{\tilde\eta}\,,\quad
    \partial_{\tilde\eta}\mapsto\tilde\eta\,,\quad
    1\mapsto\eta_1\cdots\eta_r\tilde\eta_1\cdots\tilde\eta_s\,.
  \end{gathered}
\end{align}
The replacements of the fermions here can be realized by fermionic
Fourier transformations.

Applying the transformation \eqref{eq:trafo-osc} in case of
the representations $\oscrep_c$ and together with
\eqref{eq:fermi-trafo-dual} for $\bar{\oscrep}_c$ yields the desired
relations of the generators,
\begin{align}
  \label{eq:trafo-gen}
  \begin{aligned}
    (\mathbf{J}_{\indnm{AB}})\leftrightarrow
    D
    (\mathfrak{J}_{\indnm{AB}})
    D^{-1}\,,\quad
    (\bar{\mathbf{J}}_{\indnm{AB}})\leftrightarrow
    D
    (\bar{\mathfrak{J}}_{\indnm{AB}})
    D^{-1}\,
  \end{aligned}
\end{align}
with the supermatrix
\begin{align}
  \label{eq:trafo-gen-sim}
  D=
  \left(
  \begin{array}{c:c:c:c}
    1_p&0&1_p&0\\[0.3em]
      \hdashline&&&\\[-1.0em]
    0&\sqrt{2}\,1_r&0&0\\[0.3em]
      \hdashline&&&\\[-1.0em]
    -1_p&0&1_p&0\\[0.3em]
      \hdashline&&&\\[-1.0em]
    0&0&0&\sqrt{2}\,1_s
  \end{array}\right)\,,
\end{align}
whose grading can be inferred from \eqref{eq:osc-split}. Consequently,
the relations between the central elements \eqref{eq:central-ord} and
\eqref{eq:fermi-replabel-pos} for the representations $\oscrep_c$
together with those between \eqref{eq:central-dual} and
\eqref{eq:fermi-replabel-neg} for $\bar{\oscrep}_c$ read, respectively,
\begin{align}
  \label{eq:fermi-replabel}
    \mathbf{C}
    \leftrightarrow
    \mathfrak{C}\,,\quad
    \bar{\mathbf{C}}
    \leftrightarrow
    \bar{\mathfrak{C}}\,.
\end{align}
We remark that if the $\mathfrak{gl}(n|m)$ generators from
\eqref{eq:trafo-gen} appear in Lax operators
\eqref{eq:yangian-def-lax}, the similarity transformation can be
absorbed by a redefinition of the generators $E_{\mathcal{AB}}$. Thus
we did not have to include such a similarity transformation in the
specification of the monodromy matrices in \eqref{eq:mono-osc-gen} or
\eqref{eq:mono-spinor-gen}.

Finally, we show the equivalence of the unitary Graßmannian integral
formula \eqref{eq:grass-int-unitary} for $|\Psi_{2K,K}\rangle$ and
that in \eqref{eq:grass-int-barg} for $\Psi_{2K,K}$ in case of the
algebras under consideration. This is achieved by applying the
Bargmann realization to the superoscillators
$\bar{\mathbf{A}}_{\indnm{A}}^i$ and the vacuum $|0\rangle$ in
$|\Psi_{2K,K}\rangle$, as specified by the first replacement in
\eqref{eq:trafo-osc}. Next, we perform the Bargmann transformation
\eqref{eq:barg-trafo-ordinary} at all sites $i=1,\ldots,2K$. In both
of these steps, we apply the replacement \eqref{eq:fermi-trafo-dual}
at the sites $i=1,\ldots,K$ with negative energy representations.  The
$U(K)$ integral and the function $\mathscr{F}(\mathcal{C})$ in the
Graßmannian formula \eqref{eq:grass-int-unitary} are unaffected by
these operations. The exponential function in the integrand is
transformed as
\begin{align}
  \label{eq:trafo-complete-integrand}
  (\det\mathcal{C})^re^{\tr(\mathbf{I}_\bullet \mathcal{C}^\dagger+\mathcal{C}\mathbf{I}_\circ^t)}|0\rangle
  \leftrightarrow
  \delta_{\mathbb{C}}^{pK|0}(C^\perp\boldsymbol{\lambda})  
  \delta^{0|rK}(C^\perp\boldsymbol{\eta})
  \delta^{0|sK}(C\boldsymbol{\tilde\eta})\,,
\end{align}
where we dropped an overall sign factor from rearranging the fermionic
variables. This shows the equivalence of the two unitary Graßmannian
integral formulas \eqref{eq:grass-int-unitary} and
\eqref{eq:grass-int-barg} for $\mathfrak{u}(p,p|r+s)$ Yangian
invariants. Details on this calculation are provided in
\cite{Kanning:2016eit}. Let us sketch some key aspects. We split each
of the matrices
$\mathbf{I}_{\mathrel{\ooalign{\raisebox{0.4ex}{$\scriptstyle\bullet$}\cr\raisebox{-0.4ex}{$\scriptstyle\circ$}}}}$,
recall the definition of their entries in
\eqref{eq:bullets-nc-recall}, into a sum of a matrix containing the
bosonic oscillators and one with the fermionic ones. This leads to a
factorization of the exponential function in
\eqref{eq:trafo-complete-integrand}. The Bargmann transformation of
the factor with the bosonic variables becomes a high-dimensional
Gaußian integral. Its zero modes yield Fourier representations of the
complex delta functions in \eqref{eq:trafo-complete-integrand}. The
series expansion of the exponential involving the fermionic variables
truncates, which allows us to rewrite this exponential as fermionic
delta functions.

\section{R-Matrix and Divergencies}
\label{sec:r-matrix}

Until now, we have not directly evaluated the unitary Graßmannian
integral \eqref{eq:grass-int-unitary} in the oscillator basis. We
perform such a computation here for the sample invariant
$|\Psi_{4,2}\rangle$ and establish its relation to the R-matrix. In
particular, we demonstrate that the R-matrix of the Heisenberg spin
chain and that of the planar $\mathcal{N}=4$ SYM one-loop spectral
problem are contained in our formalism. What is more, our
investigation of $|\Psi_{4,2}\rangle$ in the oscillator basis reveals
singular contributions to $\Psi_{4,2}$ in spinor helicity variables
that have been neglected so far.

\subsection{R-Matrices of Heisenberg Chain and Spectral Problem}
\label{sec:r-formula}

The evaluation of the unitary Graßmannian integral
\eqref{eq:grass-int-unitary} for the $\mathfrak{u}(p,q|r+s)$ Yangian
invariant $|\Psi_{4,2}\rangle$ parallels that in case of the bosonic
algebra $\mathfrak{u}(p,q)$ published in \cite{Kanning:2014cca}. To
begin with, we parameterize the $U(2)$ integration variable in
\eqref{eq:grass-int-unitary} as
\begin{align}
  \label{eq:para-u2}
    \mathcal{C}=
    \begin{pmatrix}
      e^{i(\gamma+\alpha)} \cos{\theta}&-e^{i\beta}\sin{\theta}\\
      e^{i(\gamma-\beta)}\sin{\theta}&e^{-i\alpha}\cos{\theta}\\
    \end{pmatrix}
\end{align}
with $\alpha,\beta,\gamma\in[0,2\pi]$ and
$\theta\in[0,\tfrac{\pi}{2}]$. The Haar measure
$[\D\mathcal{C}] =(2\pi)^{-3}\sin{(2\theta)}
\D\theta\D\alpha\D\beta\D\gamma$ can then be obtained from
\eqref{eq:measure-haar}. We rewrite the integrals in the variables
$\alpha$, $\beta$, and $\gamma$ as complex contour integrals around
unit circles and evaluate them using the residue theorem. The
remaining integral in $\theta$ then reduces to an integral
representation of the Euler beta function $\text{B}(x,y)$. This leads
to the invariant
\begin{align}
  \label{eq:gr-sample-inv42}
  \begin{aligned}
    |\Psi_{4,2}\rangle=
    \quad\quad
    \smash{\sum_{\mathclap{\substack{g_{13},\ldots,g_{24}=0\\h_{13},\ldots,h_{24}=0\\\text{with \eqref{eq:gr-sample-inv42-constraints}}}}}^\infty}\,\quad\quad
    &\frac{(1\bullet 3)^{g_{13}}}{g_{13}!}\frac{(1\bullet 4)^{g_{14}}}{g_{14}!}
    \frac{(2\bullet 3)^{g_{23}}}{g_{23}!}\frac{(2\bullet 4)^{g_{24}}}{g_{24}!}\\
    \cdot\,&\frac{(1\circ 3)^{h_{13}}}{h_{13}!}\frac{(1\circ 4)^{h_{14}}}{h_{14}!}
    \frac{(2\circ 3)^{h_{23}}}{h_{23}!}\frac{(2\circ 4)^{h_{24}}}{h_{24}!}
    |0\rangle\\
    \cdot\,&(-1)^{g_{14}+h_{14}}\text{B}(g_{14}+h_{23}+1,h_{13}+g_{24}-v_1+v_2)\,.
  \end{aligned}
\end{align}
In this formula the summation range is constrained by
\begin{align}
  \label{eq:gr-sample-inv42-constraints}
  \begin{aligned}
    g_{13}-h_{13}+g_{14}-h_{14}=g_{13}-h_{13}+g_{23}-h_{23}&=-c_1+q-s\,,\\
    g_{23}-h_{23}+g_{24}-h_{24}=g_{14}-h_{14}+g_{24}-h_{24}&=-c_2+q-s\,.
  \end{aligned}
\end{align}
These constraints assure that the eigenvalues of
$\bar{\mathbf{C}}^1,\bar{\mathbf{C}}^2,\mathbf{C}^3,\mathbf{C}^4$
acting on the invariant are, respectively, $c_1,c_2,-c_1,-c_2$.
Furthermore, we have to assume $\Real(v_1-v_2)<0$ in order for the
beta function integral to converge. In our final formula
\eqref{eq:gr-sample-inv42}, we dropped a numerical prefactor.

The Yangian invariant $|\Psi_{4,2}\rangle$ is of key importance
because its Yangian invariance condition \eqref{eq:yi-exp-1} is
equivalent to a Yang-Baxter equation. This was originally pointed out
in \cite{Ferro:2013dga} in the context of the deformed amplitude
$\mathcal{A}_{4,2}^{(\text{def\/})}$, which we encountered in the
review section~\ref{sec:grassdef}. Because the Yang-Baxter equation is
at the core of the QISM, the equivalence is most naturally established
using the Yangian invariance condition \eqref{eq:yi} in this language,
see e.g.\ the discussion in \cite{Kanning:2016eit}. As a consequence,
$|\Psi_{4,2}\rangle$ can be identified with an R-matrix, and thus we
represent it graphically as
\begin{align}
  \label{eq:r-mat-graph}
  \begin{aligned}
    \begin{tikzpicture}
      \draw[thick,rounded corners=10pt]
      (0.25,0) node[below]{$1$} -- (0.25,0.5);
      \draw[thick,rounded corners=10pt]
      (1.25,0.5) -- (1.25,0) node[below]{$3$};
      \draw[thick,rounded corners=10pt]
      (0.75,0) node[below]{$2$} -- (0.75,0.5);
      \draw[thick,rounded corners=10pt]
      (1.75,0.5) -- (1.75,0) node[below]{$4$};
      \draw (1,1) 
      node[minimum height=1cm,minimum width=2cm,draw,
      thick,rounded corners=8pt,densely dotted] 
      {$|\Psi_{4,2}\rangle$};
    \end{tikzpicture}
  \end{aligned}
  \quad
  =
  \quad
  \begin{aligned}
    \begin{tikzpicture}
      \draw[thick,rounded corners=10pt]
      (0.25,0) node[below]{$1$} -- (0.25,1.25) -- (1.25,1.25) -- (1.25,0) node[below]{$3$};
      \draw[thick,rounded corners=10pt]
      (0.75,0) node[below]{$2$} -- (0.75,1) -- (1.75,1) -- (1.75,0) node[below]{$4$};
      \draw (1,1) 
      node[minimum height=1cm,minimum width=2cm,
      thick,rounded corners=8pt,densely dotted] 
      {$$};
    \end{tikzpicture}
  \end{aligned}
  \,.
  \quad
\end{align}
Let us fill in the details of this identification. The invariant
$|\Psi_{4,2}\rangle$ from \eqref{eq:gr-sample-inv42} is a state in
$\bar{\oscrep}_{c_1}\otimes\bar{\oscrep}_{c_2}\otimes\oscrep_{-c_1}\otimes\oscrep_{-c_2}$. The
representations $\bar{\oscrep}_{c_i}$ and
$\oscrep_{-c_i}$ are dual to each other and therefore of the same
dimension. Thus computing
$|\Psi_{4,2}\rangle^{\dagger_1\dagger_2}$, where the conjugations act
only on the respective oscillators, and then identifying the site
indices
$i,j$ of oscillators in the contractions
$(i\mathrel{\ooalign{\raisebox{0.7ex}{$\bullet$}\cr\raisebox{-0.3ex}{$\circ$}}}
j)$ as $1\equiv 3$ and $2\equiv
4$ results in an operator on
$\oscrep_{-c_1}\otimes\oscrep_{-c_2}$. This is the R-matrix
$R_{\oscrep_{-c_1}\oscrep_{-c_2}}(z)$ with the spectral parameter
$z\equiv v_1-v_2$ and subscripts referring to
$\mathfrak{u}(p,q|r+s)$ representations. Our Graßmannian integral
\eqref{eq:grass-int-unitary} yields a novel
$U(2)$ integral formula for this R-matrix with oscillator
representations.

It is instructive to execute this construction of the R-matrix in the
simplest case conceivable, i.e.\ for the compact bosonic algebra
$\mathfrak{u}(2,0)\cong\mathfrak{u}(2)$ with
$c_1=c_2=-1$. Here the intricate sums in \eqref{eq:gr-sample-inv42}
reduce to just two terms,
\begin{align}
  \label{eq:gr-sample-inv42-u2}
  |\Psi_{4,2}\rangle
  =
  \frac{\big[z(1\bullet 3)(2\bullet 4)+(1\bullet 4)(2\bullet 3)\big]|0\rangle}{z(1-z)}\,.
\end{align}
The corresponding R-matrix acts on two copies of the space
$\oscrep_1=\mathbb{C}^2$. Each copy is spanned by two states
$\bar{\mathbf{a}}_{1}^i|0\rangle$ and
$\bar{\mathbf{a}}_{2}^i|0\rangle$, which are built from the creation
operators \eqref{eq:osc-split} contained in the oscillator
contractions. These states are realized in the following as $\bigl(
\begin{smallmatrix}
  1\\
  0\\
\end{smallmatrix}\bigr)$ and $\bigl(
\begin{smallmatrix}
  0\\
  1\\
\end{smallmatrix}\bigr)$, respectively. This gives rise to 
\begin{align}
  \label{eq:gr-sample-r-u2}
  R_{\oscrep_1\oscrep_1}(z)
  =
  \frac{1}{z(1-z)}
  \begin{pmatrix}
    1+z&0&0&0\\
    0&z&1&0\\
    0&1&z&0\\
    0&0&0&1+z\\
  \end{pmatrix}\,,
\end{align}
which is the R-matrix of the celebrated \emph{Heisenberg spin chain},
as reviewed e.g.\ in \cite{Faddeev:1996iy}. This connection
illustrates the relevance of the unitary Graßmannian integral approach
for integrable spin chain models with their vast associated
literature. Next, we return to the general setting with the
non-compact superalgebra
$\mathfrak{u}(p,q|r+s)$. A formula for the R-matrix corresponding to
the Yangian invariant
$|\Psi_{4,2}\rangle$ in \eqref{eq:gr-sample-inv42} was already worked
out in \cite{Ferro:2013dga}. For the algebra
$\mathfrak{u}(2,2|4)$ it is essentially\footnote{The oscillators in
  \cite{Beisert:2003jj} satisfy non-standard reality conditions. This
  difference compared to our conventions in section~\ref{sec:osc-rep},
  which follow \cite{Bars:1982ep,Gunaydin:1984fk}, does not seem to
  affect the
  $\mathfrak{gl}(4|4)$ invariant R-matrix.} the R-matrix of the spin
chain governing the planar
$\mathcal{N}=4$ SYM \emph{one-loop spectral problem}
\cite{Beisert:2003yb,Beisert:2003jj}. In view of this result, the
Bargmann transformation from section~\ref{sec:trafo-int-gen} is an
explicit change of basis from the oscillator R-matrix of the spectral
problem to the Yangian invariant
$\Psi_{4,2}$ in spinor helicity variables, which appears to be a
deformation of the amplitude
$\mathcal{A}_{4,2}$, recall section~\ref{sec:psi42u224}.

The $U(2)$ integral formula for the $\mathfrak{u}(p,q|r+s)$ R-matrix
obtained from the Graßmannian integral~\eqref{eq:grass-int-unitary},
as explained in this section, suggests several generalizations that we
deem worthy to be explored. One might wonder whether the formula can
be extended beyond the class of oscillator representations of
section~\ref{sec:osc-rep}. Even further, it might exist on the
algebraic level of the universal R-matrix, see e.g.~\cite{Boos:2010ss}
and the original references therein. This would be a path to elucidate
a possible quantum group origin of the Graßmannian integral. A
different direction would be to extend our $U(2)$ integral formula to
Beisert's R-matrix \cite{Beisert:2006qh} with centrally extended
$\mathfrak{su}(2|2)$ symmetry. Two copies of this solution to the
Yang-Baxter equation describe the asymptotic scattering of excitations
in the all-loop spectral problem of planar $\mathcal{N}=4$
SYM. Moreover, it generalizes Shastry's R-matrix \cite{Shastry:1986}
for the one-dimensional Hubbard model. This line of research could be
a stepping stone to a unitary Graßmannian integral for Yangian
invariants associated with all-loop amplitudes.

\subsection{Normalization and Divergent Terms}
\label{sec:norm-div}

Here we inspect the normalization of the $\mathfrak{u}(p,q|r+s)$
Yangian invariant $|\Psi_{4,2}\rangle$ in \eqref{eq:gr-sample-inv42},
that arises naturally from the unitary Graßmannian integral
\eqref{eq:grass-int-unitary} in oscillator variables. In the limit
$z\equiv v_1-v_2\to 0$, we will observe a divergent $\frac{1}{z}$
contribution. In fact, we recognize such a term immediately in the
simple $\mathfrak{u}(2)$ case of $|\Psi_{4,2}\rangle$ displayed in
\eqref{eq:gr-sample-inv42-u2}, see also the formula
\eqref{eq:gr-sample-r-u2} of the corresponding R-matrix. This
observation seems to \emph{clash} with the undeformed limit, $z\to 0$
and $c_i=0$, of $\Psi_{4,2}$ for $\mathfrak{u}(2,2|4)$ in spinor
helicity variables discussed in section~\ref{sec:psi42u224}. There we
did not encounter a divergent term but straightforwardly obtained the
finite amplitude $\mathcal{A}_{4,2}$ in the limit. To resolve this
apparent contradiction, we will diligently reexamine the evaluation of
the Graßmannian integral \eqref{eq:grass-int-barg} for
$\Psi_{4,2}$. This will reveal a $\frac{1}{z}$ term that is only
present for special kinematic configurations and is thereby easily
overlooked. The existence of this term substantially affects the
conceptual relation between the Yangian invariant $\Psi_{4,2}$ and the
amplitude $\mathcal{A}_{4,2}$.

Let us consider $|\Psi_{4,2}\rangle$ for $\mathfrak{u}(p,q|r+s)$ with
equal representation labels $c_1=c_2$, which is given in form of
convoluted sums in \eqref{eq:gr-sample-inv42}. To extract the leading
term as $z\to 0$, we expand the Euler beta functions in the
summands. We find it to be a divergent $\frac{1}{z}$ contribution,
\begin{align}
  \label{eq:norm-osc-div}
  |\Psi_{4,2}\rangle=
  \frac{1}{z}
  |\Upsilon\rangle_{14}|\Upsilon\rangle_{23}
  +\mathcal{O}(z^0)
  \quad\text{with}\quad
    |\Upsilon\rangle_{ij}
  =\quad\sum_{\mathclap{\substack{g,h=0\\g-h=q-s-c_i}}}^\infty\;
  \frac{(i\bullet j)^{g}}{g!}\frac{(i\circ j)^{h}}{h!}|0\rangle\,,
\end{align}
where we dropped an overall sign factor compared to
\eqref{eq:gr-sample-inv42}. This result is conveniently expressed in
terms of the state
$|\Upsilon\rangle_{ij}\in\bar\oscrep_{c_i}\otimes\oscrep_{c_j=-c_i}$. Computing
$|\Upsilon\rangle_{ij}^{\dagger_i}$ and then identifying the site
indices of oscillators from the contractions
$(i\mathrel{\ooalign{\raisebox{0.7ex}{$\bullet$}\cr\raisebox{-0.3ex}{$\circ$}}}
j)$ as $i\equiv j$ gives the identity operator on $\oscrep_{-c_i}$,
see \cite{Kanning:2016eit}. Moreover, $|\Upsilon\rangle_{ij}$ is
identical to the Yangian invariant $|\Psi_{2,1}\rangle$ defined by the
Graßmannian integral \eqref{eq:grass-int-unitary} with some relabeling
of the oscillators. Its expression as infinite sums in
\eqref{eq:norm-osc-div} matches the series expansion of a modified
Bessel function of the first kind. This Bessel function is the
simplest $1\times 1$ case of the determinant formula for
$|\Psi_{2K,K}\rangle$ with the special Graßmannian integrand
\eqref{eq:ls-model} mentioned towards the end of
section~\ref{sec:int-osc}. Eventually, we can infer from
\eqref{eq:norm-osc-div} the normalization of the R-matrix
$R_{\oscrep_{-c_1}\oscrep_{-c_1}}(z)$, which is constructed from
$|\Psi_{4,2}\rangle$ as described in
section~\ref{sec:r-formula}. Often an R-matrix is normalized such
that, for vanishing spectral parameter $z=0$, it is finite and reduces
to a permutation operator, see e.g.\ the textbook
\cite{Gomez:2005}. In contrast, $R_{\oscrep_{-c_1}\oscrep_{-c_1}}(z)$
diverges for $z\to 0$ as $\frac{1}{z}$, and the coefficient of this
divergence is the permutation operator\footnote{The identification of
  oscillator indices below \eqref{eq:r-mat-graph}, which is needed for
  the R-matrix, differs from that mentioned after
  \eqref{eq:norm-osc-div}. Thus we obtain the permutation operator and
  not the identity.} on $\oscrep_{-c_1}\otimes\oscrep_{-c_1}$. This
can be observed very explicitly for the simple $\mathfrak{u}(2)$
R-matrix $R_{\oscrep_{1}\oscrep_{1}}(z)$ in \eqref{eq:gr-sample-r-u2},
where the permutation operator on
$\mathbb{C}^2\otimes\mathbb{C}^2\cong\mathbb{C}^4$ is just a
$4\times 4$ matrix. In conclusion, the unitary Graßmannian integral
\eqref{eq:grass-int-unitary} for $|\Psi_{4,2}\rangle$ leads to an
uncommon normalization of the corresponding R-matrix.

The unitary Graßmannian integrals \eqref{eq:grass-int-unitary} for
$|\Psi_{4,2}\rangle$ in oscillator variables and
\eqref{eq:grass-int-barg} for $\Psi_{4,2}$ in spinor helicity
variables are related by a change of basis, which is implemented by
the Bargmann transformation of
section~\ref{sec:trafo-int-gen}. Therefore the $\frac{1}{z}$ term has
also to be present for $\Psi_{4,2}$. Before extracting it from the
Graßmannian integral \eqref{eq:grass-int-barg}, it is helpful to
recapitulate a similar but simpler calculation. Also the gamma
function $\Gamma(z)$ diverges for $z\to 0$ as $\frac{1}{z}$. This can
be shown using its Euler integral representation valid for
$\Real z>0$, see e.g.\ \cite{Miller:2006},
\begin{align}
  \label{eq:norm-gamma}
  \Gamma(z)
  =\int\displaylimits_0^\infty \D\tau\, e^{-\tau}\tau^{z-1}
  =\frac{1}{z}\int\displaylimits_0^\infty\D\tau\,e^{-\tau}\frac{\D}{\D\tau}\tau^{z}
  =\frac{1}{z}+\mathcal{O}(z^0)\,.
\end{align}
The first integrand can be expanded into non-negative powers of $z$
for $z\to 0$. However, because of the singularity at $\tau=0$, this
expansion does not commute with the integration. Thus we rewrite the
gamma function as the second integral in \eqref{eq:norm-gamma}. After
performing a partial integration, we are left with an integrand that
is regular in $\tau$. Its expansion in $z$ can be integrated term by
term and yields the rightmost side of \eqref{eq:norm-gamma}.

We apply a similar strategy to the $\mathfrak{u}(2,2|0+4)$ Yangian
invariant $\Psi_{4,2}$ with $c_1=c_2$ defined by the unitary
Graßmannian integral \eqref{eq:grass-int-barg}. We parameterize the
$U(2)$ integration variable $\mathcal{C}$ as in \eqref{eq:para-u2}.
To extract the divergent term, we concentrate on the integral in
$\theta$ because the integrand
\eqref{eq:eq:grass-int-unitary-integrand-final} of the Graßmannian
formula has only singularities in this variable. Proceeding
essentially along the lines of \eqref{eq:norm-gamma} for this integral
and assuming\footnote{This is consistent as the same assumption was
  required to evaluate the oscillator invariant $|\Psi_{4,2}\rangle$
  in \eqref{eq:gr-sample-inv42}.}  $\Real z<0$ to eliminate certain
boundary terms, we isolate the leading $\frac{1}{z}$
contribution. After performing the remaining trivial integrals in
$\alpha$, $\beta$, and $\gamma$, we are left with
\begin{align}
  \label{eq:norm-spin-div}
  \Psi_{4,2}=
  \frac{1}{z}\Upsilon_{14}\Upsilon_{23}
  +\mathcal{O}(z^0)
  \quad\text{with}\quad
  \Upsilon_{ij}=
  \left(\frac{\lambda^i_1}{\lambda^j_1}\right)^{\!\!c_i}\!\!\!
  \frac{\overline{\lambda^i_1}}{(\lambda^j_1)^2\lambda^i_1}
  \delta^3(p^i+p^j)\delta^{0|4}(\lambda^i_1\tilde\eta^i+\lambda^j_1\tilde\eta^j)\,,
\end{align}
where we neglected a numerical prefactor. The three-dimensional delta
function $\delta^3(P)=\delta(P_{11})\delta_{\mathbb{C}}(P_{21})$
suffices because it implies $P_{22}=0$ for $P$ being the sum of two
massless momenta. Clearly, the decisive property of the divergent
$\frac{1}{z}$ term in \eqref{eq:norm-spin-div} is its absence for
\emph{generic} momentum configurations with $p^1+p^2+p^3+p^4=0$ as it
only contributes for \emph{special} kinematics with $p^1+p^4=0$ and
$p^2+p^3=0$. Thereby it is easily overlooked in the spinor helicity
basis.

What are the implications of the divergent term derived in the present
section? The evaluation of the unitary Graßmannian integral
\eqref{eq:grass-int-barg} for the $\mathfrak{u}(2,2|0+4)$ Yangian
invariant $\Psi_{4,2}$ in section~\ref{sec:psi42u224} resulted in the
expression \eqref{eq:inv42-u2204-final}. This reduces to the amplitude
$\mathcal{A}_{4,2}$ in the undeformed limit $z\to 0$ and $c_i=0$. Our
careful reexamination of the integral here revealed that
\eqref{eq:inv42-u2204-final} has to be supplemented by the
$\frac{1}{z}$ term in \eqref{eq:norm-spin-div}. Thus, strictly
speaking, the undeformed limit of $\Psi_{4,2}$ is divergent and
$\mathcal{A}_{4,2}$ occurs as the coefficient of $z^0$ in an expansion
for small $z$. This role of the amplitude is very similar to that in
\cite{Zwiebel:2011bx}, to which we referred to already at the very end
of the review section~\ref{sec:grassdef}. In that reference,
$\mathcal{A}_{4,2}$ is constructed from the one-loop dilatation
generator of the planar $\mathcal{N}=4$ SYM spectral problem, rather
than directly from the Yangian invariant R-matrix of the one-loop spin
chain. This interpretation of the amplitude $\mathcal{A}_{4,2}$ does
not mean that it is not ``integrable'' but only changes its conceptual
place within the integrable structure.

What can we learn from the $\frac{1}{z}$ divergence of $\Psi_{4,2}$
about the general $\mathfrak{u}(2,2|0+4)$ Yangian invariant
$\Psi_{2K,K}$ defined by the unitary Graßmannian integral
\eqref{eq:grass-int-barg}? Iterating the recursion
\eqref{eq:glue-graphical} and switching from the oscillator basis
there to spinor helicity variables, we find that $\Psi_{2K,K}$ can be
constructed by gluing together $\frac{1}{2}K(K-1)$ copies of
$\Psi_{4,2}$. Focusing on $\Psi_{2K,K}$ with equal parameters $c_i$
and also equal $z_i\equiv v_i-v_{i+1}$ denoted $z$, the complex
deformation parameter of each $\Psi_{4,2}$ turns out to be a multiple
of $z$. Thus the leading term of $\Psi_{2K,K}$ as $z\to 0$ is of order
$z^{-\frac{1}{2}K(K-1)}$. In our crude evaluation of the Graßmannian
formula for $\Psi_{6,3}$ in section~\ref{sec:psi63u224}, we did not
observe any divergent terms. Hence the terms of orders
$\frac{1}{z^3}$, $\frac{1}{z^2}$, and $\frac{1}{z}$ have to be
restricted to special kinematic configurations. Consequently, we could
naively obtain the amplitude $\mathcal{A}_{6,3}$ in one kinematic
region simply by taking the undeformed limit of $\Psi_{6,3}$. Our
discussion of the Yangian invariant $\Psi_{8,4}$ in
section~\ref{sec:psi84u224} implies from the current vantage
perspective that there exist terms of orders $\frac{1}{z^6}$ through
$\frac{1}{z^3}$ for special kinematics. At $\frac{1}{z^2}$ and
$\frac{1}{z}$ there are contributions for generic kinematics, which
cause even a naive undeformed limit of $\Psi_{8,4}$ to
diverge. Knowing about the divergent terms of $\Psi_{4,2}$ and
$\Psi_{6,3}$ for special kinematic configurations makes the sudden
appearance of divergencies for generic kinematics in case of
$\Psi_{8,4}$ less surprising. Notwithstanding these new insights, it
remains both a challenging and imperative task to identify the
amplitude $\mathcal{A}_{8,4}$ in the small-$z$ expansion of the
Yangian invariant $\Psi_{8,4}$.

\section{Link to Cusp Equation}
\label{sec:cusp-equation}

Until this point, we meticulously explored the mathematical
consequences of the unitary Graßmannian integral formulas
\eqref{eq:grass-int-barg} and \eqref{eq:grass-int-unitary}. To catch a
glimpse of where this research might be headed, we proceed less
stringently in this section. This will reveal an intriguing connection
to the equation \cite{Eden:2006rx,Beisert:2006ez} that is believed to
govern the so-called \emph{cusp anomalous dimension} in planar
$\mathcal{N}=4$ SYM. This function of the 't Hooft coupling can be
extracted from the anomalous dimensions of leading-twist operators. It
is an integral part of the conjectured \emph{all-loop} expression for
MHV amplitudes \cite{Bern:2005iz}, which is known to receive
corrections starting at six particles \cite{Bern:2008ap}. Therefore
our systematic approach to Yangian invariants appears to contain
already hints of all-loop results.

To expose them, we consider the Yangian invariant $|\Psi_{4,2}\rangle$
in the oscillator basis given by the Graßmannian integral
\eqref{eq:grass-int-unitary} with a $U(2)$ contour. Specializing the
deformation parameters $v_i^\pm$ such that its integrand
$\mathscr{F}(\mathcal{C})$ is of the form \eqref{eq:ls-model} reduces
the integral to the Leutwyler-Smilga model \cite{Leutwyler:1992yt}.
Ordinarily, this integral for $|\Psi_{4,2}\rangle$ contains the
$2\times 2$ matrices
$\mathbf{I}_{\mathrel{\ooalign{\raisebox{0.4ex}{$\scriptstyle\bullet$}\cr\raisebox{-0.4ex}{$\scriptstyle\circ$}}}}$,
whose entries are built out of oscillator variables associated with
$\mathfrak{u}(p,q|m)$ representations, and a Fock vacuum
$|0\rangle$. In what follows, we neglect this oscillator structure ad
hoc, and we even impose constraints on the now numerical entries of
the matrices
$\mathbf{I}_{\mathrel{\ooalign{\raisebox{0.4ex}{$\scriptstyle\bullet$}\cr\raisebox{-0.4ex}{$\scriptstyle\circ$}}}}$. This
results in a family of integrals indexed by an integer parameter,
\begin{align}
  \label{eq:cusp-u2int}
  \hat K_\nu(t,t')=\frac{1}{2}
  \int\displaylimits_{U(2)}\!\![\D\mathcal{C}]\,
  (\det\mathcal{C})^\nu
  e^{\tr(\mathbf{I}_\bullet\mathcal{C}^\dagger+\mathcal{C}\mathbf{I}_\circ^t)}
  \quad
  \text{with}
  \quad
  \mathbf{I}_\bullet=-\mathbf{I}_\circ=\frac{1}{2}
  \begin{pmatrix}
    t&0\\
    0&t'\\
  \end{pmatrix}\,.
\end{align}
These integrals can be computed with the determinant formula
\cite{Schlittgen:2002tj} mentioned previously in
section~\ref{sec:int-osc}. In particular,
\begin{align}
  \label{eq:cusp-eval}
  \hat K_0(t,t')=\frac{t J_1(t)J_0(t')-t'J_0(t)J_1(t')}{t^2-t'^2}\,,\quad
  \hat K_1(t,t')=\frac{t'J_1(t)J_0(t')-tJ_0(t)J_1(t')}{t^2-t'^2}\,,
\end{align}
where $J_\nu(t)$ denotes the standard Bessel function.

Excitingly, precisely these two functions make up the kernel of the
cusp equation in the planar $\mathcal{N}=4$ theory. The cusp anomalous
dimension $f(g)$ depends on the 't Hooft coupling constant $\lambda$
via $g^2=\frac{\lambda}{16\pi^2}$. It is given as
$f(g)=16 g^2\hat\sigma(0)$ with the fluctuation density
$\hat\sigma(t)$ obeying the integral equation
\cite{Eden:2006rx,Beisert:2006ez}
\begin{align}
  \label{eq:cusp-eq}
  \hat\sigma(t)=
  \frac{t}{e^t-1}
  \Biggr[\hat K(2gt,0)-4g^2\int\displaylimits_0^\infty\D t'\,\hat K(2gt,2gt')\hat\sigma(t')\Biggr]\,.
\end{align}
The complete kernel for this equation, which incorporates effects from
the asymptotic Bethe ansatz and the dressing phase, was found in
\cite{Beisert:2006ez},
\begin{align}
  \label{eq:cusp-kernel}
  \hat K(t,t')
  =
  \hat K_0(t,t')+\hat K_1(t,t')
  +8g^2\int\displaylimits_0^\infty\D t''\,\hat K_1(t,2gt'')\frac{t''}{e^{t''}-1}\hat K_0(2gt'',t')\,.
\end{align}
Thus we established a link from the $U(2)$ integral expression
\eqref{eq:grass-int-unitary} for the Yangian invariant
$|\Psi_{4,2}\rangle$ to the building blocks \eqref{eq:cusp-eval} of
the cusp kernel. It is on a heuristic level at present. Presumably,
details of the $\mathfrak{u}(2,2|4)$ representations have to be
injected to overcome, for instance, the ad hoc choice of the matrices
in \eqref{eq:cusp-u2int}. Still, we are intrigued by these seeds of
all-loop results in our unitary Graßmannian integral approach to the
construction of tree-level amplitudes.

\section{Conclusions and Outlook}
\label{sec:conclusions}

We began this article by recalling aspects of our and others' earlier
work on the systematic construction of Yangian invariants from
integrability in section~\ref{sec:review}. This allows, in
contradistinction to just ``observing'' it, to take Yangian symmetry
as a starting point, to then employ the quantum inverse scattering
method, and to finally systematically construct such invariants.
Applying this methodology in section~\ref{sec:unitary-integral} for
the general symmetry algebra $\mathfrak{u}(p,p|m)$, while paying close
attention to its correct reality conditions, we end up, in the special
case of $N=2K$ points and superhelicity $K$, with a family of unitary
Graßmannian ``contour'' integrals. The construction requires a certain
deformation of the naive integrands by extra parameters.

Evaluating the deformed integrals in case of the superconformal
algebra $\mathfrak{psu}(2,2|4)$ for $N=4,6,8$ points, attempting to
remove the deformation parameters, and comparing to the corresponding,
known, physical tree-level amplitudes, one, surprisingly, finds
differences that become more and more pronounced as $N$ increases,
cf.\ section~\ref{sec:sample-invariants}. For $N=4$ at generic momenta
one reproduces the four-particle MHV amplitude. However, even here a
closer look unveils the existence of certain contact terms at
collinear momenta that diverge upon taking the deformation parameters
to zero, see section~\ref{sec:norm-div} and in particular
\eqref{eq:norm-spin-div}. For $N=6$ at generic momenta the physical
six-particle NMHV amplitude is only reproduced in one out of four
kinematical sectors, while terms divergent in the deformation
parameters again appear at various collinear momentum
configurations. An analogous toy integral for the case $N=4$ and the
algebra $\mathfrak{u}(1,1)$ reproduces the proper R-matrix of this
simpler algebra. This forces us to conclude that the physical
six-particle NMHV amplitude in $\mathcal{N}=4$ SYM is not ``as
integrable as hoped for''. These tensions between the Yangian
invariant and the physical amplitude worsen for $N=8$. Divergences in
the deformation parameters are now present even at generic
momenta. Furthermore, the residues of these terms apparently no longer
combine to the physical eight-point N$^2$MHV amplitude in any
kinematical sector \cite{Kanning:2018a}.

What does this mean, especially in the light of the fact that there
have been \emph{local} proofs of Yangian invariance of the physical
amplitudes in the planar $\mathcal{N}=4$ model, using differential
operators representing the algebra? We suspect that Yangian invariance
of the perturbative physical amplitudes is nevertheless subtly
broken. This includes, but significantly goes beyond the effects
already observed in
\cite{Bargheer:2009qu,Sever:2009aa,Bargheer:2011mm}. One explanation
would be that the physical amplitudes are only infinitesimally
invariant, and that Yangian invariance (meaning either conformal or
dual-conformal symmetry, or both) is broken by \emph{large}
transformations.  We are currently investigating this possibility
\cite{Kanning:2018}. On the positive side, we found some evidence for
Yangian invariance at all-loop level, cf.\ section
\ref{sec:cusp-equation}. Could it be that the detected problems with
the Yangian invariance of the tree-level scattering amplitudes of
$\mathcal{N}=4$ SYM may be elegantly resolved at the non-perturbative,
all-loop level?

\section*{Acknowledgments}
\label{sec:acknowledgements}

We thank Jacob Bourjaily and Gregor Richter for earlier collaboration
and help on several questions related to this work. We are thankful to
Lance Dixon, Harald Dorn, Livia Ferro, Andrew Hodges, and Lionel Mason
for valuable discussions. This work was funded in part by the Deutsche
Forschungsgemeinschaft (DFG, German Research Foundation) under project
number 270039613.

\appendix

\section{Parity Symmetry}
\label{sec:parity}

In this appendix, we investigate a discrete symmetry transformation of
the unitary Graßmannian integrals \eqref{eq:grass-int-barg} for
$\Psi_{2K,K}$ in spinor helicity variables and
\eqref{eq:grass-int-unitary} for $|\Psi_{2K,K}\rangle$ in oscillator
variables. We define this so-called \emph{parity transformation}
$\mathcal{P}$ by reversing the order of the particles $i=1,\ldots,K$
with negative energies and also that of $i=K+1,\ldots,2K$ with
positive energies. On the level of the two Graßmannian integral
formulas, this transformation acts as, respectively,
\begin{align}
  \label{eq:parity-contr}
  \begin{gathered}
    \boldsymbol{\lambda}^\pm
    \stackrel{\mathcal{P}}{\mapsto}
    \mathcal{E}\boldsymbol{\lambda}^\pm\,,\quad
    \boldsymbol{\eta}^\pm
    \stackrel{\mathcal{P}}{\mapsto}
    \mathcal{E}\boldsymbol{\eta}^\pm\,,\quad
    \boldsymbol{\tilde\eta}^\pm
    \stackrel{\mathcal{P}}{\mapsto}
    \mathcal{E}\boldsymbol{\tilde\eta}^\pm\,,\\
    \text{and}\quad
    \mathbf{I}_{\mathrel{\ooalign{\raisebox{0.4ex}{$\scriptstyle\bullet$}\cr\raisebox{-0.4ex}{$\scriptstyle\circ$}}}}
    \stackrel{\mathcal{P}}{\mapsto}
    \mathcal{E}\,\mathbf{I}_{\mathrel{\ooalign{\raisebox{0.4ex}{$\scriptstyle\bullet$}\cr\raisebox{-0.4ex}{$\scriptstyle\circ$}}}}\,\mathcal{E}\,
  \end{gathered}
  \quad\text{with}\quad
  \mathcal{E}=
  \begin{pmatrix}
    0&\cdots&0&1\\
    \vdots&\iddots&1&0\\
    0&\iddots&\iddots&\vdots\\
    1&0&\cdots&0\\
  \end{pmatrix}\in U(K)\,.
\end{align}
Here $\boldsymbol{\eta}^\pm$ are $K\times r$ blocks and
$\boldsymbol{\tilde\eta}^\pm$ are $K\times s$ blocks of fermionic variables,
which are defined analogously to the blocks $\boldsymbol{\lambda}^\pm$
of $\boldsymbol{\lambda}$ in \eqref{eq:spinors-partition}. The
transformation $\mathcal{P}$ is a symmetry of both Graßmannian
integral formulas if the deformation parameters satisfy
\begin{align}
  \label{eq:parity-constr-special}
  v_1-v_2=v_2-v_3=\cdots=v_{K-1}-v_{K}\,,\quad
  c_1=c_2=\cdots= c_K\,.
\end{align}
This can be proven using the left- and right-invariance of Haar
measure and the identity
$[1,2,\ldots,j]_{\mathcal{E}\mathcal{C}\mathcal{E}}=\overline{[1,2,\ldots,K-j]}_{\mathcal{C}}\det\mathcal{C}$
for the minors appearing in the manifestly single-valued Graßmannian
integrand in \eqref{eq:eq:grass-int-unitary-integrand-final}.

We focus on the action of $\mathcal{P}$ on the Yangian invariant
$|\Psi_{4,2}\rangle$ given by \eqref{eq:grass-int-unitary}. In this
case \eqref{eq:parity-constr-special} yields no constraints for the
parameters $v_1,v_2\in\mathbb{C}$ and imposes equal representation
labels $c_1=c_2$. In section~\ref{sec:r-matrix}, we argued that
$|\Psi_{4,2}\rangle$ can be understood as an R-matrix acting on the
tensor product $\oscrep_{-c_1}\otimes\oscrep_{-c_1}$ of two
$\mathfrak{u}(p,q|r+s)$ representations. On this level, $\mathcal{P}$
permutes the two tensor factors, which is a symmetry of this
R-matrix. In the literature on integrable models, see e.g.\
\cite{Gomez:2005}, this property is known as ``parity invariance'',
hence our name for $\mathcal{P}$.

\section{Unitary Contour from Gluing}
\label{sec:gluing}

We argued in section~\ref{sec:contour} that momentum conservation
\eqref{eq:unitary-constr} naturally \emph{suggests} to equip the
Graßmannian integral \eqref{eq:grass-int-barg}, which computes the
Yangian invariant $\Psi_{2K,K}$, with a $U(K)$ contour for the
integration variable $\mathcal{C}$. Strictly speaking, however, it
does not completely fix the contour. To illustrate this point, we
consider the matrix $\mathcal{C}(\smallunitary)$ in
\eqref{eq:psi63u224-cmatrix} from the six-particle sample invariant
$\Psi_{6,3}$ of $\mathfrak{u}(2,2|0+4)$. It obeys the momentum
condition \eqref{eq:unitary-constr} for any $\smallunitary\in\mathbb{C}$
but is unitary only for $\smallunitary\in U(1)$. Here, working in the
oscillator basis, we \emph{derive} the $U(K)$ contour of the
Graßmannian integral~\eqref{eq:grass-int-unitary} for the
$\mathfrak{u}(p,q|r+s)$ Yangian invariant $|\Psi_{2K,K}\rangle$ by
``gluing'' together multiple copies of $|\Psi_{4,2}\rangle$ with
$U(2)$ contours. There is no reason to question the contour of
$|\Psi_{4,2}\rangle$ because it leads to the correct R-matrix, as
explained in section~\ref{sec:r-formula}. Let us remark that our
gluing procedure gives rise to a parameterization of $U(K)$ in terms
of $U(2)$ matrices that can be traced back to the classic work
\cite{Hurwitz:1897}.

We proceed by induction with the assumption that $|\Psi_{2K,K}\rangle$
is obtained from a Graßmannian integral with $U(K)$ contour. We glue
onto it invariants $|\Psi_{4,2}^{(t)}\rangle$ with $t=1,\ldots,K$ and
show that the result yields a Graßmannian integral formula for
$|\Psi_{2(K+1),K+1}\rangle$ with a $U(K+1)$ contour. This is best
described graphically,
\begin{align}
  \label{eq:glue-graphical}
  \begin{aligned}
    \begin{tikzpicture}
      \draw[thick,rounded corners=10pt]
      (0.0,0) node[below]{$1$} -- (0.0,0.75) -- (2.5,0.75);
      \draw[thick,rounded corners=10pt]
      (3.5,0.75) -- (5.0,0.75) -- (5.0,0) node[below]{$K\!+\!2$};
      \draw[thick,rounded corners=10pt]
      (1.0,0) node[below]{$2$} -- (1.0,1.5);
      \draw[thick,rounded corners=10pt]
      (2.0,0) node[below]{$3$} -- (2.0,1.5);
      \draw[thick,rounded corners=10pt]
      (4,0) node[below]{$K\!+\!1$} -- (4,1.5);
      \draw[thick,rounded corners=10pt]
      (6.0,0) node[below]{$K\!+\!3$} -- (6.0,1.5);
      \draw[thick,rounded corners=10pt]
      (7.5,0) node[below]{$2(K\!+\!1)$} -- (7.5,1.5);
      \node[below] at (1.5,0.75) {$y_2$};
      \node[below] at (2.5,0.75) {$y_3$};
      \node[below] at (3.5,0.75) {$y_K$};
      \node[right] at (1,1.12) {$x_1$};
      \node[right] at (2,1.12) {$x_2$};
      \node[right] at (4,1.12) {$x_K$};
      \node[] at (6.75,0.75) {$\cdots$};
      \node[] at (3,0.75) {$\cdots$};
      \draw (4.25,2.25) 
      node[minimum height=1.5cm,minimum width=7.0cm,draw,
      thick,rounded corners=8pt,densely dotted] 
      {$|\Psi_{2K,K}\rangle$};
    \end{tikzpicture}
  \end{aligned}
  \!\!\!\!
  \propto
  \enspace
  \begin{aligned}
    \begin{tikzpicture}
      \draw[thick,rounded corners=10pt]
      (0.0,0) node[below]{$1$} -- (0.0,1.5);
      \draw[thick,rounded corners=10pt]
      (1.5,0) node[below]{$K\!+\!1$} -- (1.5,1.5);
      \draw[thick,rounded corners=10pt]
      (2.5,0) node[below]{$K\!+\!2$} -- (2.5,1.5);
      \draw[thick,rounded corners=10pt]
      (4.0,0) node[below]{$2(K\!+\!1)$} -- (4.0,1.5);
      \node[] at (0.75,0.75) {$\cdots$};
      \node[] at (3.25,0.75) {$\cdots$};
      \draw (2,2.25) 
      node[minimum height=1.5cm,minimum width=4.5cm,draw,
      thick,rounded corners=8pt,densely dotted] 
      {$|\Psi_{2(K+1),K+1}\rangle$};
    \end{tikzpicture}
  \end{aligned}
  \!\!\!\!\!\!\!.
\end{align}
Here we visualized the $|\Psi_{4,2}^{(t)}\rangle$ as in
\eqref{eq:r-mat-graph} by intersections of lines. Moreover, the labels
$x_i$ and $y_i$ are oscillator site indices associated with internal
lines. Translated into a formula, \eqref{eq:glue-graphical} becomes
\begin{align}
  \label{eq:glue-psi}
  |\Psi_{4,2}^{(1)}\rangle^{\dagger_{x_1}\dagger_{y_2}}
  |\Psi_{4,2}^{(2)}\rangle^{\dagger_{x_2}\dagger_{y_3}}
  \cdots
  |\Psi_{4,2}^{(K)}\rangle^{\dagger_{x_K}}
  |\Psi_{2K,K}\rangle
  \propto
  |\Psi_{2(K+1),K+1}\rangle\,.
\end{align}
On the left-hand side, the invariants are obtained from the unitary
Graßmannian integral formula \eqref{eq:grass-int-unitary} with some
relabeling of the oscillator contractions \eqref{eq:osc-matrix} and
the deformation parameters in the integrands
\eqref{eq:eq:grass-int-unitary-integrand-final} to be in accordance
with \eqref{eq:glue-graphical}. This reads for the invariants
$|\Psi_{2K,K}\rangle$ and $|\Psi_{4,2}^{(t)}\rangle$, respectively,
\begin{align}
  \begin{gathered}
    (i\mathrel{\ooalign{\raisebox{0.7ex}{$\bullet$}\cr\raisebox{-0.3ex}{$\circ$}}} j)
    \mapsto
    (x_i\mathrel{\ooalign{\raisebox{0.7ex}{$\bullet$}\cr\raisebox{-0.3ex}{$\circ$}}} j+2)\,,
    \quad
    \begin{array}{l}
      (v_i,c_i)\mapsto(v_{i+1},c_{i+1})
    \end{array}
    \quad\text{for}\quad
    \begin{array}{l}
      i=1,\ldots,K\,,\\
      j=K+1,\ldots,2K\,,
    \end{array}
    \\
    \begin{pmatrix}
      (1\mathrel{\ooalign{\raisebox{0.7ex}{$\bullet$}\cr\raisebox{-0.3ex}{$\circ$}}} 3)&
      (1\mathrel{\ooalign{\raisebox{0.7ex}{$\bullet$}\cr\raisebox{-0.3ex}{$\circ$}}} 4)\\
      (2\mathrel{\ooalign{\raisebox{0.7ex}{$\bullet$}\cr\raisebox{-0.3ex}{$\circ$}}} 3)&
      (2\mathrel{\ooalign{\raisebox{0.7ex}{$\bullet$}\cr\raisebox{-0.3ex}{$\circ$}}} 4)\\
    \end{pmatrix}
    \mapsto
    \begin{pmatrix}
      (y_t\mathrel{\ooalign{\raisebox{0.7ex}{$\bullet$}\cr\raisebox{-0.3ex}{$\circ$}}} y_{t+1})&
      (y_t\mathrel{\ooalign{\raisebox{0.7ex}{$\bullet$}\cr\raisebox{-0.3ex}{$\circ$}}} x_t)\\
      (t+1\mathrel{\ooalign{\raisebox{0.7ex}{$\bullet$}\cr\raisebox{-0.3ex}{$\circ$}}} y_{t+1})&
      (t+1\mathrel{\ooalign{\raisebox{0.7ex}{$\bullet$}\cr\raisebox{-0.3ex}{$\circ$}}} x_t)\\
    \end{pmatrix},
    \quad
    \begin{array}{l}
      (v_1,c_1)\quad\text{unchanged}\,,\\
      (v_2,c_2)\mapsto(v_{t+1},c_{t+1})\,
    \end{array}
  \end{gathered}
\end{align}
with $y_1\equiv 1$ and $y_{K+1}\equiv K+2$. The right-hand side of
\eqref{eq:glue-psi} is given by the Graßmannian formula
\eqref{eq:grass-int-unitary} without any replacements. In what
follows, we prove that both sides are indeed proportional.

We begin by manipulating the left-hand side of
\eqref{eq:glue-psi}. The oscillators with site indices $x_i$ and $y_i$
appear only inside of vacuum expectation values because they are
associated with internal lines in \eqref{eq:glue-graphical}. We
eliminate them by calculating these expectation values. What is more,
we combine the integrands of the $U(2)$ integrals and the $U(K)$
integral. This yields
\begin{align}
  \label{eq:glue-lhs-vevs}
  \int\displaylimits_{U(2)}\!\![\D\mathcal{D}^{(1)}]\,
  \cdots\!\!\!
  \int\displaylimits_{U(2)}\!\![\D\mathcal{D}^{(K)}]\,
  \!\!\!\!
  \int\displaylimits_{U(K)}\!\![\D\mathcal{D}]\,\,
  \big|\mathcal{D}^{(2)}_{13}\big|^{2}\cdots
  \big|\mathcal{D}^{(K)}_{13}\big|^{2(K-1)}
  \mathscr{F}(\mathcal{C})
  (\det\mathcal{C})^r
  e^{\tr(\mathbf{I}_\bullet\mathcal{C}^\dagger+\mathcal{C}\mathbf{I}_\circ^t)}
  |0\rangle\,,
\end{align}
where we denote the integration variables originating from
$|\Psi_{2K,K}\rangle$ and $|\Psi_{4,2}^{(t)}\rangle$ by $\mathcal{D}$
and $\mathcal{D}^{(t)}$, respectively. The integrand
$\mathscr{F}(\mathcal{C})$ and the exponential with the $U(K+1)$
matrix
\begin{align}
  \label{eq:glue-c-vevs}
  \mathcal{C}=
  \left(
  \begin{array}{c:c}
    \begin{matrix}
      \mathcal{D}^{(1)}_{13}&\mathcal{D}^{(1)}_{14}\\[0.3em]
      \mathcal{D}^{(1)}_{23}&\mathcal{D}^{(1)}_{24}\\
    \end{matrix}&0\\[1.4em]
    \hdashline&\\[-1.0em]
    0&1_{K-1}\vphantom{\mathcal{D}^{(X)}_{XX}}\\
  \end{array}
  \right)
  \cdots
  \left(
  \begin{array}{c:c:c}
    \mathcal{D}^{(K)}_{13}&0&\mathcal{D}^{(K)}_{14}\\[0.3em]
    \hdashline&&\\[-1.0em]
    0&1_{K-1}&0\\[0.3em]
    \hdashline&&\\[-1.0em]
    \mathcal{D}^{(K)}_{23}&0&\mathcal{D}^{(K)}_{24}\\
  \end{array}
  \right)
  \left(
  \begin{array}{c:c}
    1\vphantom{\mathcal{D}^{(X)}_{XX}}&0\\[0.3em]
    \hdashline&\\[-1.0em]
    0&\begin{matrix}
      \\[-0.3em]
      \mathmakebox[4.2em][c]{\mathcal{D}}\\[1.0em]
    \end{matrix}\\
  \end{array}
  \right)\,
\end{align}
are already those of $|\Psi_{2(K+1),K+1}\rangle$. Therefore we can
focus on the Haar measures in \eqref{eq:glue-lhs-vevs} next. Starting
out a bit more general, we introduce the complex \emph{Stiefel
  manifold} $V_L(\mathbb{C}^M)$ with $L\leq M$ which is the set of
$M\times L$ matrices $\mathcal{S}$ satisfying
$\mathcal{S}^\dagger\mathcal{S}=1_L$. It generalizes the unitary group
manifold to non-square matrices as $V_M(\mathbb{C}^M)=U(M)$. Its Haar
measure can be realized as
\begin{align}
  \label{eq:glue-stiefel}
  \int\displaylimits_{V_L(\mathbb{C}^M)}\!\!\!\![\D\mathcal{S}]
  \propto
  \int\displaylimits_{\mathbb{C}^{LM}}\!\!
  \frac{\D^{\,LM}\!\mathcal{S}^\dagger\!\D^{\,LM}\!\mathcal{S}}{(2i)^{LM}}\,
  \delta^{L^2}(\mathcal{S}^{\dagger}\mathcal{S}-1_L)\,,
\end{align}
where the delta function of a Hermitian matrix is defined as the
product of real delta functions of its diagonal elements times complex
delta functions of the upper triangular ones. We implement the Haar
measures $[\D\mathcal{D}^{(t)}]$ in \eqref{eq:glue-lhs-vevs} as in
\eqref{eq:glue-stiefel} and eliminate degrees of freedom using the
delta functions. The factors $\big|\mathcal{D}^{(t)}_{13}\big|^{2}$ in
\eqref{eq:glue-lhs-vevs} cancel with factors arising from these
measures. Further degrees of freedom can be integrated out because the
matrices $\mathcal{D}^{(t)}$ appear only in one particular combination
inside of $\mathcal{C}$ in \eqref{eq:glue-c-vevs}. Yet others can be
absorbed into $\mathcal{D}$ using the left-invariance of
$[\D\mathcal{D}]$. As a result, \eqref{eq:glue-lhs-vevs} becomes
\begin{align}
  \label{eq:glue-lhs-coset}
  \int\displaylimits_{V_1(\mathbb{C}^{K+1})}\!\!\!\!\!\!\![\D\mathcal{S}]\,
  \!\!
  \int\displaylimits_{U(K)}\!\![\D\mathcal{D}]\,\,
  \mathscr{F}(\mathcal{C})(\det\mathcal{C})^r
  e^{\tr(\mathbf{I}_\bullet\mathcal{C}^\dagger+\mathcal{C}\mathbf{I}_\circ^t)}
  |0\rangle\,,
\end{align}
where we neglected a numerical prefactor, and we have
\begin{align}
  \label{eq:glue-c-coset}
  \mathcal{C}=
  \left(
  \begin{array}{c:c}
    \\[-0.35em]
    \!\!\mathcal{S}&\mathcal{T}(\mathcal{S})\!\!\\[0.95em]
  \end{array}
  \right)
  \left(
  \begin{array}{c:c}
    1&0\\[0.3em]
    \hdashline&\\[-1.0em]    
    0&\mathcal{D}\\
  \end{array}
  \right)\in U(K+1)\,.
\end{align}
The first factor in $\mathcal{C}$ stems from multiplying the first $K$
factors in \eqref{eq:glue-c-vevs}, and the $K+1\times K$ matrix
$\mathcal{T}(\mathcal{S})$ is completely determined in terms of
$\mathcal{S}$. The Stiefel manifold in \eqref{eq:glue-lhs-coset} is
just a unit sphere and can also be interpreted as a coset space,
$V_1(\mathbb{C}^{K+1})\cong S^{2K+3}\cong U(K+1)/U(K)$. The first
factor in \eqref{eq:glue-c-coset} is a representative of the coset
space element $\mathcal{S}$ in $U(K+1)$. With this interpretation, the
product of the two integrals in \eqref{eq:glue-lhs-coset} equals
$\int\displaylimits_{U(K+1)}[\D\mathcal{C}]$, see e.g.\ the
explanation around equation (5.121) in \cite{Gilmore:2012}. Therefore
\eqref{eq:glue-lhs-coset} is nothing but the right-hand side of
\eqref{eq:glue-psi}. Q.E.D.

\bibliography{literature}{}
\bibliographystyle{utphys}

\end{document}